\documentclass[prd,11pt,onecolumn,preprintnumbers,superscriptaddress,tightenlines,nofootinbib,letterpaper,longbibliography,twoside]{revtex4-2}
\usepackage{array}[=2016-10-06]

\makeatletter
\renewcommand{\thesection}{\arabic{section}}
\renewcommand{\p@section}{}
\renewcommand{\thesubsection}{\thesection.\arabic{subsection}}
\renewcommand{\p@subsection}{}
\renewcommand{\thesubsubsection}{\thesubsection.\arabic{subsubsection}}
\renewcommand{\p@subsubsection}{}
\def\appendix{%
    \par
    \setcounter{section}\z@
    \setcounter{subsection}\z@
    \setcounter{subsubsection}\z@
    \def\thesubsection{\thesection.\arabic{subsection}}%
    \def\thesubsubsection{\thesubsection.\arabic{subsubsection}}%
    \def\p@subsection{}%
    \def\p@subsubsection{}%
    \@addtoreset{equation}{section}%
    \def\theequation@prefix{\thesection}%
    \addtocontents{toc}{\protect\appendix}%
    \@ifstar{%
    \def\thesection{\unskip}%
    \def\theequation@prefix{A.}%
    }{%
    \def\thesection{\Alph{section}}%
    }%
}%
\makeatother

\usepackage{tikz}
\usepackage{amsmath,amssymb} 
\usepackage{bbm}
\usepackage{bm}
\usepackage{graphicx} 
\usepackage[export]{adjustbox} 
\usepackage{url} 
\usepackage[dvipsnames]{xcolor} 
\usepackage{siunitx} 

\usepackage{float}
\usepackage{microtype}
\usepackage{physics}
\usepackage[normalem]{ulem}
\usepackage{enumitem}

\usepackage{hyperref} 
\usepackage[capitalize]{cleveref} 
\newcommand\myshade{80} 
\colorlet{mylinkcolor}{ForestGreen}
\colorlet{mycitecolor}{Red}
\colorlet{myurlcolor}{violet}
\hypersetup{
  linkcolor  = mylinkcolor!\myshade!black,
  citecolor  = mycitecolor!\myshade!black,
  urlcolor   = myurlcolor!\myshade!black,
  colorlinks = true
}

\usepackage{booktabs}
\usepackage{tabularx}
\usepackage{ragged2e}
\newcolumntype{P}[1]{>{\RaggedRight\arraybackslash}p{#1}}
\renewcommand{\arraystretch}{1.5}

\newcommand{\Li}[2]{\operatorname{Li}_{#1}\!\left(#2\right)}
\newcommand{\SLi}[2]{S_{#1}\!\left(#2\right)}

\newcommand{\dr}{\mathrm{d}r\,}
\newcommand{\ds}{\mathrm{d}s\,}

\newcommand{\dOmega}{\mathrm{d}\Omega\,}
\newcommand{\Omegar}{\Omega_r}
\newcommand{\Omegas}{\Omega_s}
\newcommand{\dOmegar}{\mathrm{d}\Omegar\,}
\newcommand{\dOmegas}{\mathrm{d}\Omegas\,}
\newcommand{\rvec}{\bm{r}}
\newcommand{\drth}{\mathrm{d}^3\rvec}
\newcommand{\svec}{\mathbf{s}}
\newcommand{\dsth}{\mathrm{d}^3\svec}
\newcommand{\sphY}[3]{Y_{#1}^{#2}(#3)}
\newcommand{\sphYdef}[2]{Y_{#1}^{#2}}
\newcommand{\cartY}[2]{\mathcal{Y}^{#2}_{i_1\ldots i_{#1}}}

\newcommand{\Th}{$^{229}$Th }

\newcommand{\gs}{{\mathrm{g}}}
\newcommand{\m}{{\mathrm{m}}}
\newcommand{\Isp}{I}

\newcommand{\Hamilhat}{\hat{\mathcal{H}}}
\newcommand{\krank}{{(k)}}

\newcommand{\Qhat}[2]{\hat{Q}^{(#1)}_{#2}}
\newcommand{\That}[2]{\hat{T}^{(#1)}_{#2}}
\newcommand{\rhohat}{\hat{\rho}}

\newcommand{\Vcart}[1]{V_{\langle {#1} \rangle}}
\newcommand{\Vcartz}[1]{\Vcart{\underbrace{z\ldots z}_{#1}}}
\newcommand{\Vcarti}[1]{\Vcart{i_1\ldots i_{#1}}}

\newcommand{\Mcart}[1]{\hat{M}_{\langle {#1} \rangle}}
\newcommand{\Mcarti}[1]{\Mcart{i_1\ldots i_{#1}}}
\newcommand{\Mcartz}[1]{\Mcart{\underbrace{z\ldots z}_{#1}}}

\newcommand{\Qcart}[1]{\hat{Q}_{#1}}
\newcommand{\Qcarti}[1]{\Qcart{i_1\ldots i_{#1}}}
\newcommand{\Qcartz}[1]{\Qcart{\underbrace{z\ldots z}_{#1}}}
\newcommand{\Qcartzval}[1]{Q_{\underbrace{z\ldots z}_{#1}}}
\newcommand{\Qcartspec}[2]{Q_{\underbrace{z\ldots z}_{#1}}^{(#2)}}

\newcommand{\Vcartzz}{V_{\langle zz\rangle}}
\newcommand{\Mcartzz}{\Mcart{zz}}
\newcommand{\Qcartzzspec}[1]{Q_{zz}^{(#1)}}
\newcommand{\Qcartzz}{\Qcart{zz}}

\newcommand{\Vcartzzzz}{V_{\langle zzzz \rangle}}
\newcommand{\Mcartzzzz}{\Mcart{zzzz}}
\newcommand{\Qcartzzzzspec}[1]{Q_{zzzz}^{(#1)}}
\newcommand{\Qcartzzzz}{\Qcart{zzzz}}

\newcommand{\Qsph}[1]{\mathcal{Q}_{#1}}              
\newcommand{\Qsphspec}[2]{\mathcal{Q}_{#1}^{(#2)}}  

\newcommand{\Kband}{{[K]}}
\newcommand{\Gband}{{[\mathrm{G}]}}
\newcommand{\Mband}{{[\mathrm{M}]}}

\newcommand{\chrsq}{\langle r^2 \rangle}
\newcommand{\tilchrsq}{\langle \tilde{r}^2 \rangle}
\newcommand{\RQ}{\mathcal{R}_{Q}}

\newcommand{\KNP}{\mathcal{K}}

\newcommand{\fm}{{\mathrm{fm}}}

\newcommand{\rhoo}{\rho_0}
\newcommand{\rhoii}{\rho_0^{(2)}}
\newcommand{\Roo}{R_0}
\newcommand{\Roii}{R_0^{(2)}}
\newcommand{\cO}{\mathcal{O}}
\newcommand{\cP}{\mathcal{P}}

\newcommand{\zerosym}{\mathcal{S}_0}
\newcommand{\twosym}{\mathcal{S}_2}
\newcommand{\tetrasym}{\mathcal{S}_4}
\newcommand{\cubsym}{\mathcal{S}_\Box}

\newcommand{\Ttwo}{T_2}
\newcommand{\Tzero}{T_0}

\renewcommand{\vec}[1]{{\boldsymbol{#1}}}
\newcommand{\iso}[2]{{\ensuremath{{}^{#2}}\ensuremath{\rm #1}}}

\newcommand{\orcid}[1]{\begingroup
  \hypersetup{hidelinks}\href{https://orcid.org/#1}{\includegraphics[width=10pt]{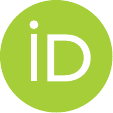}} \endgroup}

\makeatletter
\providecommand*{\diff}%
	{\@ifnextchar^{\DIfF}{\DIfF^{}}}
\def\DIfF^#1{%
	\mathop{\mathrm{\mathstrut d}}%
		\nolimits^{#1}\gobblespace}
\def\gobblespace{%
	\futurelet\diffarg\opspace}
\def\opspace{%
	\let\DiffSpace\!%
	\ifx\diffarg(%
		\let\DiffSpace\relax
	\else
		\ifx\diffarg[%
			\let\DiffSpace\relax
		\else
  			\ifx\diffarg\{%
				\let\DiffSpace\relax
			\fi\fi\fi\DiffSpace}

\begin{document}

\title{On the Structure of the Nuclear-Clock Isotope $^{229}$Th Beyond Quadratic Order}

\author{Chaja~Baruch \orcid{0000-0003-0625-0724}}
\affiliation{Physics Department, Technion–Israel Institute of Technology, Haifa 3200003, Israel}

\author{Doron~Gazit \orcid{0000-0002-0350-3266}}
\affiliation{Racah Institute of Physics, The Hebrew University,
             9190401 Jerusalem, Israel}

\author{Hans-Werner~Hammer \orcid{0000-0002-2318-0644}}
\affiliation{Technische Universit\"at Darmstadt, Department of Physics,
64289 Darmstadt, Germany}
\affiliation{Helmholtz Forschungsakademie Hessen für FAIR (HFHF) and Extreme Matter Institute EMMI, GSI Helmholtzzentrum für Schwerionenforschung GmbH, 64291 Darmstadt, Germany}

\author{Fiona~Kirk \orcid{0000-0002-2234-5216}}
\email{fiona.kirk@weizmann.ac.il}
\affiliation{Department of Particle Physics and Astrophysics,
             Weizmann Institute of Science, Rehovot 761001, Israel}

\author{Joachim~Kopp \orcid{0000-0003-0600-4996}}
\affiliation{PRISMA${}^{++}$ Cluster of Excellence \& Mainz Institute for
             Theoretical Physics, \\
             Johannes Gutenberg University, 55099 Mainz, Germany}

\author{Gil~Paz \orcid{0000-0001-5605-2043}}
\affiliation{Department of Physics and Astronomy, Wayne State University,
             Detroit, Michigan 48201, USA}

\author{Ekkehard~Peik \orcid{0000-0001-6309-2975}}
\affiliation{Physikalisch-Technische Bundesanstalt (PTB), 38116 Braunschweig, Germany}

\author{Gilad~Perez \orcid{0000-0002-3878-1821}}
\affiliation{Department of Particle Physics and Astrophysics,
             Weizmann Institute of Science, Rehovot 761001, Israel}

\author{Yotam~Soreq \orcid{0000-0003-4515-5789}}
\affiliation{Physics Department, Technion–Israel Institute of Technology, Haifa 3200003, Israel}

\author{Konstantin~Springmann \orcid{0000-0002-9617-6136}}
\affiliation{Department of Particle Physics and Astrophysics,
             Weizmann Institute of Science, Rehovot 761001, Israel}

\date{\today}

\preprint{MITP-26-062,  WSU-HEP-2602} 


\begin{abstract}
Precision measurements of the exceptionally low-energy nuclear transition of $^{229}$Th, offer an extraordinary sensitivity to a variety of new light fields. 
However, interpreting such measurements requires nuclear-structure information beyond what is currently constrained by spectroscopy.
We develop a systematic description of the electric multipole interactions governing the hyperfine structure of ${}^{229}$Th, including the hexadecapole contribution, and perform a symmetry-based analysis of the resulting hyperfine splittings in crystalline environments.
Hyperfine spectroscopy generally constrains combinations of nuclear moments and crystal-field gradients, leading to degeneracies when higher multipole interactions are allowed. 
We then combine experimental nuclear observables, particle--rotor relations, and a deformed-Fermi charge-density model to investigate the electromagnetic contribution to the nuclear clock's sensitivity to variations in the fine-structure constant.
We find that higher-order nuclear deformation and state-dependent differences in the radial charge distribution of the clock states can substantially alter the predicted sensitivity, introducing large uncertainties that are not constrained by existing measurements.
Our results highlight the importance of combining nuclear and solid-state theory with complementary measurements, including spectroscopy of higher-spin states, to constrain the nuclear structure and improve predictions of the clock's sensitivity to new physics.
Our symmetry-based analysis also provides a framework for using hyperfine spectroscopy to investigate local defect symmetries in crystalline hosts.
\end{abstract}


\maketitle 

\newpage
\makeatletter
\def\l@subsubsection#1#2{}
\makeatother

\tableofcontents

\newpage 

\section{Introduction}
The first laser excitation of the \Th isomer has transformed the nuclear clock from a distant dream into a platform for precision spectroscopy~\cite{Tiedau:2024obk, Zhang:2024ngu}. 
In particular, spectroscopy of \Th embedded in CaF$_2$ has resolved the hyperfine structure of the nuclear transition and determined its frequency relative to an atomic clock~\cite{Zhang:2024ngu}. 
The immediate challenge is now to understand and control the interaction between the nucleus and its solid-state environment well enough to obtain narrow, reproducible clock lines. 
Recent measurements have begun to characterize the dependence of the transition frequency and linewidth on temperature, Th concentration, and the local crystal environment~\cite{Higgins:2024fir, Ooi:2025hmd, Hiraki:2025uur, morawetz:2026}.

The hyperfine splitting of the clock transition, arising from the coupling of nuclear multipole moments to the electric field gradients~(EFGs) generated by the host crystal, provides a valuable probe of the nuclear charge distribution.
So far, the focus was primarily on the mean-square charge radius and the electric quadrupole moment~\cite{Zhang:2024ngu}. 
Here, we also explore higher electric multipole interactions that are accessible through hyperfine spectroscopy. 
Neglecting magnetic contributions, parity forbids static electric dipole and octupole moments, while angular-momentum selection rules forbid static spectroscopic multipole moments of rank $k > 2I$, where $I$ is the nuclear spin.
This leaves the hexadecapole as the only additional allowed electric multipole, contributing to the $5/2^+$ ground state but not to the $3/2^+$ isomer.

The intrinsic hexadecapole deformation of the \Th ground-state rotational band has already been constrained by Coulomb-excitation measurements~\cite{CEBemis_1988}.
Hyperfine spectroscopy offers a complementary route to probing nuclear hexadecapole moments, although their weak coupling to the surrounding electromagnetic environment poses considerable experimental challenges.
An early search for hexadecapole splitting in a solid-state system using ${}^{181}$Ta NMR in nominally cubic KTaO${}_3$ was hindered by residual quadrupole broadening~\cite{Doering:1986}. 
More recently, however, precision hyperfine spectroscopy of an individual ${}^{93}$Nb nuclear spin yielded a spectrum consistent with a possible hexadecapole interaction~\cite{Travesedo:2026zeh}.

Since the measured splittings depend both on nuclear moments and crystal-field gradients, extracting nuclear-structure information also requires an understanding of the local crystal environment and its symmetries.
Recent measurements in CaF${}_2$ have identified distinct Th centers, including sites with strongly suppressed EFGs~\cite{Hiraki:2025uur,morawetz:2026}. 
This motivates us to investigate the hyperfine splitting patterns associated with different crystal symmetries and the nuclear-structure information that can be extracted from them.
Using such crystals, three independent teams have successfully realized the world's first operational nuclear clocks by stabilizing a vacuum-ultraviolet laser to the low-energy nuclear transition of $^{229}$Th \cite{ToscaniDeCol:2026a,Huang:2026a,Wang:2026}.

Beyond its importance for developing a stable and precise nuclear clock, understanding the nuclear-structure of \Th is also essential for estimating its sensitivity to physics beyond the Standard Model.
The exceptionally low $\sim \SI{8}{eV}$ excitation energy of \Th is believed to arise from a near cancellation between much larger electromagnetic and strong contributions.
A perturbation that affects these contributions differently, for example a variation of Standard-Model parameters induced by ultralight dark matter coupling to the strong nuclear sector, can produce an enhanced shift in the clock frequency~\cite{FlambaumOrig,Peik:2020cwm,Beeks:2024xnc,Caputo:2024doz,Fuchs:2024xvc}, making nuclear clocks exceptionally sensitive probes of a broad range of new physics~\cite{Arvanitaki:2014faa, Arakawa:2026mls,ToscaniDeCol:2026a,Delaunay:2025lgk,Kim:2022ype}.

Quantifying this enhancement requires theoretical input on the electromagnetic or strong contributions to the nuclear transition energy, since these cannot be measured separately.
The electromagnetic contribution is generally more tractable and can be estimated by combining nuclear observables with models of the nuclear charge distribution.
Existing analyses have relied principally on the measured change of the charge radius and quadrupole moment between the two states~\cite{Berengut:2009zz,Safronova:2018fue,Beeks:2024xnc,Caputo:2024doz}.
The resulting sensitivity can be very large, with recent estimates giving  $10^4$~\cite{Caputo:2024doz} and $K_\alpha =5900\,(2300)$~\cite{Beeks:2024xnc}.
However, unlike the coupling of higher nuclear multipoles to the external crystal field, their contribution to the internal Coulomb energy is not suppressed by the large separation between nuclear and crystal length scales.
Consequently, higher-order deformation may contribute substantially to the Coulomb-energy difference, even when its spectroscopic effects are too small to be resolved.
Determining the sensitivity of the nuclear clock thus requires additional experimental constraints and theoretical input on the intrinsic nuclear structure.

This paper is organized as follows. 
We begin by developing the general electric multipole formalism for the \Th clock states (ground state and isomer) in \cref{sec:general_multipole_formalism}, before investigating the role of crystal symmetry and interpreting existing hyperfine measurements in  \cref{sec:crystal_sym}.
In \cref{sec:theory}, we review complementary theoretical descriptions of nuclear structure and its interaction with external electric fields, including NRQED (non-relativistic quantum electrodynamics), nuclear energy-density-functional calculations, a halo-inspired description, the particle--rotor model, and a deformed Fermi charge distribution. 
In the latter approach, we use experimental input to constrain the charge distribution, with the particle--rotor model providing the connection between its intrinsic multipole moments and the measured spectroscopic moments.
In \cref{sec:K} we examine the implications of these approaches for the clock's sensitivity to new physics. 
Our main quantitative estimate is based on the deformed Fermi charge distribution, whose parameters are constrained by experimental input using the particle--rotor model to relate intrinsic and spectroscopic multipole moments.
We also assess the uncertainties associated with this description.
We summarize our conclusions in \cref{sec:Summary}.

\section{General Multipole Formalism}
\label{sec:general_multipole_formalism}

In this section we develop the general multipole formalism for the hyperfine structure of the \Th clock states. 
We construct the rank-$k$ multipole Hamiltonians (\cref{sec:H_multipole}), define the spectroscopic moments (\cref{sec:spectroscopic_multipoles}), relate the spectroscopic and intrinsic moments (\cref{sec:rotorspec}), and
apply the formalism to the nuclear clock states (\cref{sec:clockhamiltonians}).

\subsection{Electric Multipole Hamiltonians}
\label{sec:H_multipole}

The electrostatic interaction between the charge distribution of a nucleus with spin $I$ and the surrounding crystal field can be described by a multipole expansion of the electrostatic potential about the nuclear position.
The interaction Hamiltonian takes the form, see, e.g.~\cite{Schwartz:1954zz},
\begin{align}
    \label{eq:MultipolExp}
    \Hamilhat
    =
    \sum_{k=0}^{\infty} \Hamilhat^\krank \, ,
\end{align}
where each term couples nuclear and crystal tensors of equal rank. 
The rank-$k$ contribution follows from the Taylor expansion and is given by,
\begin{align}
    \label{eq:Hcart}
    \Hamilhat^\krank 
    = 
    \frac{1}{k!} \Vcarti{k}\Mcarti{k}\,,
\end{align}
where $\Vcarti{k}$ and $\Mcarti{k}$ are the symmetric traceless parts of the tensors
\begin{align}
    \label{eq:MEFG}
    V_{i_1\ldots i_k} 
    \equiv 
    \left.\frac{\partial^k V(\rvec)}{\partial x_{i_1}\cdots \partial x_{i_k}}\right\vert_{\rvec=0}\,,
    \qquad\qquad
    \hat{M}_{i_1\ldots i_k} 
    \equiv 
    \int\!\drth \rhohat(\rvec) \, x_{i_1}\cdots x_{i_k}\,.
\end{align}
$V_{i_1\ldots i_k}$ is the rank-$k$ EFG tensor generated by the crystal environment and $\hat{M}_{i_1\ldots i_k}$ is the corresponding Cartesian moment of the nuclear charge-density operator $\rhohat(\rvec)$ with $\int\!\drth \rhohat(\rvec)=Z$.
The matrix elements of the operator $\hat{M}_{i_1\ldots i_k}$ in a state of definite angular momentum $I$ are fixed by the angular-momentum algebra after projecting onto its irreducible rank-$k$ component (see \cref{app:multipoletranslations}).

A simple dimensional estimate suggests that the contribution of higher multipoles decreases rapidly because the crystal field varies on length scales comparable to the crystal lattice spacing, $s$, whereas the nuclear charge distribution extends only over distances of order $\sqrt{\chrsq}$, the nuclear charge radius.
Estimating the crystal EFG as $V_{i_1\ldots i_k} \sim e Z_{\rm eff} / s^{k+1}$, where $Z_{\rm eff}$ characterizes the local electrostatic potential, and assuming nuclear multipole matrix elements of the orders $e Z_{\rm Th} \chrsq^{k/2}$, we obtain, up to angular momentum factors,
\begin{align}
    \label{eq:multipole_scaling}
    \left. \vert \bra{I,m}\Hamilhat^\krank \ket{I,m}\right.\vert
    \sim 
    \frac{e^2 Z_{\rm Th} Z_{\rm eff}}{s} \bigg( \frac{\chrsq^\frac12}{s} \bigg)^k\,.
\end{align}
For CaF${}_2$, we take the characteristic length scale of the local crystal environment to be approximately half the lattice constant, $s \sim a/2\approx 3\,\mathrm{\AA}$~\cite{Rizescu2018}, while the charge radius of ${}^{229}$Th is $\chrsq\approx (6\,\fm)^2$~\cite{Angeli:2013epw}, yielding an expansion parameter $\sqrt{\chrsq}/s \sim 2 \times 10^{-5}$.

To make the angular-momentum structure of the interaction explicit, it is convenient to express the interaction in terms of irreducible spherical tensors,
\begin{align}
    \label{eq:Hsph}
    \Hamilhat^\krank 
    = 
    \Qsph{k}
    \sum_{q=-k}^k (-1)^q V_q^\krank \That{k}{-q} \,,
\end{align}
where $V_q^\krank$ are the spherical components of the EFG tensor.
The Taylor factor $1/k!$ appearing explicitly in the Cartesian form \cref{eq:Hcart} has been absorbed into the normalization of the spherical EFG components $V_q^\krank$.
The operators $\That{k}{q}$ are dimensionless irreducible tensor operators carrying the angular-momentum structure of the interaction, while $\Qsph{k}$ contains the dimensions and nucleus-dependent multipole information.
(See \cref{app:multipoletranslations} for the relation between the spherical and Cartesian components, and \cref{tab:paramdefs} for a summary of introduced parameters.)

The matrix elements of a rank-$k$ spherical tensor operator, which has $2k+1$ components $q=-k,\ldots, k$, between angular-momentum eigenstates obey the Wigner--Eckart theorem~\cite{Wigner1993,Racah:1942gsc,Edmonds:1955fi}, 
\begin{align}
    \label{eq:WignerEckart}
    \bra{I'm'} \That{k}{q} \ket{I m}
    = 
    (-1)^{I'-m'}
    \langle I' || \That{k}{}|| I\rangle 
    \begin{pmatrix}
        I' & k & I\\
        -m' & q & m
    \end{pmatrix}\,,
\end{align}
which separates the angular dependence, described by the Wigner $3j$-symbol, from the reduced matrix element $\langle I' || \That{k}{} || I\rangle$ carrying the information about nuclear structure. 
The triangle rule $|I'-I| \leq k \leq I'+I$ which the $3j$-symbols obey bounds the rank; within a spin-$I$ state it gives $k \leq 2I$, so all static moments of rank $k>2I$ vanish.

\subsection{Spectroscopic Multipole Moments}
\label{sec:spectroscopic_multipoles}

The spectroscopic multipole moment is the nuclear factor that, together with the electric field gradient, determines the hyperfine shift of the stretched state, i.e., the state of maximal projection $m=I$ onto the quantization axis.
By the angular-momentum selection rules of the Wigner $3j$-symbol in \cref{eq:WignerEckart}, only the $q=0$ term in \cref{eq:Hsph} contributes to the matrix element $\bra{I,I} \Hamilhat^\krank \ket{I,I}$. 
The rank-$k$ spherical spectroscopic multipole moment, $\Qsphspec{k}{\Isp}$, is now defined through
\begin{align}
    \label{eq:Qspect-Qintrinsic-relation}
    \bra{I,I} \Hamilhat^\krank \ket{I,I} 
    \equiv 
    \Qsphspec{k}{\Isp} V_0^\krank \, .
\end{align}
Factoring out the electric field gradients and applying the Wigner--Eckart theorem, it can also be expressed as (see also Eq.~(3--25) in Bohr and Mottelson, Vol.~1~\cite{BohrMottelsonVol1})
\begin{align}
    \label{eq:Qsphspecdef}
    \Qsphspec{k}{\Isp} 
    = \Qsph{k} \, \bra{I, I} \That{k}{q=0} \ket{I, I}
    = \Qsph{k} \, \langle I || \That{k}{} ||I \rangle 
    \begin{pmatrix}
         I & k & I \\
        -I & 0 & I
    \end{pmatrix}\,.
\end{align}
In the literature, spectroscopic multipole moments are often quoted not in terms of the spherical moments $\Qsphspec{k}{\Isp}$, but in terms of the \emph{Cartesian spectroscopic multipole moments}, which are defined as the stretched-state expectation values of the symmetric traceless nuclear moments defined in \cref{eq:MEFG}:
\begin{align}
    Q^{(I)}_{z \ldots z}
    &\equiv 
    c_k \bra{I,I} \Mcart{z \ldots z} \ket{I,I} \, .
\end{align}
The prefactors $c_k$ are convention-dependent. 
They are usually chosen such that the polynomial under the integral defining $\Mcart{z \ldots z}$ (\cref{eq:MEFG}) has integer coefficients. 
For the quadrupole and hexadecapole, this leads to
\begin{align}
    \Qcartzzspec{I}
    &\equiv 3\bra{I,I}\Mcartzz\ket{I,I}\,,
    &
    \Qcartzzzzspec{I}
    &\equiv 35\bra{I,I}\Mcartzzzz\ket{I,I}\,.
\end{align}
As shown in \cref{app:multipoletranslations}, the relations between the Cartesian and spherical spectroscopic multipole moments are
\begin{align}
    \label{eq:Qcarts24def}
    \Qcartzzspec{I}
    &=\sqrt{\frac{16\pi}{5}}\Qsphspec{2}{I}\,,
    &
    \Qcartzzzzspec{I}
    &=\frac{16\sqrt{\pi}}{3}\Qsphspec{4}{I}\,.
\end{align}
The corresponding relation between the spherical and Cartesian EFG tensors is (see \cref{app:multipoletranslations})
\begin{align}
    V_0^\krank = \frac{1}{k!} \sqrt{\frac{4\pi}{2k+1}} \Vcartz{k}\,.
\end{align}
Specifically,
\begin{align}
    V_0^{(2)} = \sqrt{\frac{\pi}{5}}\Vcartzz\,, \qquad\qquad
    V_0^{(4)} = \frac{\sqrt{\pi}}{36} \Vcartzzzz\,.
\end{align}
Combining these with \cref{eq:Qcarts24def}, the stretched-state interaction can equivalently be written in the spherical or Cartesian convention as
\begin{align}
    \langle I,I | \Hamilhat^{(2)}|I,I\rangle 
    &=
    \Qsphspec{2}{I}V_0^{(2)}
    = \frac{1}{4} \Qcartzzspec{I} \Vcartzz \, ,
\end{align}
\begin{align}\label{eq:streched}
    \langle I,I | \Hamilhat^{(4)}|I,I\rangle 
    &=
    \Qsphspec{4}{I}V_0^{(4)}
    = \frac{1}{192} \Qcartzzzzspec{I} \Vcartzzzz \, .
\end{align}
%

\subsection{Spectroscopic Multipole Moments in Rotational Bands}
\label{sec:rotorspec}

Up to this point, no assumption has been made about the internal structure of the nucleus.
We now consider an axially symmetric nucleus described by the particle--rotor model (see \cref{sec:particle-rotor,app:rotormultipoles} for details), which leads to an additional quantum number $K$ defining the rotational bands. 
$K$ is the projection of the nucleus spin ($I$) on the nucleus symmetry axis. 
Each band consists of a tower of rotational states, labeled by the spin quantum number $I$. 
For each rotational band, the Hamiltonian may be decomposed into multipoles as before:
\begin{align}
    \label{eq:MultipolExpRotor}
    \Hamilhat_K 
    = 
    \sum_{k=0}^\infty \Hamilhat_K^\krank\,, \quad\text{with}\quad 
    \Hamilhat_K^\krank 
    = \Qsph{k}^\Kband \sum_{q=-k}^k (-1)^q V_q^\krank \That{k}{-q}\,.
\end{align}
where $\Qsph{k}^\Kband$ is the intrinsic moment of the band $\Kband$ and $\That{k}{-q}$ is the spin tensor containing the angular-momentum factors.

As we show in \cref{app:rotormultipoles}, the spectroscopic moment of a band member with angular momentum $I$ is obtained by projecting the intrinsic moment $\Qsph{k}^\Kband$ onto the corresponding laboratory-frame state
\begin{align}
    \label{eq:QsphsQsph}
    \Qsphspec{k}{I, K}
    =&
    \cP_k^{(I,K)}
    \Qsph{k}^\Kband\,.
\end{align}
where $\cP_k^{(I,K)}$ is a purely geometrical projection factor that takes the form
\begin{align}
    \label{eq:PkIK}
    \cP_k^{(I,K)}
    \equiv
    (-1)^{I-K}  (2I+1)
    \begin{pmatrix}
        I & k & I\\
        -K & 0 & K
    \end{pmatrix}
    \begin{pmatrix}
        I & k & I \\
        -I & 0 & I
    \end{pmatrix}\,.
\end{align}
Thus, a single intrinsic moment $\Qsph{k}^\Kband$ determines the spectroscopic moments of all members of the band $\Kband$.

For the \Th clock states, the ground state $(\gs) \equiv (I, K) = (5/2,5/2)$ and the isomer $(\m) \equiv (I, K) = (3/2,3/2)$ are band heads of two distinct rotational bands, denoted $\Gband$ and $\Mband$, respectively.
\Cref{eq:QsphsQsph} gives
\begin{align}
    \begin{aligned}
    \Qsphspec{2}{\gs} &= \frac{5}{14} \Qsph{2}^\Gband\, , \qquad
    & 
    \Qsphspec{2}{\m} &= \frac{1}{5} \Qsph{2}^\Mband\, , 
    \\
    \Qsphspec{4}{\gs} &= \frac{1}{42} \Qsph{4}^\Gband\, , \qquad
    & 
    \Qsphspec{4}{\m} &= 0\, .
    \end{aligned}
    \label{eq:QsphsQsphNuclearClock}
\end{align} 
The vanishing isomeric spectroscopic hexadecapole follows from the angular-momentum selection rule $k\leq 2I$.
Note that this does not imply that the intrinsic hexadecapole moment $\Qsph{4}^\Mband$ vanishes. 
For the ratio of quadrupole moments, we have
\begin{align}
    \label{eq:rotorQrat}
    \frac{\Qsphspec{2}{\m}}{\Qsphspec{2}{\gs}} = \frac{14}{25} \frac{\Qsph{2}^\Mband}{\Qsph{2}^\Gband}\,. 
\end{align}
This relation does not constrain the ratio of the two intrinsic moments, since they belong to different bands, but if they are similar, the expected ratio of the spectroscopic quadrupole moments is $14/25 = 0.56$.
This is remarkably close to the measured value $0.57003(1)$~\cite{Zhang:2024ngu}.

\subsection{Hyperfine Hamiltonians of the Nuclear Clock States}
\label{sec:clockhamiltonians}

We now apply the general formalism to the hyperfine structure of the \Th nuclear clock states and discuss how their structure is constrained by the crystal environment.

\subsubsection{Quadrupole Hamiltonians}
\label{sec:Thquadrupole}

The quadrupole interaction is especially simple because the rank-2 EFG-tensor of~\cref{eq:MEFG} is a real symmetric traceless $3\times 3$ matrix. 
Diagonalizing it defines its principal-axis frame. 
By convention, $\Vcartzz$ is chosen to be the eigenvalue with the largest absolute value. 
In this frame~\cite{abragam1961principles},
\begin{align}
    \Vcart{xy} = \Vcart{xz} = \Vcart{yz} = 0\,,
\end{align}
while tracelessness requires
\begin{align}
    \label{eq:Qtr}
    \Vcart{xx} + \Vcart{yy} + \Vcartzz = 0 \, .
\end{align}
The quadrupole EFG is therefore fully specified by $\Vcartzz$ and the asymmetry parameter
\begin{align}
    \label{eq:etadef}
    \eta 
    \equiv 
    \frac{\Vcart{xx}-\Vcart{yy}}{\Vcartzz}\,.
\end{align}
In the spherical representation, \cref{eq:Hsph}, $V_{\pm 1}^{(2)} = 0$ in the principal-axis frame, while the $q=0,\pm2$ components of the EFG tensor generally remain nonzero.
This restriction follows from the diagonal form of the rank-2 tensor and does not necessarily extend to higher-rank EFG tensors.

Consequently, the quadrupole Hamiltonian contains only matrix elements with $\Delta m= 0, \pm 2$ and separates into two independent blocks. 
For the isomer, $I=3/2$, these blocks contain the subspaces
\begin{align}
    m \in \left\lbrace 
    \frac{3}{2}, -\frac{1}{2}
    \right\rbrace\,,
    \qquad
    m \in \left\lbrace 
    -\frac{3}{2}, \frac{1}{2}
    \right\rbrace\,,
    \label{eq:m-sector-m}
\end{align}
while for the ground state, $I=5/2$,
\begin{align}
    m \in \left\lbrace 
    \frac{5}{2}, \frac{1}{2}, -\frac{3}{2}
    \right\rbrace\,,
    \qquad
    m \in \left\lbrace 
    - \frac{5}{2}, -\frac{1}{2}, \frac{3}{2}
    \right\rbrace\,.
    \label{eq:m-sector-g}
\end{align}
The two blocks are related by time reversal and therefore have identical spectra, so it is sufficient to focus on one of them for each nuclear state.
Using the matrix elements derived in \cref{app:multipoletranslations}, we obtain
\begin{align}
    \label{eq:Hquadmgeneral}
    \mathcal{H}^{(2)}_{\left\lbrace\frac{3}{2},-\frac{1}{2}\right\rbrace}
    &=\mathcal{H}^{(2)}_{\left\lbrace-\frac{3}{2},\frac{1}{2}\right\rbrace}  
    = 
    \frac{1}{4}\Qcartzzspec{\m} \Vcartzz
    \begin{pmatrix}
       1 & \frac{\eta }{\sqrt{3}} \\
        \frac{\eta }{ \sqrt{3}} & -1 \\
    \end{pmatrix}
    \,,\\
    \label{eq:Hquadgsgeneral}
    \mathcal{H}^{(2)}_{\lbrace\frac{5}{2},\frac{1}{2},-\frac{3}{2}\rbrace}
    &=\mathcal{H}^{(2)}_{\lbrace-\frac{5}{2},-\frac{1}{2},\frac{3}{2}\rbrace}  
    = 
    \frac{1}{4}\Qcartzzspec{\gs} \Vcartzz
    \begin{pmatrix}
         1 & \frac{\eta }{ \sqrt{10}} & 0 \\
         \frac{\eta }{\sqrt{10}} & -\frac{4}{5} & \frac{3 \eta }{5 \sqrt{2}} \\
         0 & \frac{3 \eta }{5 \sqrt{2}} & -\frac{1}{5} 
    \end{pmatrix}\,,
\end{align}
where we again used the notation $(\gs) \equiv (I, K) = (5/2,5/2)$ and $(\m) \equiv (I, K) = (3/2,3/2)$.

The same interaction can be expressed in terms of spin operators~\cite{abragam1961principles}
\begin{align}
    \Hamilhat_I^{(2)}=
    \frac{\Qcartzzspec{I} \Vcartzz}{4I(2I-1)}\left(3 \hat{I}_z^2-\hat{\mathbf{I}}^2+\frac{\eta}{2}(\hat{I}_+^2+\hat{I}_-^2)\right) \,,
    \label{eq:HQspec}
\end{align}
where $I=5/2$ for the ground state and $I=3/2$ for the isomer, and $\hat{I}_\pm \equiv \hat{I}_x\pm i \hat{I}_y$. 
The equivalence of the different representations is shown explicitly in \cref{app:multipoletranslations}.

\subsubsection{Octupole Hamiltonian}
\label{sec:Thoctupole}

The \Th nucleus is expected to possess an octupole deformation that is important for the nuclear transition rates, see~\cite{Minkov:2026vlc}.
The clock states are, however, parity eigenstates, so their static spectroscopic electric-octupole moments vanish in the absence of parity violation. 
We therefore drop the rank-3 (octupole) electrostatic hyperfine interaction from our discussion.

\subsubsection{Hexadecapole Hamiltonian}
\label{sec:Thexadexapole}

We next consider the rank-4 (hexadecapole) electric interaction. 
For the isomeric state, $I=3/2$, the spectroscopic hexadecapole moment vanishes by the angular-momentum selection rule $k\leq 2I$.
The rank-4 interaction therefore contributes only to the hyperfine structure of the $I=5/2$ ground-state .
For studies of the nuclear hexadecapole in crystal environments see e.g.~\cite{Travesedo:2026zeh}.

In contrast to the quadrupole EFG, a symmetric traceless rank-4 EFG tensor in three spatial dimensions has nine independent real components and cannot in general be reduced to a principal-axis form characterized by only two parameters.
We parametrize its spherical components by (see also \cref{app:multipoletranslations})
\begin{align}
    \label{eq:zetadefs}
    \xi_q^{(4)} 
    \equiv 
    \frac{V_q^{(4)}}{V_0^{(4)}} = 
    \begin{cases}
        1 & q=0\\
        \zeta_q & q=1,2,3,4\\
        -\zeta_{-q}^* & q =-1,-3\\
        \zeta_{-q}^* & q= -2,-4\,.
    \end{cases}
\end{align}
For an axially symmetric EFG, only the $q=0$ components survive, corresponding to $\eta=0$ at rank two and $\zeta_1=\zeta_2=\zeta_3=\zeta_4=0$ at rank four.
Note that unlike the quadrupole interaction in its principal-axis frame, the general rank-4 interaction permits matrix elements with odd $\Delta m$ in addition to $\Delta m= 0, \pm 2, \pm 4$.
Consequently, the two sets of states that are decoupled under the quadrupole interactions (\cref{eq:m-sector-m,eq:m-sector-g}) are in general coupled by the hexadecapole interaction.
Crystal symmetries can nevertheless simplify the Hamiltonian, as we will discuss below.

The general $I=5/2$ hexadecapole Hamiltonian is obtained from the matrix representation derived in \cref{sec:matrix-representations}. 
In the standard descending-$m$ basis, $m = \lbrace 5/2, 3/2, 1/2, -1/2, -3/2, -5/2\rbrace$,
\begin{align}
    \mathcal{H}^{(4)}_{I=5/2}
    = \frac{1}{192}\Vcartzzzz \Qcartzzzzspec{5/2}
    \begin{pmatrix}
        1
        & -2\zeta_1^*
        & 3\zeta_2^*
        & -\sqrt{14}\zeta_3^*
        & \sqrt{14}\zeta_4^*
        & 0
        \\
        -2\zeta_1
        & -3
        & \sqrt{10}\zeta_1^*
        & -\sqrt{5}\zeta_2^*
        & 0
        & \sqrt{14}\zeta_4^*
        \\
        3\zeta_2
        & \sqrt{10}\zeta_1
        & 2
        & 0
        & -\sqrt{5}\zeta_2^*
        & \sqrt{14}\zeta_3^*
        \\
        -\sqrt{14}\zeta_3
        & -\sqrt{5}\zeta_2
        & 0
        & 2
        & -\sqrt{10}\zeta_1^*
        & 3\zeta_2^*
        \\
        \sqrt{14}\zeta_4
        & 0
        & -\sqrt{5}\zeta_2
        & -\sqrt{10}\zeta_1
        & -3
        & 2\zeta_1^*
        \\
        0
        & \sqrt{14}\zeta_4
        & \sqrt{14}\zeta_3
        & 3\zeta_2
        & 2\zeta_1
        & 1
    \end{pmatrix}\,.
\end{align}
Here, $\Vcartzzzz$ sets the overall scale of the rank-4 EFG, while the four generally complex parameters $\zeta_q$ characterize its angular structure.

\section{Crystal Symmetries and Connection to Data}
\label{sec:crystal_sym}

The local crystal environment determines the EFG tensors and thereby the nuclear multipole interactions accessible through hyperfine spectroscopy.
Theoretical studies of \Th-doped CaF${}_2$ have considered thorium substitution at Ca${}^{2+}$ sites~\cite{Dessovic_2014}, while X-ray absorption measurements provide experimental evidence that \Th occupies these sites in the Th${}^{4+}$ charge state~\cite{Takatori:2025}.
The associated charge compensation and microscopic defect configurations, however, remain under debate. 
Proposed configurations include interstitial fluoride ions and calcium vacancies, while possible dopant arrangements include isolated \Th ions and pairs of neighboring thorium dopants (dimers)~\cite{Pimon22,Takatori:2025,rehn2026defect,Li:2026izx}.
Charge neutrality must be maintained overall, but the compensating defects need not necessarily be located immediately adjacent to the thorium ions.

Experimental studies have identified several distinct local thorium environments, including D-centers exhibiting a substantial quadrupole asymmetry parameter $\eta$ (\cref{eq:etadef})~\cite{Zhang:2024ngu,Hiraki:2025uur,morawetz:2026} and O-centers for which no hyperfine splitting has been resolved~\cite{Hiraki:2025uur,morawetz:2026}.
The D-centers have been associated with a thorium-dimer configuration of $D_{2h}$ symmetry~\cite{Hiraki:2025uur,morawetz:2026}, while the O-centers are consistent with approximately cubic symmetry.
Additional spectroscopic sites have also been reported~\cite{Hiraki:2025uur}.

Rather than assuming a particular microscopic defect configuration, we consider three effective EFG symmetry classes: 
twofold ($\twosym$), tetragonal ($\tetrasym$) and cubic ($\cubsym$).
These classes impose progressively stronger constraints on the EFG tensors, reducing the number of independent parameters entering the hyperfine Hamiltonian.
The corresponding EFG constraints and symmetry-restricted Hamiltonians are derived in \cref{app:crystal_syms} and summarized in \cref{tab:symclasses}, together with representative crystallographic point groups. 
For a discussion of the crystallographic point groups and their symmetry operations, see Ref.~\cite{glazer2012space}.

\begin{table}[t]
    \centering
    \setlength{\tabcolsep}{4pt}
    \renewcommand{\arraystretch}{1.5}
    \begin{tabular*}{\textwidth}{l@{\extracolsep{\fill}}c@{\extracolsep{\fill}}c@{\extracolsep{\fill}}c@{\extracolsep{\fill}}c}
        \toprule
        Symmetry class
        &  Constraints
        & $k=2$
        & $k=4$
        & Point groups\\
        \hline
        Generic ($\zerosym$)
        & Reality \eqref{eq:Vreal}
        & $q=0, \pm 1, \pm 2$
        & $q=0, \pm 1, \ldots, \pm 4$
        & $C_1, C_i$\\
        &
        & $\Vcartzz, \eta$ \eqref{eq:etadef}
        & $\Vcartzzzz, \zeta_{1,\ldots ,4}$ \eqref{eq:zetacartgeneral}
        &
        \\
        \hline
        Twofold ($\twosym$)
        & $C_2(z), \sigma_x$ \eqref{eq:twosymconds}
        & $q=0, \pm 2$
        & $q=0, \pm 2, \pm 4$
        & $D_2, C_{2v}, D_{2h}$
        \\
        &
        & $\Vcartzz, \eta$ \eqref{eq:etadef}
        & $\Vcartzzzz, \zeta_2, \zeta_4$ \eqref{eq:zetatwosym}
        &
        \\
        \hline
        Tetragonal ($\tetrasym$)
        & $+C_4(z)$ \eqref{eq:tetrasymcond}
        & $q=0$
        & $q=0, \pm 4$
        & $C_4$, $S_4$, $C_{4h}$, $D_4$, \\
        &
        & $\Vcartzz$
        & $\Vcartzzzz, \zeta_4$ \eqref{eq:zetatetra}
        & $C_{4v}$, $D_{2d}$, $D_{4h}$
        \\
        \hline
        Cubic ($\cubsym$)
        & $+C_3([111])$ \eqref{eq:cubsymcond}
        & none
        & fixed $q=0, \pm 4$ 
        & $T, T_h, O, T_d, O_h$
        \\
        &
        & -
        & $\Vcartzzzz$ \eqref{eq:zetacub}
        &
        \\
        \bottomrule
    \end{tabular*}
    \caption{Effective symmetry classes and constraints they impose on the rank-2 and rank-4 EFG tensors.
    Crystallographically distinct point groups are grouped together when they impose equivalent constraints on the even-rank EFG components in appropriately chosen coordinate frames.
    Symmetries are ordered such that the constraints on the EFG tensors become progressively more restrictive from top to bottom. 
    Each rank column gives the allowed spherical components and independent EFG parameters.
    The asymmetry parameters $\eta_1, \eta_2$ and $\zeta_1, \ldots , \zeta_4$ are generally complex, but those surviving in the twofold and higher-symmetry classes can be chosen to be real without loss of generality.}
    \label{tab:symclasses}
\end{table}

\subsection{Experimental Observables}
\label{sec:observables}

\begin{figure}[t]
\centering

\begin{minipage}[t]{0.49\textwidth}
\vspace{0pt}
\includegraphics[]{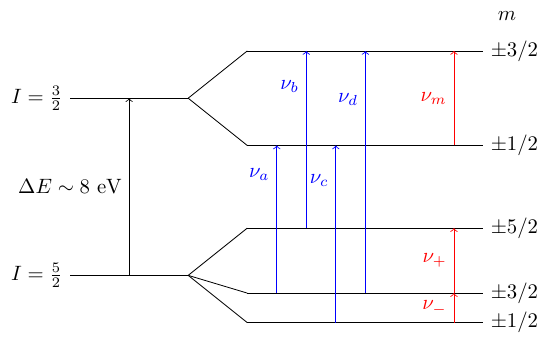}
\end{minipage}
\hfill
\begin{minipage}[t]{0.49\textwidth}
\vspace{15pt}
\centering
\begin{tabular}{c}
\hline
$\nu_a$ = 2,020,407,283.847(4)\,MHz\\
$\nu_b$ = 2,020,407,298.727(4)\,MHz\\
$\nu_c$ = 2,020,407,446.895(4)\,MHz\\
$\nu_d$ = 2,020,407,530.918(4)\,MHz\\
\hline
$\nu_m = \nu_d-\nu_a$ = 247.071(5)\,MHz\\
$\nu_+ = \nu_d-\nu_b$ = 232.191(5)\,MHz\\
$\nu_- = \nu_c-\nu_a$ = 163.048(5)\,MHz\\
\hline
\end{tabular}
\end{minipage}

\caption{
Hyperfine level structure of the nuclear clock states (left) and measured transition frequencies (right), taken from Ref.~\cite{Zhang:2024ngu}.
The three independent hyperfine splittings $\nu_{\m}$, $\nu_+$ and $\nu_-$ (red) are obtained from the measured transition frequencies $\nu_{a,b,c,d}$ (blue), as defined in \cref{eq:numpm}. 
The ground state and isomer are separated by $\Delta E \sim $ 8~eV.
}
\label{fig:ThSplit}
\end{figure}

The hyperfine splitting scheme of \Th is shown in \cref{fig:ThSplit}.
Owing to Kramers degeneracy, the ground ($I=5/2$) and isomeric ($I=3/2$) states can split into at most three and two distinct hyperfine levels, respectively. 
An enhanced local crystal symmetry can reduce the number of non-degenerate states.

The experimentally measured transition frequencies $\nu_{a, \ldots, d}$ can therefore be expressed in terms of three independent hyperfine splittings,
\begin{align}
    \label{eq:numpm}
    \nu_m \equiv \nu_d - \nu_a 
    \, , 
    \qquad
    \nu_- \equiv \nu_c - \nu_a 
    \, , 
    \qquad
    \nu_+ \equiv \nu_d - \nu_b \, .
\end{align}
The remaining independent frequency scale is the unperturbed nuclear transition energy $\Delta E$.

Neglecting magnetic interactions, the hyperfine splittings considered here arise entirely from the nuclear quadrupole and hexadecapole moments and their couplings to the crystal's EFG tensors.
Odd-rank electric multipole moments (in particular the octupole) are inaccessible with the definite-parity nuclear clock states; higher even-rank moments are forbidden by their angular momenta and the selection rule $k\leq 2I$.

Since the multipole Hamiltonian involves products of nuclear moments and EFG components, spectroscopy alone cannot in general separate the two.
Determining the absolute nuclear moments therefore requires additional information about the EFGs, for instance from theoretical models of the crystal environment.
Crystal symmetries can help disentangle the nuclear multipole moments from the EFGs by reducing the number of independent parameters entering the hyperfine spectrum.

\subsection{Twofold Symmetry ($\twosym$)}

\begin{figure}
    \centering
    \includegraphics[width=0.35\linewidth]{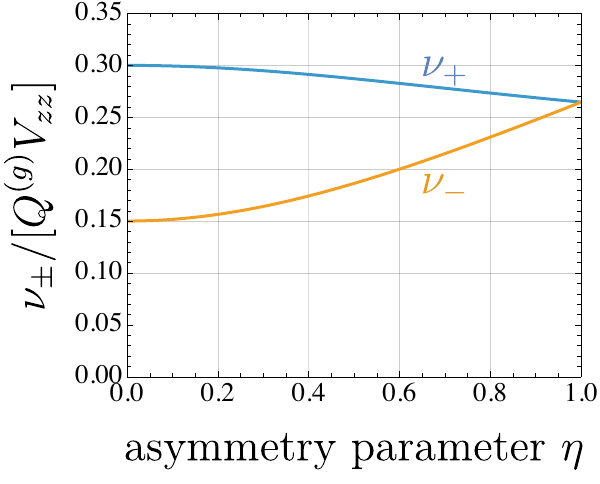}
    \hspace{10mm}
    \includegraphics[width=0.35\linewidth]{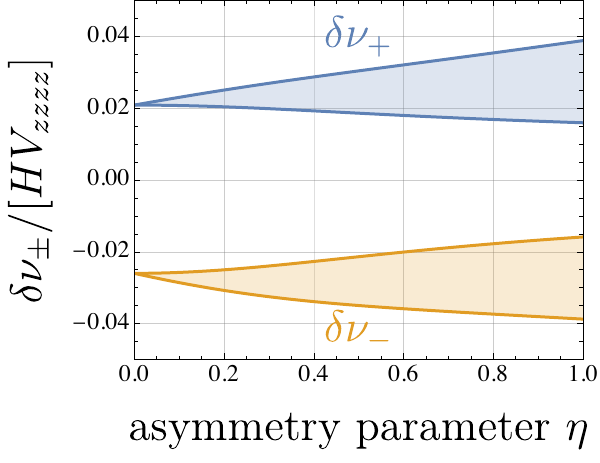}
    \caption{Quadrupole contributions (left) and linear in $h$ hexadecapole corrections (right) to the ground-state hyperfine splittings in the twofold symmetry class, as a function of the asymmetry parameter $\eta$.
    Each contribution is normalized by the respective interaction scale, $\Qcartzzspec{\gs}\Vcartzz$ and $\Qcartzzzzspec{\gs}\Vcartzzzz$. 
    The shaded bands are obtained by varying $\zeta_2$ and $\zeta_4$ $\in [0,1]$.
    }
    \label{fig:twofold_sym}
\end{figure}

The twofold symmetry class ($\twosym$) is particularly relevant for the experimentally observed D-centers~\cite{Zhang:2024ngu,Hiraki:2025uur,morawetz:2026}, which exhibit quadrupole splitting with an asymmetry parameter $\eta \simeq 0.6$~\cite{Zhang:2024ngu,Hiraki:2025uur,morawetz:2026}.
We define this class by a twofold rotation axis and a mirror plane containing this axis.
These symmetries restrict the EFG tensors to even-$q$ components.
This does not have an effect on the quadrupole parameters $\Vcartzz$ and $\eta$ (see \cref{eq:etadef})
It does, however, set the hexadecapole parameters $\zeta_1 = \zeta_3 = 0$, leaving $\Vcartzzzz$, $\zeta_2$ and $\zeta_4$ free and simplifying the ground-state Hamiltonians to \cref{eq:Hquadgstwosym,eq:Hexagstwo}.
To quantify the hexadecapole interaction, we introduce the dimensionless ratio
\begin{align}
    \label{eq:hdef}
    h\equiv \frac{\Qcartzzzzspec{\gs}\,\Vcartzzzz}{\Qcartzzspec{\gs}\,\Vcartzz}\,.
\end{align}
The ground-state hyperfine splittings depend on $\eta$, $\zeta_2$, $\zeta_4$ and $h$.
Their dependence on the quadrupole asymmetry parameter $\eta$ is illustrated in \cref{fig:twofold_sym}, showing separately the quadrupole contributions to $\nu_\pm$ (left) and linear in $h$ hexadecapole corrections $\delta \nu_\pm$ (right).
The former are normalized by $\Qcartzzspec{\gs}\Vcartzz$, while the latter are normalized by $\Qcartzzzzspec{\gs}\Vcartzzzz$.
The shaded bands are obtained by varying $\zeta_2$ and $\zeta_4$ $\in [0,1]$.

\begin{figure}
    \centering
    \includegraphics[width=0.6\linewidth]{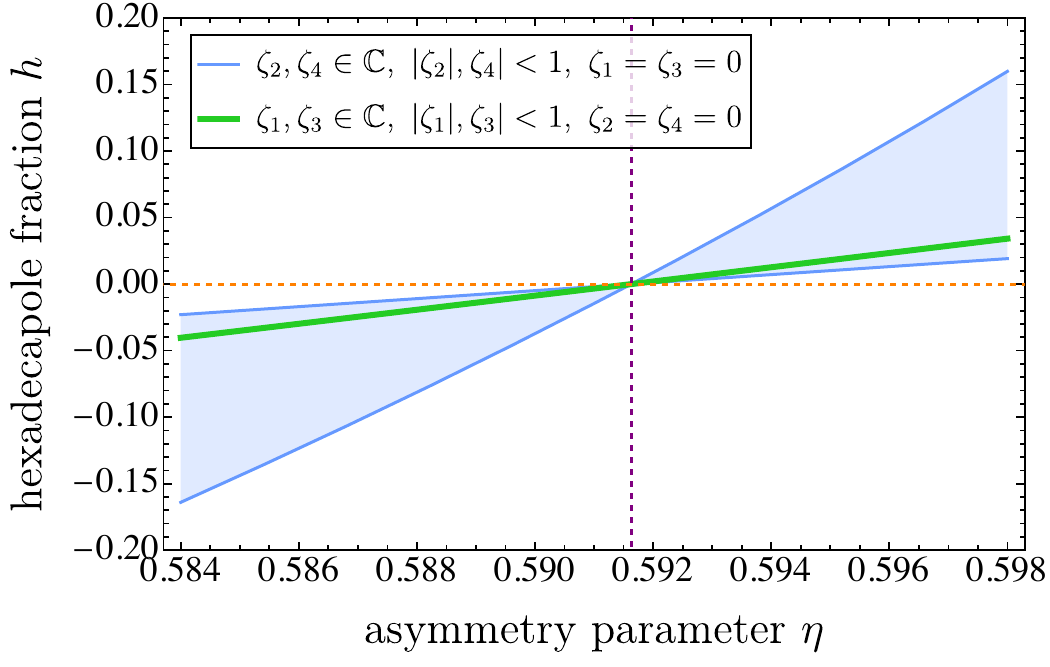}
    \caption{The hexadecapole fraction $h\equiv \Qcartzzzzspec{\gs} \Vcartzzzz/(\Qcartzzspec{\gs} \Vcartzz)$ as a function of the rank-2 asymmetry parameter $\eta$ defined in~\cref{eq:etadef}, obtained by fitting the measured hyperfine splittings of Ref.~\cite{Zhang:2024ngu}.
    Blue shaded region: $\zeta_{1,3}=0$, with the colored region spanning the solutions obtained for $|\zeta_2|,|\zeta_4|\leq 1$. 
    Thick green line: $\zeta_{2,4}=0$, spanning the solutions obtained for $|\zeta_1|,|\zeta_3|\leq 1$. 
    (This region is very narrow and therefore appears approximately as a line.)
    The vertical purple dashed line indicates the $\eta$ value obtained when the hexadecapole interaction is neglected, $h=0$, indicated by the horizontal orange line. 
    This corresponds to the value reported in Ref.~\cite{Zhang:2024ngu}. }
    \label{fig:heta}
\end{figure}

Neglecting the hexadecapole interaction, as is often done in the literature, the three measured splittings $\nu_m$, $\nu_+$ and $\nu_-$ uniquely determine the two quadrupole interaction scales $\Qcartzzspec{\gs}\Vcartzz$ and $\Qcartzzspec{\m}\Vcartzz$, and the dimensionless asymmetry parameter $\eta$.
But allowing a nonzero hexadecapole contribution introduces additional free parameters, notably the interaction scale $\Qcartzzzzspec{\gs}\Vcartzzzz$ as well as the dimensionless parameters $\zeta_2$ and $\zeta_4$.
Consequently, the number of unknown parameters exceeds the number of independent observables and the determination is no longer unique.
We illustrate this degeneracy in \cref{fig:heta} using the hyperfine spectroscopy measurements of Ref.~\cite{Zhang:2024ngu}, listed in \cref{fig:ThSplit}.
Varying $\eta$, $\zeta_2$ and $\zeta_4$ generates a continuous family of solutions for $\Qcartzzspec{\gs}\Vcartzz$, $\Qcartzzspec{\m}\Vcartzz$, and $\Qcartzzzzspec{\gs}\Vcartzzzz$.
The left panel of \cref{fig:heta} assumes twofold symmetry, $\zeta_1 = \zeta_3 = 0$, while the right panel explores the case $\zeta_2=\zeta_4=0$ but nonzero $\zeta_1$ and $\zeta_3$.
The value of $\eta$ extracted in Ref.~\cite{Zhang:2024ngu} under the assumption of a purely quadrupole interaction corresponds to the special case of $h=0$, indicated by the dashed lines.

\subsection{Tetragonal Symmetry ($\tetrasym$)}

While the dominant quadrupole-split sites (D-centers) exhibit a substantial nonzero asymmetry parameter $\eta\simeq 0.6$~\cite{Zhang:2024ngu, Hiraki:2025uur, morawetz:2026}, other sites have considerably smaller fitted asymmetries, in some case consistent with $\eta=0$ \cite{Hiraki:2025uur}.
Although this does not directly establish tetragonal symmetry, it motivates us to consider this more restrictive symmetry class.

The tetragonal symmetry class ($\tetrasym$) inherits the symmetries of the $\twosym$ class (invariance under rotations by 180$^{\circ}$ about an axis, reflection symmetry on a plane containing that axis), but additionally imposes invariance under fourfold rotations about the symmetry axis. 
These requirements restrict the spherical EFG components to $q=0, \pm 4$.
Consequently, only the $q=0$ component survives at rank two, so that the quadrupole EFG is axially symmetric, $\eta=0$ (see \cref{eq:etadef}).
At rank four, the EFG is characterized by $\Vcartzzzz$ and the single asymmetry parameter $\zeta_4$.
The corresponding Hamiltonians are given in \cref{eq:Htetra}.

The quadrupole Hamiltonians are diagonal in the $\ket{I,m}$ basis, while the off-diagonal hexadecapole terms proportional to $\zeta_4$ couple states with $\Delta m=\pm 4$.
For a nonzero quadrupole interaction, these terms do not contribute to the linear in $h$ hexadecapole energy corrections (see also \cref{app:crystal_syms}).
Consequently, the three interaction scales $\Qcartzzspec{\gs}\Vcartzz$, $\Qcartzzspec{\m}\Vcartzz$ and $\Qcartzzzzspec{\gs}\Vcartzzzz$ can be determined from the hyperfine splittings, up to corrections of second order in the hexadecapole interaction, $\mathcal{O}(h^2)$.
\begin{align}
    \label{eq:hfstetra}
    \Qcartzzspec{\m}\Vcartzz =& 2[\nu_{\m}]_{\tetrasym} \,,\\
    \Qcartzzspec{\gs}\Vcartzz =& \frac{10}{21} \left(5[\nu_+]_{\tetrasym}+4[\nu_-]_{\tetrasym}\right)\, , \\
    \Qcartzzzzspec{\gs}\Vcartzzzz =& \frac{96}{7} \left( [\nu_+]_{\tetrasym}-2[\nu_-]_{\tetrasym} \right)\,.
\end{align}
Extracting the corresponding nuclear moments still requires independent knowledge of the EFGs.
For $\Vcartzz$ theoretical estimates are available~\cite{Dessovic_2014,Pimon22}.

\subsection{Cubic Symmetry ($\cubsym$)}

Resonances without resolved hyperfine splitting in \Th-doped CaF${}_2$~\cite{Hiraki:2025uur,morawetz:2026} suggest there exist defect sites with a strongly suppressed quadrupole EFG, consistent with approximately cubic local symmetry.
Assuming the nuclear quadrupole moment inferred from the resolved spectra, Ref.~\cite{morawetz:2026} obtains an upper limit of approximately $|\Vcartzz|\lesssim 0.1~\mathrm{V/\AA}^2$, compared with EFGs of order 100 V/\AA${}^4$ at the quadrupole s-split sites. 

For an ideal cubic environment ($\cubsym$), the rank-2 EFG vanishes identically, while the rank-4 EFG retains a single independent parameter.
Cubic symmetry fixes the relative weights of its $q=0$ and $q=\pm 4$ spherical components via $\zeta_4=5/\sqrt{70}$ (see \cref{eq:zetacub}).
The corresponding Hamiltonian is given in \cref{eq:Hexacub}.
The hexadecapole interaction splits the ground-state manifold into a twofold- and a fourfold-degenerate level, while the isomeric manifold remains unsplit.
Inverting the ground-state hyperfine splitting yields
\begin{align}
    |\Qcartzzzzspec{\gs}\Vcartzzzz| = 32 [\nu_\pm]_{\cubsym}
\end{align}
where $[\nu_\pm]_{\cubsym} \equiv [\nu_+]_{\cubsym}=[\nu_-]_{\cubsym}$.

Although the observed O-centers do not establish exact cubic symmetry, assuming such symmetry and requiring the predicted splitting to remain below the observed O-center linewidth $\Gamma \approx 1.1(1)~$MHz reported in Ref.~\cite{morawetz:2026} gives
\begin{align}
    \left| \Qcartzzzzspec{\gs}\Vcartzzzz \right| \lesssim 32 \Gamma \simeq 34~\mathrm{MHz}\,.
\end{align}
This bound is not expected to constrain the hexadecapole interaction at its natural scale.
Indeed, combining the naive dimensional analysis (NDA) estimate of \cref{eq:multipole_scaling} with the characteristic quadrupole interaction scale $\Qcartzzspec{\gs}\Vcartzz\sim 339 ~e\mathrm{bV\AA}^{-2} \simeq 821~$MHz inferred from the resolved hyperfine spectrum of Ref.~\cite{Zhang:2024ngu}, we obtain
\begin{align}
    \Qcartzzzzspec{\gs}\Vcartzzzz 
    & \sim \left(\frac{\sqrt{\chrsq}}{s}\right)^2 \Qcartzzspec{\gs} \Vcartzz 
    \sim 0.4~\mathrm{Hz}\,.
\end{align}
The corresponding cubic hyperfine splitting is $[\nu_{\pm}]_{\cubsym} \approx 13~$mHz, approximately eight orders of magnitude below the sensitivity inferred from the observed O-center linewidth and below the spectroscopic resolution currently demonstrated in optical atomic clocks.
Note, however, that this estimate uses the quadrupole interaction measured at the D-centers as a benchmark for the O-centers, implicitly assuming comparable characteristic length scales and effective charges in their local electrostatic environments.
Differences in the microscopic defect configurations could lead to substantial deviations from the NDA estimate.
Looking ahead, single-ion nuclear-clock experiments may offer substantially improved spectroscopic resolution and control over the local electromagnetic environment, potentially opening a route to probing the nuclear hexadecapole interaction.
Nevertheless, achieving the sensitivity required to resolve the hexadecapole splitting will remain a considerable experimental challenge.

\section{Theoretical Approaches}
\label{sec:theory}

The nuclear observables relevant for the $^{229}$Th clock probe rather different aspects of the ground and isomeric states. 
While the charge radius and spectroscopic quadrupole moments are experimentally
constrained, higher multipole moments and the difference in Coulomb energy between the two clock states are not directly known. 
Their interpretation therefore necessarily involves assumptions about the underlying nuclear structure. 
It is useful to examine these assumptions from several complementary viewpoints, including microscopic nuclear-structure calculations, effective descriptions and phenomenological charge-density models.

In this section, we review a number of such approaches and identify the relations among the nuclear observables that follow within each of them.
We begin with the model-independent NRQED description, which organizes the relevant nuclear information in terms of low-energy electromagnetic moments. 
We then turn to microscopic energy-density-functional calculations of $^{229}$Th, including recent studies that incorporate pairing, configuration mixing, symmetry restoration, and reflection-asymmetric deformation. 
These are followed by the halo-inspired description and the particle--rotor model, the latter providing the connection between intrinsic structure and laboratory-frame observables. 
Finally, we introduce a deformed Fermi charge distribution, whose parameters are constrained by experimental input using these relations. 
This provides the basis for our estimate of the Coulomb-energy difference and the corresponding new-physics sensitivity in \cref{sec:K}.

These approaches should not be viewed as successive approximations to a single underlying calculation. 
Rather, they emphasize different aspects of the problem and involve different theoretical inputs. 
Of particular interest for the present work is the extent to which the experimentally accessible charge radius and quadrupole moments constrain higher moments of the charge distribution. 
Conversely, comparing the different descriptions allows us to identify which conclusions are robust consequences of symmetry and measured observables, and which depend more strongly on assumptions concerning deformation, collective dynamics, pairing, or the microscopic nuclear interaction.

\subsection{Nonrelativistic QED (NRQED)}
\label{subsec:eft-nrqed}

As discussed in \cref{sec:H_multipole}, the \iso{Th}{229} nucleus is probed by an ``electronic" system, namely the atomic shells and crystal environment, see, e.g.\ Ref.~\cite{Thielking:2017qet}.
The typical size of the lattice spacing is $a \sim \si{\AA}$, which is much larger than the nuclear size $r_N \approx \SI{6}{fm}$. 
The large hierarchy $r_N / a \sim 10^{-5}\ll 1$ suggests utilizing effective field theory methods, in particular Non-Relativistic QED~(NRQED).

The NRQED operators are constructed from the electric and magnetic fields, the time and space components of the covariant derivative, and the spin operator of the particle. 
The Wilson coefficients depend on nuclear properties such as charge, magnetic moment, and charge radius. 
In the literature, the Wilson coefficients are often written as dimensionless numbers divided by the appropriate power of the particle mass $M$. 

As was shown in Ref.~\cite{Assi:2025nim}, these dimensionless numbers might scale as a power of the mass times the nuclear size. 
As a result, the Wilson coefficients scale as an appropriate power of $r_N$.
Since the EM fields vary on the scale of $a$, higher dimensional operators are expected to be suppressed by higher powers of $r_N/a$. 
We also emphasize that the Wilson coefficients of the spin operators are directly related to the \emph{spectroscopic} moments defined in~\cref{eq:Qsphspecdef}.
This is because the internal structure of the nucleus is captured by the higher dimensional operators. 

The most well-studied NRQED Lagrangian is the one for spin-1/2 fermions, see~Ref.~\cite{Paz:2015uga} for a pedagogical discussion. 
For the spin-1/2 case the allowed operators in the Lagrangian are known up to and including dimension 8~\cite{Hill:2012rh}, which corresponds to Wilson coefficients of length dimension $\si{fm}^4$.
They can be easily and explicitly constructed using the method presented in~\cite{Gunawardana:2017zix}.  
The spin-1 case was discussed, for example, in Ref.~\cite{PhysRevA.82.052520} and~\cite{Davoudi:2015zda}. 
Although it is not immediately clear that all the possible operators of a given dimension were listed in these references, both contain the quadrupole operator. 
Studies involving higher-spin particles, such as spin $3/2$ and $5/2$, are less common in the literature, perhaps because they cannot be easily compared to the corresponding relativistic Lagrangians. 
Next we discuss how to construct NRQED operators for spin $3/2$ and $5/2$. 
We limit ourselves to operators with Wilson coefficients up to and including dimension of the fourth power of length.

For a particle of spin $s$, corresponding to the nuclear angular momentum $I$ used elsewhere in this work, the Hamiltonian in spin space is a $(2s+1)\times(2s+1)$ Hermitian matrix. 
It can be written as a linear combination of products of spin operators that are fully symmetric and traceless in their spatial indices.
Such a rank-$r$ tensor has $2r+1$ independent components \cite{Edmonds:1955fi} and transforms as a rank-$r$ spherical
tensor. 
By the Wigner--Eckart theorem, it vanishes identically within a spin-$s$ multiplet for $r>2s$, the same condition as obtained below \cref{eq:WignerEckart}. 
Since $\sum_{r=0}^{2s}(2r+1)=(2s+1)^2$, the tensors of rank $r=0,\dots,2s$ form a complete basis for the $(2s+1)\times(2s+1)$ Hermitian matrices.

As an example, for a spin-1/2 fermion, all operators can be built from the $2\times2$ unit matrix and the rank-1 tensor, which is just the vector of  Pauli matrices, $\sigma^i$. 
As another example, for a spin-1 particle, all operators can be built from the $3\times3$ unit matrix, the three-component rank-1 tensor $s^i$, and the five-component rank-2 tensor $(s^i s^j)^{(2)}$.
Taken together, these tensors form a basis of nine matrices in spin space for the NRQED Lagrangian of a spin-1 particle. 
In particular, for spin-1 there is no magnetic octupole operator. 
Indeed, in~\cite{PhysRevA.82.052520} the Wilson coefficient at tree level of the magnetic octupole operator for an elementary spin-1 particle is zero. 
In~\cite{Davoudi:2015zda} there is no magnetic octupole operator for a \emph{composite} spin-1 particle either.
For the operators of interest in the context of the \Th ground state and isomer, we conclude that the spin-3/2 isomer cannot have a hexadecapole operator, while a spin-5/2 particle can have a hexadecapole operator.

For a charged particle the NRQED operators can be constructed from rotationally-invariant combinations of gauge-invariant ``building blocks": 
the temporal and spatial components of the covariant derivative, $iD_t$ and $i\vec{D}$, and the spin $\vec{s}$, see Ref.~\cite{Paz:2015uga}. The commutator of $iD_t$ and $i\vec{D}$ is proportional to the electric field $\vec{E}$, which is gauge invariant.\footnote{Unlike in the rest of the paper, in this section we do not use the ``hat" notation above operators.} 
In principle, some of the operators constructed this way can be removed by field redefinitions or by using the equations of motion. Furthermore, we are interested in operators whose Wilson coefficients are finite in the infinite mass limit. 
We will discuss operators with no magnetic fields, linear in $\vec{E}$ that we believe cannot be removed by a field redefinition and whose Wilson coefficients are finite in the infinite mass limit, as we explain below.

We limit ourselves to operators that contain commutators of the form $[D^{i_1},[\dots,[D^{i_n}, E^j]\dots]=\nabla^{i_1}\dots\nabla^{i_n}E^j$. 
These will be contracted with a traceless completely symmetric combination of $\vec{s}$'s. 
Requiring that both the crystal environment and the nucleus are invariant under time-reversal rules out any operator that contains an even number of derivatives, in particular,  the electric dipole and electric octuple operators.  

Consider first contractions with no spin operators. 
Using one derivative we have the operator $\bm\nabla\!\cdot\!\bm E$ whose Wilson coefficient in the infinite mass limit is proportional to the charge radius squared \cite{Hill:2012rh,Assi:2025nim}. 
Using three derivatives we find three possible operators \cite{Hill:2012rh,Gunawardana:2017zix}. 
Only one of them has a Wilson coefficient that does not vanish in the $M\to\infty$ limit: the operator $\nabla^2(\bm\nabla\!\cdot\!\bm E)$ \cite{Assi:2025nim}.

For two $\vec{s}$'s define 
\begin{align}
    \left(s^is^j\right)^{(2)}
    \equiv
    \left( s^is^j+s^js^i\right)/2-\bm{s^2}\delta^{ij}/3\,, 
    \label{eq:2spins}
\end{align}
where we follow the notation of~\cite{PhysRevA.82.052520}. 
This can be combined with one derivative operator to form the quadrupole operator $\big(s^is^j \big)^{\!(2)}\,\nabla^iE^j$. 
For the case of three derivatives we can expect several possible operators. 
But in analogy to the zero-spin case above and as we explain below, we expect only one operator that is finite in the $M\to\infty $ limit: $\big(s^is^j \big)^{\!(2)}\bm{\nabla^2}(\nabla^iE^j)$.

The case of four $\vec{s}$'s is more involved because of the non-commutative nature of the spin operators. 
We are not aware of an expression for the tensor in the literature. 
We find that it is
\begin{align}
    \left(s^is^js^ks^l\right)^{(4)} 
    \equiv&
    \dfrac1{24}\big(
    s^is^js^ks^l+\mbox{all possible permutations of }i,j,k,l\big)\nonumber\\
   &
   -\dfrac1{14}\left[
   \delta^{ij}\Ttwo(k,l)+\delta^{kl}\Ttwo(i,j)
   +\delta^{ik}\Ttwo(j,l)+\delta^{jl}\Ttwo(i,k)
   +\delta^{il}\Ttwo(j,k)+\delta^{jk}\Ttwo(i,l)
   \right] \nonumber\\
   &
   +\frac1{35}\left( \delta^{ij}\delta^{kl}+\delta^{ik}\delta^{jl}+\delta^{il}\delta^{jk}\right)\Tzero
   \,,
   \label{eq:4spins}
\end{align}
where
\begin{align}
    \Ttwo(a,b)
    =
    \left(s^as^bs^rs^r+s^as^rs^bs^r+s^as^rs^rs^b+s^rs^as^bs^r+s^rs^as^rs^b+s^rs^rs^as^b\right)/6+a\leftrightarrow b\,,
\end{align}
and
\begin{align}
\Tzero=\left(s^ps^ps^rs^r+s^ps^rs^ps^r+s^ps^rs^rs^p\right)/3\,.
\end{align}
The combination $\left(s^is^js^ks^l\right)^{(4)}$ can only be contracted with three derivatives of $\bm E$ to give the hexadecapole operator $\big(s^is^js^ks^\ell\big)^{\!(4)}\,\nabla^i\nabla^j\nabla^kE^\ell$.

Adding the Wilson coefficients, the allowed operators are for the isomer
\begin{eqnarray}
    \label{eq:HNRQED32}
    {\cal H}^{\mbox{\scriptsize NRQED}}_{3/2}
    &\ni&
    eZ\,\frac{\langle r_{E,3/2}^2\rangle}{3!}\,(\bm\nabla\!\cdot\!\bm E)
-\frac{e}{2}\frac{Q^E_{3/2}}{3} \big(s^is^j \big)^{\!(2)}\,\nabla^iE^j \nonumber\\ 
&+& eZ\,\frac{\langle r_{E,3/2}^4\rangle}{5!}\,\bm\nabla^2(\bm\nabla\!\cdot\!\bm E)-  \frac{e}{12}\,\frac{Q^E_{3/2}}{3} \langle r_{Q,3/2}^2\rangle\big(s^is^j \big)^{\!(2)}\bm{\nabla^2}(\nabla^iE^j)+\cdots\,,
\end{eqnarray}
and for the ground state 
\begin{eqnarray}\label{eq:HNRQED52}
   && {\cal H}^{ \mbox{ \scriptsize NRQED}}_{5/2}
    \ni
    eZ\,\frac{\langle r_{E,5/2}^2\rangle}{3!}\,(\bm\nabla\!\cdot\!\bm E)
-\frac{e}{2}\frac{Q^E_{5/2}}{10} \big(s^is^j \big)^{\!(2)}\,\nabla^iE^j+\nonumber\\ 
&&+ eZ\,\frac{\langle r_{E, 5/2}^4\rangle}{5!}\,\bm\nabla^2(\bm\nabla\!\cdot\!\bm E)-\frac{e}{12}\,\frac{Q^E_{5/2}}{10} \langle r_{Q,5/2}^2\rangle\big(s^is^j \big)^{\!(2)}\bm{\nabla^2}(\nabla^iE^j)-\frac{e}{4!}\frac{H^E_{5/2}}{60}\big(s^is^js^ks^\ell\big)^{\!(4)}\,\nabla^i\nabla^j\nabla^kE^\ell\nonumber\\
&&+\cdots\,.
\end{eqnarray}
In the equations above $\langle r_{E}^2\rangle$ is the square of the charge radius, called $\ev{r^2}$ in the rest of the paper. 
$Q^E$ is the spectroscopic electric quadrupole moment, $\langle r_{E}^4\rangle$ is the ``averaged" fourth power of the charge radius, and $\langle r_{Q}^2\rangle$, the square of the quadrupole radius, is determined by the slope of the electric quadrupole form factor at zero ($q^2=0$). 
The notation is analogous to the spin-1 case discussed in \cite{PhysRevA.82.052520}. 
$H^E_{5/2}$ is the hexadecapole moment that does not appear in \cite{PhysRevA.82.052520}.

Using $\bm{E=-\nabla}V$ and comparing \eqref{eq:HNRQED32} and \eqref{eq:HNRQED52} to \eqref{eq:HQspec} gives, \smash{$Q^E_{3/2}=\Qcartzzspec{3/2}$} and \smash{$Q^E_{5/2}=\Qcartzzspec{5/2}$}. Calculating the stretched-state matrix element \eqref{eq:streched} from \eqref{eq:HNRQED52} and using the Laplace equation for $V$, and its implications to the fourth derivatives of $V$ gives \smash{$H^E_{5/2}=\Qcartzzzzspec{5/2}$}.
We emphasize again that \smash{$Q^E_{3/2}$}, \smash{$Q^E_{5/2}$}, and \smash{$H^E_{5/2}$} are \emph{spectroscopic} moments. 
In the EFT they are independent coefficients. 
Given a microscopic theory of $^{229}$Th there could be relations between different Wilson coefficients. 

Let us comment on the completeness of the basis shown in \eqref{eq:HNRQED32} and \eqref{eq:HNRQED52}. 
Based on \cite{Lorce:2009br}, we expect that both the spin 3/2 and spin 5/2 particles have an electric charge form factor and an electric quadrupole form factor. 
The value of the electric charge form factor at zero ($q^2=0$) determines the total charge $Z$, the slope at zero determines $\langle r_E^2 \rangle$, and the second derivative at zero determines $\langle r_E^4 \rangle$, see , e.g.,~\cite{Assi:2025nim}. 
The value of electric quadrupole form factor at zero ($q^2=0$) determines the quadrupole moment $Q^E$ and the slope at zero determines $Q^E\langle r_Q^2 \rangle$. 
For spin 5/2 particles we also have the hexadecapole form factor, whose value at zero ($q^2=0$) determines the hexadecapole moment $H^E_{5/2}$. Thus the Wilson coefficients of possible operators linear in $\bm{E}$ with one spatial derivative can depend only on $\langle r_E^2 \rangle$ and $Q^E$ and, as a result of Lorentz invariance \cite{Heinonen:2012km}, Wilson coefficients of lower dimensional operators.
Operators with three spatial derivatives can depend only on $\langle r_E^4 \rangle$, $Q^E\langle r_Q^2 \rangle$, and $H^E_{5/2}$ and, as a result of Lorentz invariance \cite{Heinonen:2012km}, Wilson coefficients of lower dimensional operators. 
In the $M\to \infty$ limit the contribution of Wilson coefficients of lower dimensional operators to higher dimensional operators vanish. See, for example, the spin-independent operators in \cite{Hill:2012rh} with the Wilson coefficients $c_{X1}, c_{X2}, c_{X3}$.
Thus we expect the Hamiltonians shown in equations \eqref{eq:HNRQED32} and \eqref{eq:HNRQED52} to contain all the NRQED operators linear in $\bm E$ that are finite in the $M\to\infty$ limit.

Note that with only one lattice site the \Th hyperfine splitting probes only a linear combination of $Q^E_{5/2,3/2}+Q^E_{5/2,3/2}\langle r^2_{Q,5/2,3/2} \rangle$ and not each operator alone. 
In order to disentangle them, at least two sites or different crystals and theory inputs of the fields gradient are required.
Alternatively, one can use atomic spectroscopy (not in a lattice), see relevant discussion in~\cite{Assi:2025nim}.

\subsection{Energy-Density-Functional (EDF) Descriptions}
\label{sec:edf}
We briefly review two recent energy-density-functional~(EDF) studies
that provide a complementary microscopic perspective on the structure
of the $^{229}$Th clock states.
In these approaches, nuclear deformation and electromagnetic properties
are calculated self-consistently from an underlying nuclear EDF.
A central challenge is the exceptionally small energy splitting between
the ground and isomeric states: the observed $\sim 8\,\mathrm{eV}$
transition energy is many orders of magnitude below typical nuclear
energy scales and is sensitive to pairing, polarization, and collective
correlations.
The two studies discussed below examine, respectively, the electromagnetic
properties of the clock states and the origin of their near-degeneracy.

Restrepo-Giraldo \textit{et al.}~\cite{Restrepo-Giraldo:2026blh}
employ a multi-reference density functional theory (MR-DFT) framework.
Their calculations start from self-consistent blocked quasiparticle
configurations that can break parity and rotational symmetry,
allowing for octupole deformation.
Parity and angular momentum are subsequently restored, and
different configurations are mixed.
Several Skyrme functionals are considered, including both time-even
and time-odd contributions to the mean field.
To reduce uncertainties associated with the description of octupole
deformation, the authors correlate their predictions with the intrinsic
octupole moments of the neighboring even-even nuclei
$^{226}$Ra and $^{230}$Th, using the measured moments to constrain
the results.

The calculation provides a reasonably good simultaneous description
of the M1 transition strength $B({\rm M1})$ and several electromagnetic
moments of the two clock states.
In particular, the predicted spectroscopic electric quadrupole moments,
\begin{equation}
    Q(5/2^+) = 2.98(5)\,\mathrm{b},
    \qquad
    Q(3/2^+) = 1.728(12)\,\mathrm{b},
\end{equation}
agree within a few percent with the measured values
$3.11(2)\,\mathrm{b}$ and $1.77(1)\,\mathrm{b}$, respectively.
The calculated $B({\rm M1})$ is also consistent with experiment,
although its theoretical uncertainty is considerably larger.
Comparing different variants of the calculation, the authors find
that configuration mixing has a relatively small effect,
while time-odd core polarization is important for magnetic observables.
Most importantly for our discussion, imposing reflection symmetry,
and thus suppressing octupole deformation, produces a significantly
larger M1 transition strength:
the predicted $B({\rm M1})$ increases to approximately
$0.17$--$0.19\,\mu_N^2$, about five times the measured value
of $0.0388(12)\,\mu_N^2$.
This demonstrates the importance of reflection-asymmetric correlations, such as octuple deformation,
for reproducing the observed electromagnetic properties.

A complementary EDF study by Minkov and P\'alffy~\cite{Minkov:2026vlc}
uses a self-consistent Skyrme--Hartree--Fock--BCS approach with the
SIII functional to investigate the near-degeneracy of the
$5/2^+$ and $3/2^+$ states.
By varying the pairing strengths, the authors find that
reflection-symmetric solutions retain an energy splitting at the
keV scale.
In contrast, allowing axial octupole deformation leads to a crossing
of the two levels, making sub-keV splittings possible.
While sufficient fine tuning of the pairing strengths can produce
energies close to the observed $\sim 8\,\mathrm{eV}$,
the estimated numerical accuracy is only $\sim 0.1\,\mathrm{keV}$.
Thus, the calculation demonstrates a possible microscopic mechanism
for the near-degeneracy but does not provide an eV-level prediction.

The calculated Coulomb-energy difference between the clock states
also depends strongly on reflection asymmetry.
Its implications for the sensitivity enhancement factor $K$
are discussed separately in \cref{sec:edf-sens}.

Together, these studies indicate that reflection-asymmetric
correlations play an important role in both the electromagnetic
properties and the near-degeneracy of the $^{229}$Th clock states.
However, quantitative predictions remain sensitive to the choice
of EDF, pairing treatment, and numerical implementation.
This motivates complementary approaches that can be constrained
more directly by experimentally accessible observables.

\subsection{Halo-Inspired Model}
\label{subsec:halo-model}

In Ref.~\cite{Caputo:2024doz}, $^{229}$Th was described by a model inspired by halo effective field theory. 
The nucleus is separated into a spin-zero \iso{Th}{228} core and a single neutron in an $l=2$ orbital. 
Coupling the neutron spin to the orbital angular momentum produces the $I=5/2$ ground state and the $I=3/2$ isomer, which differ by the relative orientation of the two angular momenta. 
This is a simple description of the clock states, but not a systematically controlled halo EFT for \iso{Th}{229}: 
the separation of scales is modest and $d$-wave observables can retain short-distance sensitivity.

An important difference from the particle--rotor model (see \cref{sec:rotorspec,sec:particle-rotor}) is that the halo model constructs only laboratory-frame states $\ket{Ilm}$, but does not make reference to a body-fixed intrinsic state. 
Its multipole matrix elements must therefore be interpreted as spectroscopic quantities, and the value conventionally quoted by experimentalists is the diagonal matrix element in the stretched state $m=I$, as defined in \cref{eq:Qsphspecdef}. 

Here, we follow the analysis of Ref.~\cite{Caputo:2024doz} for the charge radius difference and correct the quadrupole calculation.
In addition, we give an estimate of the hexadecapole moment in this halo model. 

Consider a normalized one-neutron wave function,
\begin{align}
    \label{eq:halo-wavefunction}
    \Phi_{Ilm,\alpha}(\rvec)
    &= 
    f(r) \, \Omega_{Ilm,\alpha}(\theta,\phi) \,
\end{align}
with angular part $\Omega_{Ilm,\alpha}$ and radial part $f(r)$, with the normalization $\int_0^\infty \mathrm{d}r\,r^2 f(r)^2 = 1$ and with $\alpha,\beta=1,2$ spinor indices.  $\Omega_{Ilm,\alpha}$ is a spinor that can be expressed in terms of spherical harmonics (see \cref{sec:halo-calc}), while $f(r)$ is in general unknown.
The magnetic moment of the neutron generates the one-body spin--orbit charge density~\cite{Ong:2010gf} 
\begin{align}
    \label{eq:halo-charge-density}
    \rho_{\mathrm{SO}}[\Phi_{Ilm}](\rvec)
    = 
    \frac{\mu_n}{2m_n^2}
    i\vec{\sigma}_{\alpha\beta}\cdot
    \left[
        (\vec{\nabla}\Phi_{Ilm,\alpha}^{*})
        \times
        (\vec{\nabla}\Phi_{Ilm,\beta})
    \right].
\end{align}
Here $\mu_n=-1.91$ is the dimensionless neutron magnetic moment~\cite{Ong:2010gf}.

The spin--orbit contribution to the mean-square charge radius is given by
\begin{align}
    \delta\langle r^2\rangle_{\mathrm{SO}}^{j}
    \equiv \frac{1}{Z}\int \mathrm{d}^3\rvec\,r^2
    \rho_{\mathrm{SO}}[\Phi_{Ilm}](\rvec).
    \label{eq:halo-charge-radius-def}
\end{align}
This is a scalar under rotations and hence independent of $m$.
Following Ref.~\cite{Caputo:2024doz}, the explicit calculation shown in \cref{sec:halo-calc} gives 
\begin{align}
    \delta\langle r^2\rangle_{\mathrm{SO}}^{(5/2)}
    = 2\frac{\mu_n}{Z m_n^2} 
     \, , \qquad\qquad
    \delta\langle r^2\rangle_{\mathrm{SO}}^{(3/2)}
    = -3\frac{\mu_n}{Zm_n^2}
     \, ,
     \label{eq:delta-r2-halo-52-32}
\end{align}
such that
\begin{align}
    \label{eq:halo-charge-radius-results}
    \Delta\langle r^2\rangle_{\mathrm{SO}}
    \equiv
    \delta\langle r^2\rangle_{\mathrm{SO}}^{(3/2)}
    -\delta\langle r^2\rangle_{\mathrm{SO}}^{(5/2)}
    = -5\frac{\mu_n}{Zm_n^2}
    = +0.0047\,\mathrm{fm}^2\,.
\end{align}
The first laser-spectroscopic extraction found $\Delta\langle r^2\rangle=0.012(2)\,\mathrm{fm}^2$ \cite{Thielking:2017qet}, while the refined atomic-structure analysis of Ref.~\cite{Safronova:2018fue} gives
\begin{align}
    \Delta\langle r^2\rangle_{\mathrm{exp}}
    =0.0105(13)\,\mathrm{fm}^2\,.
    \label{eq:halo-charge-radius-exp}
\end{align}
The halo prediction is factor of $\sim 2$ smaller than the observed central value, which can be considered as a good agreement.
Note that the expressions in \cref{eq:delta-r2-halo-52-32} do not depend on the details of the (in general unknown) radial wave function $f(r)$.

The charge-normalized spin--orbit contribution to the Cartesian spectroscopic quadrupole moment $\Qcartzzspec{I}$ is
\begin{align}
    q_{zz}^{(I)}
    \equiv \frac{\Qcartzzspec{I}}{Ze}
    = \frac{1}{Z} \int \! \mathrm{d}^3\rvec\,
      r^2(3\cos^2\theta-1) \,
      \rho_{\mathrm{SO}}[\Phi_{Ilm}](\rvec)
      \bigg|_{m=I}\,.
    \label{eq:halo-quadrupole-def}
\end{align}
Evaluating this for the maximally stretched $I=5/2$ and $I=3/2$ states gives
\begin{align}
    q_{zz}^{(5/2)} 
    = 
    -2\frac{\mu_n}{Zm_n^2} \, , \qquad\qquad
    q_{zz}^{(3/2)}
    = 
    \frac{3}{5}\frac{\mu_n}{Zm_n^2} \, ,
\end{align}
such that
\begin{align}
    \label{eq:halo-quadrupole-results}
    \Delta q_{zz}
    \equiv 
    q_{zz}^{(3/2)}-q_{zz}^{(5/2)}
    = 
    \frac{13}{5}\frac{\mu_n}{Zm_n^2}
    = 
    -0.0024\,\mathrm{fm}^2\,.
\end{align}
Note that these expressions are again independent of $f(r)$.

The ground state quadrupole moment was measured as $\Qcartzzspec{\gs}=3.11(2)\,e\,\mathrm{b}$ \cite{Porsev:2021tjc}, and the ratio $\Qcartzzspec{\m}/\Qcartzzspec{\gs}$ was found to be $\Qcartzzspec{\m}/\Qcartzzspec{\gs}=0.57003(1)$~\cite{Zhang:2024ngu}, which implies $\Qcartzzspec{\m} = 1.77(1)\,e\,\mathrm{b}$.
We can then extract
\begin{align}
    \label{eq:halo-quadrupole-exp}
    \Delta q_{zz}^{\mathrm{exp}}
    &=\frac{\Qcartzzspec{\m}-\Qcartzzspec{\gs}}{eZ}
     =-1.49(1)\,\mathrm{fm}^2 \, .    
\end{align}
which is significantly larger than the value predicted in the halo model.
More precisely, the ratio between the predicted and measured values is
\begin{align}
    \label{eq:halo-quadrupole-ratio}
    \frac{\Delta q_{zz}}{\Delta q_{zz}^{\mathrm{exp}}}
    \sim
    2\times10^{-3} \, .
\end{align}
The apparently closer agreement reported in Ref.~\cite{Caputo:2024doz} was due to a missing normalization factor $1/Z$, the use of the intrinsic rather than spectroscopic quadrupole moment, and considering the $m=1/2$ matrix elements rather than the fully stretched ones in the conventional experimental normalization. 
Note that in the standard Halo EFT for $l=2$ states, a local counterterm contribution to the quadrupole moment appears at leading order in the power counting \cite{Braun:2018hug}. 
In this theory the quadrupole moment is generated by photons coupling to the core in a UV sensitive loop diagram with $l=2$ coupling of the neutron to the core. 
While the theory is not directly applicable to the Thorium case, such a counterterm could accommodate the above discrepancy. 
This same counterterm would also contribute to E2 transition matrix elements.

We can also estimate the charge-normalized Cartesian spectroscopic hexadecapole moments in the halo model: 
\begin{align}
    \label{eq:halo-hexadecapole-def}
    q_{zzzz}^{I}
    &\equiv \frac{\Qcartzzzzspec{I}}{eZ}
    = 
    \left. \frac{1}{Z}\int \mathrm{d}^3\rvec\, r^4
    (35\cos^4\theta-30\cos^2\theta+3)
    \rho_{\mathrm{SO}}[\Phi_{Ilm}](\rvec)
    \right|_{m=I} \, .
\end{align}
Unlike its counterparts for the charge radius and the quadrupole moment, this expression does depend on the radial wave function $f(r)$.
This dependence enters through the radial integral
\begin{align}
    \label{eq:halo-radial-moment}
    \mathcal R_4
    \equiv \int_0^\infty \! \mathrm{d}r\,r^4f(r)^2
    \equiv
    \tilde p\,\mathrm{fm}^2,
\end{align}
where the last equality defines $\tilde p$ as a UV sensitive order one number. 
Evaluating the angular integral (see \cref{sec:halo-calc}), we find
\begin{align}
    q_{zzzz}^{(5/2)}
    =
    \frac{24}{7}\frac{\mu_n}{Zm_n^2}\mathcal R_4 \, , 
    \qquad \qquad 
    q_{zzzz}^{(3/2)}
    =
    0 \, ,
\end{align}
and therefore
\begin{align}
    \label{eq:halo-hexadecapole-results}
    \Delta q_{zzzz}
    \equiv 
    q_{zzzz}^{(3/2)}-q_{zzzz}^{(5/2)}
    =
    3.2\times10^{-3}\,\tilde p\,\mathrm{fm}^4 \, .    
\end{align}
Recall that the hexadecapole moment for the $I=3/2$ isomer vanishes identically because a rank-four diagonal matrix element cannot exist in a $I=3/2$ multiplet according to the Wigner--Eckart theorem, in agreement with the selection rule discussed in \cref{sec:Thexadexapole}. 

As we discussed above, the hexadecouple moment cannot be extracted from hyperfine spectroscopy as there are not enough independent observables to fix both the quadrupole and the hexadecouple moments.
Moreover, as \cref{eq:halo-hexadecapole-results} is UV sensitive via the parameter $\tilde p$, it should be regarded as an order of magnitude estimate only. 
In principle, $\tilde{p}$ could be estimated from nuclear hexadecapole measurements obtained by spectroscopy of muonic atoms~\cite{Zumbro:1984zz, Zumbro:1986qcf}.

\subsection{Particle--Rotor Model}
\label{sec:particle-rotor}

Most heavy nuclei are deformed. 
Although the underlying nuclear interactions are rotationally invariant, a nucleus with sufficiently many valence nucleons lowers its energy by assuming a deformed, non-spherical equilibrium shape.
The \emph{axial rotor model} takes this shape to be rigid and axially symmetric, reorienting as a whole without changing its internal structure; shape vibrations and triaxial deformations are neglected.
This rigidly rotating body is the \emph{rotor}.

The deformation singles out a body-fixed orientation and thereby spontaneously breaks rotational symmetry.
It is therefore convenient to work in two frames:
In the \emph{intrinsic} (body-fixed) frame the nucleus has a fixed, axially symmetric charge distribution whose multipole moments encode the nuclear structure.
Because such a configuration singles out a spatial orientation, it is not a state of definite angular momentum.
Definite angular momentum is recovered in the \emph{laboratory frame}, where the orientation of the rotor is specified by the Euler angles $\omega$, and quantizing the collective rotation of the rotor generates the physical states with definite total spin $I$.
All observables, including the multipole interaction \cref{eq:Hsph}, are defined in the laboratory frame, whereas the nuclear structure is simplest in the intrinsic frame. 
The particle--rotor model bridges the two.

\begin{figure}
    \centering
    \includegraphics[width=0.45\linewidth]{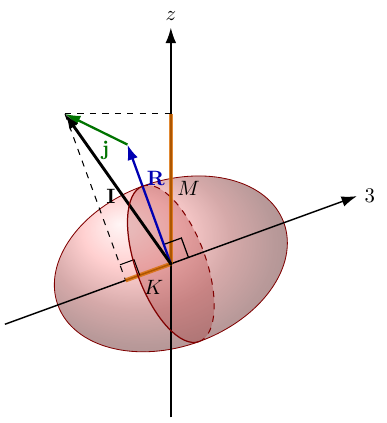}
    \caption{Illustration of the particle--rotor picture. 
    The total nuclear angular momentum $\mathbf{I}$ is the sum of the rotor angular momentum $\mathbf{R}$ and the intrinsic angular momentum $\mathbf{j}$ of the unpaired nucleon. 
    $K$ denotes the projection of $\mathbf{I}$ onto the symmetry axis (3-axis) and $M$ its projection onto the laboratory-frame quantization axis ($z$-axis). 
    When \Th is embedded in a crystal, the latter is determined by the crystal environment.}
    \label{fig:particlerotor}
\end{figure}

An odd-$A$ nucleus such as \Th can be described as an even--even core, which carries the collective rotation, and a single unpaired nucleon that, in the strong-coupling limit, adiabatically follows the deformed core. 
This is the so-called particle--rotor picture
illustrated in \cref{fig:particlerotor}.
The total angular momentum is $\mathbf{I}=\mathbf{R}+\mathbf{j}$, where $\mathbf{R}$ denotes the collective angular momentum of the core and $\mathbf{j}$ the intrinsic angular momentum of the unpaired nucleon.
Since the core is unchanged by rotation about its symmetry axis, $\mathbf{R}$ has no component along that axis.
The projection $K$ of $\mathbf{I}$ onto the symmetry axis is therefore supplied entirely by the unpaired nucleon, $K=j_3$.
Axial symmetry implies that $K$ is conserved, so it labels the intrinsic configuration.
Different rotational bands correspond to different intrinsic configurations characterized by $K$.

Quantizing the collective rotation turns a single intrinsic configuration $K$ into a rotational band, a sequence of laboratory-frame states of total spin $I=|K|,|K|+1,|K|+2,\ldots$, beginning at the band head $I=|K|$.\footnote{Even--even nuclei are the special case in which the ground state configuration has all nucleons paired, so $\mathbf{j}=0$. 
The ground-state band then has $K=0$ and is built by collective rotation alone.}
For deformed odd-mass nuclei such as \Th, the level energies are~\cite{Papenbrock:2020zhh},
\begin{align}
    \label{eq:rotorEn}
    E^{(I,K)}=E_{|K|} + \frac{I(I+1)-K^2}{2C_0}\,,
\end{align}
plus a correction which is relevant only for states with $|K|=1/2$. 
Here $C_0$ is the moment of inertia and $E_{|K|}$ sets the band-head energy, the intrinsic energy of the band.

The moment of inertia $C_0$ sets the scale of the rotational excitations.
It is useful to consider its two limiting cases:
As $1/C_0\to \infty$, the rotational excitation energies diverge for every $I>|K|$, pushing all rotational excitations to infinity and leaving only the band head at finite energy.

As $1/C_0\to 0$, the rotational term vanishes and the whole band collapses to a single degenerate level at $E_{|K|}$.
Any superposition of the band members is then stationary, including  oriented configurations $\mathcal{R}(\omega)|\chi_K\rangle$, which form a continuous family of degenerate ground states related by the rotations $\mathcal{R}(\omega)=e^{-i\boldsymbol{\omega}\cdot \mathbf{I}}$.
The selection of a particular orientation corresponds to spontaneous breaking of rotational symmetry.
The rotational band $I=|K|, |K|+1, \ldots$ thus forms a tower of excitations whose energy gaps vanish in the limit $1/C_0\to 0$, analogous to Goldstone modes associated with the broken continuous symmetry.

\Th lies between these two limits.
The band members are split by nonzero but finite excitation energies $\sim 1/C_0$ and therefore evolve with different phases.
The oriented intrinsic configuration $|\chi_K\rangle$, represented by a superposition of band members, is consequently not stationary and conversely, the intrinsic deformation is not directly observable as a stationary orientation of the nucleus.
What is observable is what the deformation generates in the laboratory frame: the pattern of rotational excitations built on each band head and the relations between the multipole moments of the individual band members.

For an axially symmetric rotor, the intrinsic multipole moment $\Qsph{k}^\Kband$ is related to the spectroscopic moment of a state with angular momentum $I$ by geometrical projection factor \smash{$\mathcal{P}_k^{(I,K)}$}, given in \cref{eq:PkIK}. 
This relation is given in \cref{eq:QsphsQsph}.
The projection factor \smash{$\mathcal{P}_k^{(I,K)}$} is fixed entirely by angular-momentum geometry; its derivation from the rotor wave function is given in \cref{app:rotormultipoles}.

\subsubsection{Application to the Clock States}

The applicability of this description to the clock states is supported by the observed rotational-band structure of $^{229}$Th (and $^{228}$Th)~\cite{KONDEV2026397}. 
In \Th, the $I=5/2$ ground state and the $I=3/2$ isomer are the band heads of the $K=5/2$ and $K=3/2$ bands, respectively, which we denote by $\Gband$ and $\Mband$.
While the particle--rotor model does not predict the band-head energies, the rotational excitations built on them are well reproduced: the $\Gband$ levels $I=7/2$, $9/2$, $11/2,...$ and the $\Mband$ levels $I=5/2$, $7/2$, $9/2$, $11/2,...$ up to about $\SI{200}{keV}$ follow a common set of parameters $1/C_0 = \SI{12.2}{keV}$, $E_{|5/2|}= \SI{-15.6}{keV}$, and $E_{|3/2|}=\SI{-9.9}{keV}$.

The two bands are built on different intrinsic configurations and therefore carry distinct intrinsic moments \smash{$\Qsph{k}^\Gband$} and \smash{$\Qsph{k}^\Mband$}.
Applying the relation between the intrinsic and spectroscopic moments in \cref{eq:QsphsQsph} to the two band heads, $I=K=5/2$ and $I=K=3/2$, yields the coefficients quoted in \cref{eq:QsphsQsphNuclearClock}.
In particular, the spectroscopic hexadecapole moment of the isomer vanishes because $k=4 > 2I^\m =3$.
The intrinsic hexadecapole moment of the $\Mband$ band need not vanish; it is simply inaccessible through the static hyperfine interaction of its $I=3/2$ band head.
It may instead be probed through static moments of higher-spin band members or electromagnetic transitions, for example in Coulomb excitation~\cite{CEBemis_1988}.

For the quadrupole moments, \cref{eq:QsphsQsph} predicts $\Qsphspec{2}{\m}/\Qsphspec{2}{\gs}=14/25 \, \Qsph{2}^\Mband/\Qsph{2}^\Gband$, where the factor $14/25=0.56$ is fixed by angular-momentum geometry. 
The rotor model does not constrain the ratio $\Qsph{2}^\Mband/\Qsph{2}^\Gband$, since the two moments belong to separate intrinsic configurations.
Nevertheless, the measured spectroscopic ratio $0.57003(1)$~\cite{Zhang:2024ngu} is close to $14/25=0.56$, the value obtained for equal intrinsic quadrupole moments.
Within the particle--rotor description, the measurement thus implies $\Qsph{2}^\Mband \approx \Qsph{2}^\Gband$.
A possible microscopic interpretation for this similarity is provided by the halo--model picture discussed in \cref{subsec:halo-model}, in which the ground state and isomer differ by a spin-flip of the valence neutron. 
Combining this picture with a quadrupole-deformed core may therefore be worth exploring.

The particle--rotor model also predicts transition matrix elements between different states in a band \cite{BohrMottelsonVol2}. 
In Ref.~\cite{CEBemis_1988}, these relations were used to analyze the Coulomb excitation of $^{229}$Th by alpha particles with an energy of 17 MeV. 
The authors extracted an intrinsic quadrupole moment of $(8.816\pm 0.090)$ $e$b for the ground state band, consistent with the spectroscopic value of Ref.~\cite{Porsev:2021tjc} when the particle--rotor relations \eqref{eq:QsphsQsph} and \eqref{eq:PkIK} are used to relate the spectroscopic and intrinsic moments.
Ref.~\cite{CEBemis_1988} also obtained an intrinsic hexadecapole moment of $(3.69\pm 0.72)$ $e$b$^2$ for the ground state band. 
Although this extraction is model-dependent since it relies on particle--rotor relations, it provides the only experimental information on the hexadecapole moment of the ground state. 

The relations used here constitute the leading-order description common to the traditional particle--rotor model~\cite{BohrMottelsonVol1,BohrMottelsonVol2} and modern collective effective field theories~\cite{Papenbrock:2020zhh}.
More detailed descriptions may additionally include inter-band Coriolis couplings and triaxial corrections, which are neglected above.

\subsubsection{Beyond the Rigid Particle--Rotor Description}

A more elaborate realization of the particle--core picture is provided by the Coherent Quadrupole--Octupole Motion (CQOM) framework \cite{Minkov:2017kju,Minkov:2018qqm,Minkov:2021ovq}. 
In this approach, the odd neutron moves in a reflection-asymmetric deformed mean field and is coupled to collective quadrupole and octupole degrees of freedom of the even-even core. 
In contrast to a rigid particle--rotor description, the quadrupole and octupole deformations are treated as dynamical collective coordinates, whose coupled motion gives rise to rotational--vibrational excitations. 
Pairing and Coriolis interactions are included, with the latter inducing mixing between different single-particle configurations. For $^{229}$Th, the $5/2^+$ ground state
and the $3/2^+$ isomer are associated predominantly with the $[633]5/2$ and $[631]3/2$ neutron configurations, respectively, coupled to the quadrupole--octupole collective motion of the core.

The CQOM framework has been applied in a series of studies to the low-energy spectrum and electromagnetic properties of $^{229}$Th~\cite{Minkov:2017kju,Minkov:2018qqm,Minkov:2021ovq}. 
The model parameters are constrained phenomenologically by the observed rotational-band structure, while the interplay between the single-particle configurations, quadrupole--octupole collectivity, and Coriolis mixing determines the properties of the clock states. 
These studies calculated the reduced $M1$ and $E2$ transition probabilities between the isomer and ground state, their magnetic dipole moments, and the structure of the associated rotational bands. 
In particular, the inclusion of Coriolis mixing was found to play an important role in reproducing the strongly suppressed isomeric transition and the magnetic moment of the isomer. Overall, CQOM provides a phenomenological description in which the observed spectroscopic properties of the two clock states emerge from the combined single-particle and collective quadrupole--octupole dynamics.

\subsection{Deformed Fermi Model}
\label{subsec:geometrical}

The simplest, and perhaps the most naive, approach to the electromagnetic properties of the nucleus is to model it as an extended charge distribution using classical electrostatics. 
Despite that, geometric models have been widely used to estimate the sensitivity of the nuclear clock to new physics~\cite{Berengut:2009zz,Fadeev:2020bgt,Beeks:2024xnc,Caputo:2024doz}.
In such models, the nuclear charge distribution is characterized by its overall size and a set of multipole
deformations.  
Such a parametrization is not a systematic expansion:
there is no underlying power counting that establishes a hierarchy
among the different multipoles or controls the truncation of the expansion. 
The deformation parameters must therefore be fixed by
experimental input whenever possible. 
When the corresponding observable is not known, the model can quantify the dependence on that deformation but it does not predict its magnitude.

We adopt a deformed Fermi charge distribution to calculate the nuclear Coulomb energy and intrinsic multipole moments,
\begin{align}
    \rho(r,\Omega) 
    = 
    \frac{\rho_0}{1+e^{\frac{r-R(\Omega)}{a}}}\,,
    \qquad 
    R(\Omega) 
    = 
    R_0 \bigg(1+\sum_{k=1}^\infty \beta_k \sphY{k}{0}{\Omega} \bigg) \, ,
    \label{eq:rho_WS}
\end{align}
where $R_0$ sets the overall nuclear size, $a$ is the surface diffuseness, and $\beta_k$ are axial deformation parameters.
We then convert the moments to spectroscopic ones measured for states of definite angular momentum (see \cref{eq:QsphsQsph}). 
The restriction to axial deformations is motivated by the description of the low-lying spectrum of \Th in terms of rotational bands built on approximately axially symmetric intrinsic configurations (see \cref{sec:particle-rotor}).
We place the origin at the center of charge, so that $\beta_1=0$ (see \cref{app:K}).

Since the ground state and isomer belong to different rotational bands (see \cref{sec:particle-rotor}), we allow each state to have its own set of deformation parameters, $\{\beta_k^\gs\}$ and $\{\beta_k^\m\}$.
As discussed in \cref{sec:observables}, the available spectroscopic information constrains only a limited set of spectroscopic moments, which in turn provide access to only a subset of the intrinsic multipole moments.

Assuming deviations from sphericity are small, i.e.\ $\beta_k \ll 1$, we expand $\rho$, $\rhoo$ and $\Roo$ up to second order in the deformation parameters $\beta_k$ (see
\cref{app:rho_exp}).  
The zeroth-order parameters are chosen such that the total charge $Ze$ and mean-square charge radius $\chrsq$ take their measured values already at $\mathcal{O}(\beta_k^0)$:
\begin{align}
    \rhoo = -\frac{Ze}{8\pi a^3 \Li{3}{-b}}\,, \qquad\qquad
    \Roo = \sqrt{\frac{5}{3}\chrsq - \frac{7}{3}(\pi a)^2}
    \label{eq:rhooRoo}
\end{align}
with $b \equiv e^{\Roo/a}$, and where $\Roo$ follows from a Sommerfeld expansion of $\chrsq$ in $a/\Roo$.
When deformations are taken into account, shifting $R(\Omega)$, $\mathcal{O}(\beta_k^2)$, corrections $\rhoii$ and $\Roii$ need to be applied to $\rhoo$ and $\Roo$ to hold $Ze$ and $\chrsq$ fixed at their observed values (see \cref{app:rhoRo}).

Note that our approach differs from the constant-volume prescription (or ``liquid-drop'' model) often found in the literature~\cite{BohrMottelsonVol2,HasseMyers}, which imposes that the nuclear \emph{volume} be the same with and without deformations and leads to a different $\mathcal{O}(\beta_k^2)$ Coulomb correction.
The advantage of our approach is that it is better suited for a data-driven analysis because nuclear charge radii are measurable, while the nuclear volume is not. 
The relation between the two prescriptions is discussed in \cref{app:scheme-dependence}.

Numerically, we use the measured ground-state mean-square charge radius~\cite{Angeli:2013epw}, its isomer shift~\cite{Thielking:2017qet, Safronova:2018fue}, and the spectroscopic quadrupole information summarized in \cref{tab:expinput}.  
The surface diffuseness is not directly constrained by data, and we choose $a^\gs = a^\m = \SI{0.5}{fm}$, representative of heavy nuclei~\cite{DeVriesDeVries}.  
\Cref{eq:rhooRoo} then fixes $\rhoo$ and $\Roo$ for each clock state, while the quadrupole deformations $\beta_2^\gs$ and $\beta_2^\m$ are determined from the ground-state spectroscopic quadrupole moment and its isomeric ratio.
To compute the Coulomb energy to $\mathcal{O}(\beta_k^2)$, the relation between the deformation parameters and the intrinsic multipole moments is needed only to
linear order,
\begin{align}
    \Qsph{k>0} = \beta_k C_k + \mathcal{O}(\beta^2)\,,
\end{align}
with $C_k$ a function of $Z$, $\chrsq$ and $a$ (see \cref{app:relation_to_observables}).  
Thus, within this parametrization, the Coulomb energy difference can be expressed directly in
terms of measured properties of the two clock states. 
Its dependence on higher multipoles, for which no spectroscopic measurement is available, is kept explicit rather than replaced by additional nuclear-structure assumptions.
These unconstrained deformations remain a source of model uncertainty in the determination of the new physics sensitivity of the nuclear clock.

\begin{table}[t]
    \centering
    \begin{tabular}{c@{\qquad}c@{\qquad}c@{\qquad}c}
    \toprule
    Parameter & Value & Reference & Notation in Ref.\\
    \midrule
    $\langle r^2\rangle^\gs~[\mathrm{fm}^2]$ & $(5.7557(143))^2$  & \cite{Angeli:2013epw} & $R$\\
    $\delta\langle r^2\rangle~[\mathrm{fm}^2]$   &  $0.0105(13)$  & \cite{Thielking:2017qet,Safronova:2018fue} & $\delta \langle r^2\rangle^{229\mathrm{m},229}$  
    \\
    $\Qcartzzspec{\gs}~[e\mathrm{b}]$    & $3.11(2)$ & \cite{porsev2021precision} 
       &  $Q$\\
    $\RQ \equiv \Qcartzzspec{\m}/\Qcartzzspec{\gs}$ & 0.57003(1) & \cite{Zhang:2024ngu}
     & $Q_{\rm is}/Q_{\rm g}$\\
    \bottomrule
    \end{tabular}
    \caption{Experimental input used in the numerical evaluation of $K$.
    Here we use the notation of the respective references. 
    A translation into our conventions is given in Table~\ref{tab:derivedinput}.}
    \label{tab:expinput}
\end{table}
\vspace*{2cm}

\section{Estimating the New Physics Sensitivity}
\label{sec:K}

We now turn to the sensitivity of the nuclear clock transition to variations of fundamental constants.
The exceptionally small ground-state--isomer splitting, $\Delta E\sim 8\,\mathrm{eV}$, is generally understood to arise from a cancellation between much larger electromagnetic and strong-interaction contributions~\cite{FlambaumOrig,FlambaumAuerbachDmitriev},
\begin{align}
    \label{eq:DeltaE}
    \Delta E
    =
    \Delta E_{\rm em}+\Delta E_{\rm nuc}\,.
\end{align}
The electromagnetic and nuclear contributions cannot be determined separately from the measured transition energy and must be estimated within a theoretical description. 
Since the total transition energy is precisely measured, it is sufficient to model theoretically either $\Delta E_{\rm em}$ or $\Delta E_{\rm nuc}$.

A variation of $\alpha_{\rm em}$ perturbs the cancellation between electromagnetic and nuclear contributions, producing a fractional shift in the transition frequency.
Within a microscopic nuclear theory, the sensitivity of the transition energy to such variations can be calculated using the Hellmann--Feynman theorem~\cite{Hellmann:1933,Feynman:1939zza,Litvinova:2009vp}.
For variations of the fine-structure constant, the corresponding enhancement factor is
\begin{align}
    \label{eq:K}
    \KNP
    &\equiv
    \frac{1}{\Delta E} 
    \pdv{\Delta E}{\log\alpha_{\rm em}}
    \simeq
    \frac{1}{\Delta E}
    \pdv{\Delta E_{\rm em}}{\log\alpha_{\rm em}}
    \simeq
    \frac{\Delta E_{\rm em}}{\Delta E}\,,
\end{align}
where the approximation assumes that the dominant electromagnetic contribution is linear in $\alpha_{\rm em}$. 
Previous nuclear-structure estimates suggest electromagnetic energy differences of order $100~\mathrm{keV}$~\cite{FlambaumAuerbachDmitriev,Litvinova:2009vp,Beeks:2024xnc}, corresponding to an enhancement $|\KNP| \sim 10^4$.

The central theoretical challenge is therefore not the small denominator $\Delta E$, which is experimentally determined with high precision, but to refine the prediction of the non-observable quantity $\Delta E_{\rm em}$.
The nuclear descriptions introduced in \cref{sec:theory} provide different theoretical approaches to estimating  $\KNP$.
We briefly discuss their implications for the sensitivity before turning to the deformed Fermi charge-density model, which allows us to estimate $\Delta E_{\rm em}$ using measured nuclear observables and to identify the remaining model-dependent contributions.

\subsection{Sensitivity in NRQED}
\label{subsec:sensitivity-nrqed}

To get a rough estimate of the electromagnetic energy difference we can look at the analogous problem of the electromagnetic contribution to the nucleon mass. 
In particular, we consider the so-called Born term from~\cite{Tomalak:2018dho} which is proportional the electromagnetic form factors $G_E$, and take the limit of an infinite proton mass. 
While for a proton the scale of the mass ($\simeq \SI{1}{GeV}$) and the scale of the inverse of the charge radius ($\simeq \SI{0.2}{GeV}$) are close \cite{ParticleDataGroup:2026mpi}, for a heavy nucleus such as $^{229}$Th, the scale of its mass ($\simeq \SI{210}{GeV}$) \cite{ParticleDataGroup:2026mpi} and the inverse of the charge radius ($\simeq \SI{0.34}{GeV}$) are well-separated and the infinite mass limit is justified. 
In this limit we find from~\cite{Tomalak:2018dho}
\begin{equation}
    E_{\mbox{\scriptsize Born}}=\frac{\alpha_{\mbox{\scriptsize em}}}{2\pi}\int_0^\infty dQ^2\,\dfrac{G_E(Q^2)^2}{Q}\,.
\end{equation}
We now expand $G_E(Q^2)=Z(1-Q^2\langle r^2\rangle/3!+\cdots)$ and keep only the first two terms. 
Integrating over $Q$ with a cutoff $\Lambda$ gives 
\begin{equation}
    \label{eq:Born}
    E_{\mbox{\scriptsize Born}}
    = \frac{\alpha_{\mbox{\scriptsize em}}}{\pi}\int_0^\Lambda dQ \left(Z^2-\frac13Q^2Z^2\langle r^2\rangle+\cdots\right)
    = \frac{\alpha_{\mbox{\scriptsize em}}}{\pi}Z^2\Lambda-\frac{\alpha_{\mbox{\scriptsize em}}}{9\pi}Z^2\Lambda^3\langle r^2\rangle+\cdots\,.
\end{equation}
We can obtain the same $\langle r^2\rangle$ dependence by calculating the lowest order contribution of the charge radius to the NRQED self energy diagram. 
We use the NRQED Feynman rules from \cite{Dye:2018rgg} with the Wilson coefficient of the Darwin term taken as $c_D/8M^2=Z\langle r^2\rangle/6$ in the $M\to\infty$ limit. This relation holds for both spin-0 \cite{Assi:2025nim} and spin-1/2 \cite{Hill:2012rh} and we assume it holds for higher spins too. 
The relevant contribution involves a time-like photon exchange between the minimal coupling and the Darwin vertices. 
Using Feynman gauge, we have in the $M\to\infty$ limit 
\begin{equation}
    -i\Sigma
    = \frac{e^2Z^2\langle r^2\rangle}3\int \dfrac{d^4q}{(2\pi)^4}\dfrac{\vec {q}^{2}}{(q^2+i\epsilon)(q^0+i\epsilon)}\,,
\end{equation}
where $-i\Sigma$ is the one-loop expression for the self energy (the sum of all one-particle-irreducible diagrams) see, e.g. Ref.~\cite{Grozin:2005yg}.
Integrating over $q^0$ using the residue theorem we find
\begin{equation}\label{eq:NRQED_Self_Energy}
    \Sigma
    = -\frac{\alpha Z^2\langle r^2\rangle}{3\pi}\int_0^\Lambda d|\vec{q}|\,\vec {q}^{2}
    = -\frac{\alpha Z^2\langle r^2\rangle\Lambda^3}{9\pi}\,.
\end{equation}
This is exactly the second term in Eq.~\eqref{eq:Born}. 

Since the ground state and isomer have the same charge,  the EM energy difference from \eqref{eq:Born} or \eqref{eq:NRQED_Self_Energy} is 
\begin{equation}
    \Delta E_{\mbox{\scriptsize em}}
    = E^{5/2}_{\mbox{\scriptsize em}}-E^{3/2}_{\mbox{\scriptsize em}}
    = \dfrac{\alpha_{\mbox{\scriptsize em}}}{9\pi}Z^2\Lambda^3\left(\langle r_{3/2}^2\rangle-\langle r_{5/2}^2\rangle\right)\equiv\dfrac{\alpha_{\mbox{\scriptsize em}}}{9\pi}Z^2\Lambda^3\Delta\langle r^2\rangle\,.
\end{equation}
Using $\Delta\langle r^2\rangle\approx 0.01$ fm$^2$, $Z=90$, $\alpha_{\mbox{\scriptsize em}}\approx1/137$ we have 
\begin{equation}
    \Delta E_{\mbox{\scriptsize em}}
    = E^{5/2}_{\mbox{\scriptsize em}}-E^{3/2}_{\mbox{\scriptsize em}}\simeq 520 \mbox{ keV}\left(\dfrac{\Delta\langle r^2\rangle}{0.01 \mbox{ fm}^2}\right)\left(\dfrac{\Lambda}{100\mbox{ MeV }}\right)^3\,.
\end{equation}
For example, if we take $\Lambda=\hbar c/\sqrt{\langle r^2\rangle}= \hbar c/(5.76 \mbox{ fm})=34.2 \mbox{ MeV}$, we have $\Delta E_{\mbox{\scriptsize em}}\approx 20$ keV, , corresponding to an enhancement $|\KNP| \sim 10^3$.

We emphasize that \eqref{eq:Born} was obtained from the spin-1/2 case. 
Since it only depends on the electric form factor $G_E$, it seems plausible that a similar expression would hold for higher spins. 
Also, the NRQED calculation in \eqref{eq:NRQED_Self_Energy} did not use any spin-dependent interactions, so it seems plausible that the same result would hold for higher spins.

To estimate the contributions of the quadrupole form factor to the EM energy one needs to extend the analysis of e.g. Ref.~\cite{Tomalak:2018dho} to the spin-1 case, since it is the lowest spin where quadrupole form factor appears. 
We leave that, as well as the contribution of the hexadecapole form factor, for a future study.

\subsection{Sensitivity in EDF Descriptions}
\label{sec:edf-sens}

Microscopic EDF calculations can in principle evaluate the electromagnetic contribution to the total energy for self-consistent ground-state and isomeric solutions. 
As discussed in \cref{sec:edf}, the multi-reference DFT calculation of Restrepo-Giraldo \textit{et al.}~\cite{Restrepo-Giraldo:2026blh} was primarily aimed at electromagnetic moments and the reduced magnetic-dipole transition probability. 
It does not quote a Coulomb-energy difference between the clock states and consequently cannot be directly used to predict the sensitivity factor $\KNP$.
Nevertheless, the spread among different Skyrme functionals and the dependence of the calculated observables on the treatment of octupole correlations and time-odd fields illustrate the nuclear-structure uncertainty that a microscopic determination of $\KNP$ must ultimately control.

Minkov and P\'alffy~\cite{Minkov:2026vlc} calculated this difference within a Skyrme--Hartree--Fock--BCS approach.
For reflection-symmetric solutions, they find $|\Delta V_C|\simeq 17$--$32\,\mathrm{keV}$.
Allowing octupole deformation increases this difference to $|\Delta V_C|\simeq 50$--$300\,\mathrm{keV}$.
In the vicinity of the $3/2^+$--$5/2^+$ level crossing (where the transition energy is particularly small, as observed), the Coulomb-energy difference remains of order $100\,\mathrm{keV}$, even as the calculated transition energy falls below the keV scale.
Combined with the measured transition energy, this suggests a sensitivity enhancement $|\KNP|\sim 10^4$.
However, the location of the level crossing depends on the pairing strengths, and the numerical accuracy of the calculated transition energy is limited to about
$0.1\,\mathrm{keV}$.
Thus, the calculation supports a large sensitivity enhancement, but does not provide a precise, parameter-free prediction of $\KNP$.

\subsection{Sensitivity in the Halo-Inspired Model}

The halo-inspired description provides an estimate of the electromagnetic energy difference. 
In the model of Ref.~\cite{Caputo:2024doz}, the $5/2^+$ ground state and $3/2^+$ isomer are described as the two possible couplings of an $l=2$ neutron to the
$^{228}$Th core. 
Since the neutron is electrically neutral, the difference in electromagnetic energy arises predominantly from the electromagnetic spin--orbit interaction associated with its anomalous magnetic moment. 
This mechanism is closely related to the spin--orbit charge density used in Sec.~\ref{subsec:halo-model} to estimate the differences of electromagnetic moments.

Ref.~\cite{Caputo:2024doz} finds for the difference of the spin--orbit energies
\begin{align}
    \label{eq:halo-sensitivity-energy}
    \Delta E_{\rm em}^{\rm halo}
    \simeq
    \Delta E_{\rm SO}
    \simeq
    1.4 \times 10^2\,p~\mathrm{keV},
\end{align}
where $p$ is an order-one coefficient encoding short-distance physics.
In particular, it depends on the asymptotic normalization of the $d$-wave neutron wave function and on its continuation into the nuclear interior. 
The halo description therefore predicts the characteristic energy scale of the effect, but not its normalization independently of UV input.

Taking \cref{eq:halo-sensitivity-energy} at face value gives
\begin{align}
    \label{eq:halo-sensitivity-K}
    |\KNP|_{\rm halo}
    \simeq
    \frac{1.4 \times 10^2\,p~\mathrm{keV}}{8.4~\mathrm{eV}}
    \simeq
    1.7\times10^4\,p\,.
\end{align}
Thus, for a natural value $p=\mathcal{O}(1)$, the halo model points to an enhancement of order $10^4$. 
This estimate is noteworthy because it arrives at the same characteristic scale as the microscopic EDF and deformed Fermi charge-density calculations, despite relying on rather different physics. 
At the same time, the explicit appearance of the UV-sensitive coefficient $p$ makes clear that the halo model does not provide a precision determination of $\KNP$ without additional short-distance information.

As discussed in Sec.~\ref{subsec:halo-model}, the halo description can be confronted with measured differences of charge radii and multipole moments. 
Such comparisons provide useful tests of the long-distance structure assumed by the model, but do not by themselves determine the short-distance coefficient entering
\cref{eq:halo-sensitivity-energy}. 
Measurements sensitive to the asymptotic normalization of the neutron wave function, such as neutron capture observables proposed in Ref.~\cite{Caputo:2024doz}, could therefore help reduce this uncertainty.

\subsection{Sensitivity in the Particle--Rotor Model}
As discussed in Sec.~\ref{sec:particle-rotor}, the particle--rotor model relates the spectroscopic multipole moments of the clock states to multipole moments of their intrinsic charge distributions through
angular-momentum projection. 
This allows measured laboratory-frame quadrupole moments, and potentially higher moments, to be translated into information about the intrinsic nuclear shape.

The particle--rotor model alone, however, does not determine the electromagnetic contribution to the energy difference between the two band heads. 
In particular, the ground state and isomer belong to different rotational bands and may correspond to different intrinsic configurations. 
Their intrinsic multipole moments are therefore independent quantities unless additional dynamical assumptions are introduced. 
Knowledge of the projection factors relating spectroscopic and intrinsic moments is not sufficient to determine the Coulomb self-energy of either intrinsic state.

A prediction for $\KNP$ consequently requires supplementing the particle--rotor description with a model for the intrinsic charge density. 
In the next subsection we use a deformed Fermi charge distribution with intrinsic deformation parameters. 
The rotor relations can be used to relate the intrinsic deformation parameters to the experimentally measured spectroscopic moments.

\subsection{Sensitivity in the Deformed Fermi Model}

Following Refs.~\cite{FlambaumOrig,FlambaumAuerbachDmitriev}, we estimate the electromagnetic contribution to the transition energy by the difference in the classical Coulomb self-energies of the ground-state and isomeric charge distributions,
\begin{align}
    \Delta E_{\rm em} 
    \approx \Delta E_{\rm C}\left[\rho\right] 
    = E_{\rm C}\big[\rho^{(m)}\big] - E_{\rm C}\big[\rho^{(g)}\big]\,.
\end{align}
This approximation neglects Coulomb exchange, many-body correlations, magnetic interactions, relativistic corrections, and radiative effects such as vacuum polarization.

To go beyond Refs.~\cite{FlambaumOrig, FlambaumAuerbachDmitriev}, we calculate the Coulomb energy using the deformed Fermi charge distribution of \cref{eq:rho_WS},
\begin{align}
    E_{\rm C} 
    = 
    \frac{1}{2} \int \drth \int \dsth \frac{\rho(\rvec)\rho(\svec)}{4\pi|\rvec-\svec|}\,.
\end{align}
Expanding to second order in the deformation parameters $\beta_k$, we obtain
\begin{align}
    E_{\rm C} = E^{(0,0)} + \sum_{k>0}\beta_k^2 \mathcal{E}_k^{(2)}
                        + \mathcal{O}(\beta^3)\,,
    \label{eq:ECexp}
\end{align}
where $E^{(0,0)}$ is the spherical (monopole) contribution, and $\mathcal{E}_k^{(2)}$ are the leading deformation corrections. 
Terms linear in $\beta_k$ and mixed terms $\beta_k\beta_{k'}$ with $k'\neq k$ vanish upon angular integration by orthogonality of the spherical harmonics.
The coefficients \smash{$\mathcal{E}_k^{(2)}$} depend only on the spherical-profile parameters $\Roo$, $a$ and $Z$. 
We evaluate them using the full Fermi distribution rather than via a thin-skin expansion (expansion in $a$)~\cite{HasseMyers, Beeks:2024xnc}. 
Details and closed analytic expressions are given in \cref{app:EC-Fermi,app:EC-deformation}.

In the $a\to 0$ limit, the spherical contribution $E^{(0,0)}$ reduces to the Coulomb energy of a uniformly charged sphere.
In contrast to the constant-volume liquid-drop model, where deformation lowers the Coulomb energy~\cite{BohrMottelsonVol2,HasseMyers},
the deformation corrections $\mathcal{E}_k^{(2)}$ are positive.
This difference follows from holding the mean-square charge radius $\chrsq$, rather than the volume, fixed when relating the deformation parameters to nuclear observables.
We adopt this fixed-$\chrsq$ scheme because $\chrsq$ is an experimental observable while the nuclear volume is not. 
The relation between the two schemes is discussed in more detail in \cref{app:scheme-dependence}.

Following Refs.~\cite{Berengut:2009zz,Beeks:2024xnc,Caputo:2024doz}, we express the $\KNP$ in terms of nuclear observables, separating experimentally constrained contributions from those requiring additional model assumptions.
Linearizing the Coulomb-energy differences in the isomer shifts gives
\begin{align}
    \KNP = \sum_X \KNP_X \,,
    \qquad
    \KNP_X \equiv \frac{1}{\Delta E} \pdv{E_{\rm C}}{X} \bigg\vert_{X=\bar{X}} \delta X
    \label{eq:KNPsectorsum}
\end{align}
where $X \in \lbrace \chrsq,\,\Qsph{2},\, a\,, \ldots\rbrace$, $\delta X=X^\m - X^\gs$,  and $\bar{X}$ denotes the reference value of $X$ specified by the experimental inputs of \cref{tab:expinput}.
Each derivative is evaluated while holding the other observables (including the charge radius, where applicable) fixed. 
 
The charge-radius and quadrupole contributions are
\begin{align}
    \KNP_{\chrsq} = -1.8 \times 10^4\,, \qquad\qquad
    \KNP_{\Qsph{2}} = + 2.6 \times 10^4\,.
\end{align}
Their partial cancellation\footnote{A similar sign difference, but with very different magnitudes, was first pointed out in \cite{Berengut:2009zz}.} amplifies the relative uncertainty of the resulting sensitivity: the approximately 12\% uncertainty on $\delta \chrsq$ translates into an uncertainty of about 30\% on their sum.

Retaining only the charge-radius and quadrupole contributions and assuming  $a^\gs = a^\m = \SI{0.5}{fm}$ (i.e.\ $\delta a = 0$), we obtain (see also Refs.~\cite{Beeks:2024xnc,Caputo:2024doz}) 
$\KNP_{\leq 2} \sim 8 \times 10^3$ with a $\sim 30\%$ uncertainty from the experimental data. 
The model uncertainties are, however, much larger than this (see below).

The logarithmic sensitivities to the measured nuclear inputs and their contributions to the propagated experimental uncertainty on $\KNP$ are summarized in \cref{tab:K_sigprop}.
If one is to assume only quadratic deformation (for instance neglecting the octupole, the hexadecapole contributions etc.) then the current experimental uncertainty is dominated by the isomer shift $\delta \chrsq$, which accounts for nearly $99\%$ of it (see also discussion in Refs.~\cite{Beeks:2024xnc,Caputo:2024doz}).
The uncertainty related to the quadrupole ratio $\RQ$ (see table \ref{tab:expinput}) is much smaller because it has been determined with high precision using hyper-fine spectroscopy~\cite{Zhang:2024ngu, morawetz:2026}.

\subsubsection*{Hexadecapole Moment}

The hexadecapole contribution cannot be treated in the same way as the monopole and quadrupole contributions because the spectroscopic hexadecapole moment of the isomer is identically zero by angular momentum selection rules (see \cref{eq:QsphsQsphNuclearClock}) while the one of the ground state can be non-zero, but has not yet been measured.
To our knowledge, the only available experimental input is the measurement of the intrinsic hexadecapole moment of the ground-state rotational band, extracted in Ref.~\cite{CEBemis_1988} as $Q_{40}=3.69(72)\,e\mathrm{b}^2$ from E4 Coulomb excitation within the axial rotor model\footnote{Both Ref.~\cite{CEBemis_1988} and our \cref{eq:Qhatdef} use the same definition of the electric multipole operator, that of Bohr and Mottelson, Vol.~II~\cite{BohrMottelsonVol2}, Eq.~(6-60). Ref.~\cite{CEBemis_1988} additionally separates out a factor $\sqrt{(2k+1)/(16\pi)}$ in their definition of $Q_{k0}$ (their Eq.~(1)), such that $Q_{k0}=\sqrt{16\pi/(2k+1)}\Qsph{k}$. 
At $k=2$ this reproduces the Bohr--Mottelson intrinsic quadrupole moment $Q_0=\sqrt{16\pi/5}\Qsph{2}$.
}.
This value is equivalent\footnote{
Converting $Q_{40}$ to our conventions gives $\Qsph{4}^\Gband = \sqrt{9/(16\pi)}Q_{40}=1.56(30)\,e\mathrm{b}^2$ and using \cref{eq:QsphsQsphNuclearClock,eq:Qcarts24def}, we obtain the Cartesian spectroscopic moment \smash{$\Qcartzzzzspec{\gs}=0.35(7)e\mathrm{b}^2$}.
} to a Cartesian spectroscopic moment \smash{$\Qcartzzzzspec{\gs}=0.35(7)e\mathrm{b}^2$}.
The intrinsic hexadecapole moment of the isomeric band, \smash{$\Qsph{4}^\Mband$}, remains unconstrained.

Linearizing in the difference \smash{$\delta \Qsph{4}\equiv\Qsph{4}^\Mband-\Qsph{4}^\Gband$}, the sensitivity becomes
\begin{align}
    \label{eq:K_result}
    \KNP_{\Qsph{4}} \sim 3 \times 10^4\times
        \bigg( \frac{\Qcartzzzzspec{\gs}}{0.35\,e\mathrm{b}^2}\bigg)^{\!2} \,
        \frac{ \delta \Qsph{4}/\Qsph{4}^\Gband}{2\delta \Qsph{2}/\Qsph{2}^\Gband} \, ,
\end{align}
where $\smash{2\delta \Qsph{2}/\Qsph{2}^\Gband\simeq 3.6\%}$ is the naive expectation under the assumption that $\smash{\Qsph{4}^\Gband}$ is of the order of $\smash{\big(\Qsph{2}^\Gband\big)^2/Ze}$ .
The large hexadecapole sensitivity coefficient shows that even a small relative difference between the intrinsic hexadecapole moments could substantially modify the sensitivity factor.
Since the intrinsic hexadecapole moment of the isomeric band is unknown, the magnitude of this correction cannot presently be determined. 
For comparison, if one takes $\delta \Qsph{4}$ from the halo model, see \cref{eq:halo-hexadecapole-results}, the resulting $\KNP_{\Qsph{4}}$ is smaller by a factor of $\sim 40$ assuming $\tilde p=1$.

The contributions discussed above do not account for unconstrained changes in surface diffuseness, octupole and higher  multipole moments, or higher-order deformation effects beyond the linearized treatment. 
As shown in \cref{tab:Kroutes}, the latter can be substantial even when $\delta \Qsph{4}=0$.

\subsubsection*{Surface Diffuseness}

In the above estimates we have assumed equal surface diffuseness parameters for the two clock states, i.e., $a^\m = a^\gs$. 
Under this assumption, the sensitivity depends only weakly on $a$.
It is, however, much more sensitive to a possible difference $\delta a = a^\m - a^\gs$.
At the reference point $a^\gs = a^\m = \SI{0.5}{fm}$ and $\delta a=0$, evaluating the Coulomb-energy difference without linearizing in the other state differences ($\delta X$), we find
\begin{align}
    \frac{\partial \KNP}{\partial a} \simeq  - 1.1 \times 10^2 ~\mathrm{fm}^{-1}\,, \qquad 
    \frac{\partial \KNP}{\partial (\delta a)} \simeq  2.4 \times 10^6 ~\mathrm{fm}^{-1}\,.
\end{align}
The sensitivity to $\delta a$ reflects the dependence of the Coulomb energy on changes in the radial charge-density profile that are not fixed by the measured $\delta\chrsq$, and should therefore be interpreted as an indication of the importance of unconstrained higher radial moments rather than as a model-independent prediction.

The strong dependence on the unconstrained parameter $\delta a$ limits the determination of the sensitivity factor $\KNP$ within the deformed Fermi model. 
Additional measurements sensitive to higher radial moments, such as $\delta \langle r^4\rangle$, could constrain this parameter, potentially through higher-order field-shift contributions extracted from isomer-shift measurements across atomic transitions with different sensitivities to higher radial moments.
Further independent measurements would allow the two-parameter Fermi description to be tested.

The sensitivity to the intrinsic hexadecapole difference and to $\delta a$ shows that the experimental uncertainties in the charge radius and quadrupole moments do not, by themselves, determine the uncertainty in $\KNP$.

\begin{table}[t]
    \centering
    \renewcommand{\arraystretch}{1.3}
    
    \begin{tabular}{l@{\qquad}c@{\qquad}c}
    \toprule
    Observable $X$ & $\partial \KNP/\partial (\log X)$ & Contribution to $\sigma_{\KNP}$ \\[2pt]
    \midrule
    $\chrsq$          & $-3.8\times 10^4$ & $\pm 0.02\times 10^4$ \\
    $\delta\chrsq$       & $-1.8\times 10^4$ & $\pm 0.23\times 10^4$ \\
    $\Qcartzzspec{\gs}$        & $+5.2\times 10^4$ & $\pm 0.03\times 10^4$ \\
    $\RQ\equiv \Qcartzzspec{\m}/\Qcartzzspec{\gs}$ &  $+1.5\times 10^6$ & $\pm 0.003\times 10^4$ \\
    \midrule
    $\beta^2$ truncation &  --- & $\pm 0.06\times 10^4$ \\
    \bottomrule
    \end{tabular}
    \caption{
    Logarithmic sensitivities of $\KNP$ to the measured nuclear observables and their contributions to the experimental uncertainty, evaluated at the reference values of \cref{tab:expinput}. 
    The second column gives the response of $\KNP$ to a fractional change in each observable ($\delta \KNP \simeq \partial \KNP/\partial (\log X) \delta X/X$), while the third gives its contribution to the propagated standard uncertainty $\sigma_{\KNP}$.
    The final row gives an estimate of the theoretical uncertainty associated with the $\mathcal{O}(\beta^2)$ truncation, obtained by comparing the Coulomb-energy difference calculated to quadratic order in the deformation parameters with the result obtained using the full (unexpanded) deformed Fermi charge distribution.}
    \label{tab:K_sigprop}
\end{table}
%

\section{Summary}
\label{sec:Summary}

In this work, we investigated the electric multipole structure of \Th and its implications for hyperfine spectroscopy and the sensitivity of the nuclear clock to physics beyond the Standard Model. 
We developed a systematic multipole description of the interaction between the nucleus and its crystal environment, including the electric hexadecapole contribution.
Allowing for hexadecapole interactions introduces degeneracies in the extraction of nuclear quadrupole moments from spectroscopic measurements.
Although the hexadecapole interaction is expected to be suppressed relative to the quadrupole interaction by a factor of order $5 \times 10^{-10}$, higher multipoles can have a significant impact on the new physics sensitivity. 

We further investigated the effects of different crystal symmetries, considering twofold, tetragonal, and cubic environments. 
The symmetry constraints determine which components of the electric-field gradients contribute to the hyperfine Hamiltonian and can help disentangle different multipole interactions. 
In particular, for ideal cubic symmetry, the quadrupole interaction vanishes, while the hexadecapole interaction remains allowed and splits the ground-state manifold into two distinct energy levels.

Connecting the spectroscopic observables to the underlying nuclear structure requires theoretical input.
We considered several complementary theoretical approaches, including NRQED, nuclear energy-density-functional calculations, a halo-inspired description, the particle--rotor model and considered a phenomenological deformed Fermi model of the nuclear charge distribution.
In particular, we constructed the relevant NRQED operators for the nuclear clock states, providing an effective description of their electromagnetic interactions that includes finite-size and higher multipole contributions.

We then examined the implications of these approaches for the clock's sensitivity to variations of fundamental constants. 
We express this sensitivity through the enhancement factor $\KNP$, which quantifies the enhancement arising from the near cancellation between electromagnetic and strong contributions to the nuclear transition energy.
Combining the rotor-model relations between spectroscopic and intrinsic moments with the phenomenological deformed Fermi model, we connected measured nuclear observables to the Coulomb-energy difference between the clock states.
Retaining only the charge-radius and quadrupole contributions,
we find that they partially cancel, making the resulting sensitivity particularly susceptible to additional nuclear-structure effects. 
In particular, the hexadecapole contribution can have a considerable impact on $\KNP$, even though the corresponding spectroscopic interaction is strongly suppressed.
The intrinsic hexadecapole difference and the surface diffuseness remain experimentally unconstrained, preventing a reliable determination of $\KNP$ within this model.

These results highlight the complementary roles of hyperfine spectroscopy, nuclear-structure measurements, and theoretical modeling in determining the properties of the \Th nuclear clock.
While crystal symmetry can help disentangle nuclear multipole interactions, spectroscopy alone cannot fully constrain the intrinsic nuclear charge distributions relevant for estimating the clock's sensitivity to new physics.
Further experimental information on higher multipole moments and radial charge distributions, together with improved theoretical descriptions, will therefore be important for reducing the uncertainties in the predicted sensitivity.

\section{Acknowledgments}
We are grateful to Andrea Caputo for collaboration during the early stages of this work; we thank Wooyong Han, and Oleksandr Tomalak for useful discussions. 
The authors acknowledge the use of OpenAI's ChatGPT and Anthropic's Claude to assist with manuscript preparation.
JK and YS would like to thank the CERN Theory Department for hospitality during large parts of this project.
CB and YS are supported by the ISF (grant No.~597/24), by BSF (grant No. 2024091) and by NSF-BSF (grant No.~2025824).
DG is supported by the ISF, grant no 889/23.
HWH was supported in part by the Deutsche Forschungsgemeinschaft (DFG, German Research Foundation) -- Project ID 279384907 -- SFB 1245, by the BMFTR Contract No.~05P24RDB, and the European Research Council (ERC, DM-Dawn, Grant Agreement No. 101199868).
FK was supported by a Postdoc.Mobility fellowship from the Swiss National Science Foundation (SNSF; grant no.~P500-2\_235391).
JK has received support from the Cluster of Excellence \href{https://www.prisma.uni-mainz.de/}{``Precision Physics, Fundamental Interactions and Structure of Matter``}, PRISMA${}^{++}$ (EXC 2118/2, Project ID 390831469) funded by the German Research Foundation (DFG).
G.\ Paz was supported by the U.S. Department of Energy
Grant No. DE-SC0007983.
G.\ Perez is supported by the Israel Science Foundation (ISF), Minerva, the NSF-BSF, and the European Research Council (ERC, DM-Dawn, Grant Agreement No.~101199868). 
KS is supported by a research grant from Mr. and Mrs. George Zbeda and by the Minerva foundation.

\newpage

\appendix

\section{Multipole Moment Conventions}
\label{app:multipoletranslations}

Nuclear multipole moments may be expressed as irreducible tensors in either spherical or Cartesian conventions, and both conventions are used in this work.
In addition, we distinguish intrinsic moments defined in the body-fixed frame from spectroscopic moments associated with states of definite angular momentum.
Here we collect the relations between these quantities and derive the corresponding normalization of the electric-field gradients.
An overview of the multipole quantities and electric-field-gradient tensors used in this work can be found in \cref{tab:paramdefs}.

\subsection{Spherical-to-Cartesian Multipole Relations}

The electric multipole moments of the nuclear charge distribution are given by the operators
\begin{align}
    \Qhat{k}{q} = \int\!\drth \, \rhohat(\rvec) \, r^k \, \sphY{k}{q}{\hat{\rvec}}\,,
    \label{eq:Qhatdef}
\end{align}
where $\rhohat$ is the nuclear charge-density operator and $\sphY{k}{\mu}{\hat{\rvec}}$ denotes the spherical harmonics evaluated in the direction $\hat{\rvec} = \rvec/r$, where $r \equiv |\vec{r}|$. 
For the spherical harmonics, we use the normalization convention $\int d\Omega \, \sphY{k}{q}{\hat{r}} \, \sphY{k'}{q'*}{\hat{r}} = \delta_{kk'} \delta_{qq'}$.
We write
\begin{align}
    \label{eq:Qhatsplit}
    \Qhat{k}{q} = \That{k}{q} \Qsph{k}\,,
\end{align}
where $\Qsph{k}$ carries the dimensional nuclear multipole information and $\That{k}{q}$ is a dimensionless irreducible tensor operator.
In the Cartesian representation, the same charge distribution defines the rank-$k$ Taylor-moment operator (see also \cref{eq:MEFG})
\begin{align}
    \label{eq:Mdef}
    \hat{M}_{i_1 \ldots i_k} 
    \equiv  
    \int \drth \rhohat(\rvec) \, x_{i_1} \cdots x_{i_k}\,.
\end{align}
This symmetric tensor is in general reducible to contributions of ranks $k, k-2, \dots$. 
Its irreducible rank-$k$ component is the symmetric traceless part,
\begin{align}
    \label{eq:Mcartidef}
    \Mcarti{k} \equiv \hat{M}_{i_1\ldots i_k} - (\text{trace terms})\,.
\end{align}
For the ranks relevant here, 
\begin{align*}
    \Mcart{ij}
    =& \hat{M}_{ij} - \delta_{ij} \frac{1}{3}\sum_a \hat{M}_{aa}
    \,,\\
    \Mcart{ijkl} 
    =& \hat{M}_{ijkl} \nonumber\\
    & - \frac{1}{7} \sum_a \left( 
               \delta_{ij} \hat{M}_{aakl} + \delta_{ik} \hat{M}_{aajl} + \delta_{il} \hat{M}_{aajk}
               + \delta_{jk} \hat{M}_{aail} +  \delta_{jl} \hat{M}_{aaik} + \delta_{kl} \hat{M}_{aaij}
               \right) \nonumber\\
    & + \frac{1}{35} \sum_{a,b} \left( \delta_{ij} \delta_{kl} + \delta_{ik} \delta_{jl} + \delta_{il} \delta_{jk}\right) \hat{M}_{aabb}\,,
\end{align*}
where $a,b \in \lbrace x,y,z \rbrace$.

The irreducible spherical and Cartesian tensors are related through the Cartesian representation of the solid spherical harmonics.
We define the tensors $\cartY{k}{q}$ by
\begin{align}
    \label{eq:cartYdef}
    r^k \sphY{k}{q}{\hat{\rvec}} \equiv \cartY{k}{q}x_{i_1}\ldots x_{i_k}\,.
\end{align}
Since the Cartesian coordinates commute, the tensors $\cartY{k}{q}$ may be chosen symmetric without loss of generality. 
Since the solid spherical harmonics $r^k\sphY{k}{q}{\hat{\rvec}}$ satisfy the Laplace equation \smash{$\nabla^2 [r^k Y_k^q(\hat{\vec{r}})] = 0$}, \cref{eq:cartYdef} implies that the irreducible Cartesian tensors are traceless.
Substituting this definition into \cref{eq:Qhatdef} gives
\begin{align}
    \label{eq:QhatoM}
    \Qhat{k}{q} = \That{k}{q} \Qsph{k} = \cartY{k}{q}\hat{M}_{i_1\ldots i_k}= \cartY{k}{q}\Mcarti{k}
\end{align}
where the second equality follows because the trace terms vanish upon contraction with the traceless tensor $\cartY{k}{q}$.
Thus the $2k+1$ spherical components \smash{$\Qhat{k}{q}$} and the $2k+1$ independent components of \smash{$\Mcarti{k}$} are two representations of the same irreducible rank-$k$ multipole tensor.

For the static spectroscopic moments considered below, only the $q=0$ component is required.
In this case the angular dependence is particularly simple,
\begin{align}
    \label{eq:sphYPk}
    \sphY{k}{0}{\hat{\rvec}}=\sqrt{\frac{2k+1}{4\pi}}P_k(\cos\theta) \,,
\end{align}
where $P_k$ is the $k^{\rm th}$ Legendre polynomial.
Writing $P_k(x) = a_k x^k+ \ldots$, with
\begin{align}
    \label{eq:akdef}
    a_k =& \frac{1}{k!}\frac{\dd^k P_k(x)}{\dd x^k}
    = \frac{(2k-1)!!}{k!} \,,
\end{align}
the axial component takes the form
\begin{align}
    \label{eq:McartiPk}
    a_k\Mcartz{k} = \int \drth \rhohat(\rvec) r^k P_k\left(\frac{z}{r}\right)\,,
\end{align}
while the general transformation \cref{eq:QhatoM} between the axial contributions to the spherical and the Cartesian moments reduces to
\begin{align}
    \label{eq:QhatOMzz}
    \Qhat{k}{q=0} 
    \equiv \That{k}{0} \Qsph{k} 
    =a_k \sqrt{\frac{2k+1}{4\pi}}\Mcartz{k}\,.
\end{align}
Spectroscopic moments of a state of definite angular momentum $I$ are defined as stretched-state matrix elements (see \cref{eq:Qsphspecdef}).
Since these are diagonal matrix elements, the Wigner--Eckart theorem, \cref{eq:WignerEckart}, implies that only the $q=0$ component contributes.
Thus
\begin{align}
    \label{eq:Qsphspecdef_app}
    \Qsphspec{k}{I}
    \equiv \bra{I,I} \Qhat{k}{q=0} \ket{I,I} 
    = \Qsph{k} \langle I || \That{k}{} || I \rangle
    \begin{pmatrix}
        I & k & I\\
        -I & 0 & I
    \end{pmatrix}\,.
\end{align}
Combining this result with \cref{eq:QhatOMzz} gives
\begin{align}
    \label{eq:QsphstoIIMII}
    \Qsphspec{k}{I} 
    = a_k \sqrt{\frac{2k+1}{4\pi}} \bra{I,I} \Mcartz{k} \ket{I,I}\,.
\end{align}
%

\begin{table}[t]
    \centering
    \renewcommand{\arraystretch}{1.5}
    \begin{tabular}{c|c|p{10cm}|c}
        \hline
        Quantity & Basis & Description & Definition \\
        \hline

        \multicolumn{4}{c}{\textit{Nuclear multipole operators}} \\
        \hline

        $\Qhat{k}{q}$
        & Spherical
        & Electric multipole operator of rank $k$ and component $q$
        & \eqref{eq:Qhatdef} \\

        $\That{k}{q}$
        & Spherical
        & Dimensionless irreducible spin tensor operator of rank-$k$ and component $q$
        & (\ref{eq:Hsph},\ref{eq:Qhatsplit}) \\

        $\Qcarti{k}$
        & Cartesian
        & Irreducible Cartesian multipole operator of rank $k$
        & \eqref{eq:QcartMcart} \\

        $\Qcartz{k}$
        & Cartesian
        & Axial component of the Cartesian multipole operator of rank $k$
        & \eqref{eq:QcartMcart} \\

        \hline
        \multicolumn{4}{c}{\textit{Spectroscopic multipole moments}} \\
        \hline

        $\Qsphspec{k}{I}$
        & Spherical
        & Stretched-state multipole moment of rank $k$
        & (\ref{eq:Qsphspecdef},\ref{eq:Qsphspecdef_app}) \\

        $\Qcartspec{k}{I}$
        & Cartesian
        & Cartesian stretched-state multipole moment of rank $k$
        & (\ref{eq:Qcarts24def},\ref{eq:QcartsQsphs}) \\

        \hline
        \multicolumn{4}{c}{\textit{Intrinsic multipole moments}} \\
        \hline

        $\Qsph{k}^{\Kband}$
        & Spherical
        & Intrinsic moment of rank $k$ and rotational band $\Kband$
        & (\ref{eq:QsphsQsph},\ref{eq:Qsphdef}) \\

        $\Qcartzval{k}^{\Kband}$
        & Cartesian
        & Cartesian intrinsic moment of rank $k$ and rotational band $\Kband$
        & \eqref{eq:QspecQint} \\

        \hline
        \multicolumn{4}{c}{\textit{Crystal-field tensors and asymmetry parameters}} \\
        \hline

        $\Vcarti{k}$
        & Cartesian
        & Irreducible rank-$k$ electric-field-gradient tensor
        & (\ref{eq:MEFG},\ref{eq:Vcartidef}) \\

        $V_q^\krank$
        & Spherical
        & Spherical electric-field-gradient of rank $k$ and component $q$
        & (\ref{eq:Hsph},\ref{eq:VsphVcart}) \\

        $\xi_q^\krank$
        & Spherical
        & Ratio $V_q^\krank/V_0^\krank$
        & \eqref{eq:xidef} \\

        $\eta$
        & Cartesian
        & Quadrupole EFG asymmetry parameter
        & (\ref{eq:etadef},\ref{eq:etadef_app}) \\

        $\zeta_q$
        & Spherical
        & Hexadecapole EFG asymmetry parameters, $q=1,\ldots,4$
        & (\ref{eq:zetadefs},\ref{eq:zetacartgeneral}) \\

        \hline
    \end{tabular}
    \caption{Summary of the multipole quantities and electric-field-gradient tensors used in this work.
    Hatted quantities denote operators, while intrinsic and spectroscopic moments are numerical quantities.}
    \label{tab:paramdefs}
\end{table}

\subsection{Multipole Hamiltonians and Electric-Field Gradients}

The electric multipole interaction Hamiltonian follows from the Taylor expansion of the crystal potential about the nuclear position.
At rank $k$, the corresponding Cartesian electric field gradient tensor is (see also \cref{eq:MEFG})
\begin{align}
    V_{i_1\ldots i_k} \equiv \left. \frac{\partial V(\rvec)}{\partial x_{i_1}\cdots \partial x_{i_k}}\right\vert_{\rvec =0}\,.
\end{align}
Like the Cartesian nuclear moment $M_{i_1\ldots i_k}$ (see \cref{eq:Mcartidef}), this tensor is symmetric but in general reducible.
Its irreducible rank-$k$ component is the symmetric traceless rank-$k$ part
\begin{align}
    \label{eq:Vcartidef}
    \Vcarti{k} \equiv V_{i_1\ldots i_k} - (\text{trace terms})\,.
\end{align}
The irreducible rank-$k$ electric multipole interaction is therefore (see also \cref{eq:Hcart})
\begin{align}
    \label{eq:Hcartapp}
    \Hamilhat^\krank
    =\frac{1}{k!} \Vcarti{k} \Mcarti{k} \,.
\end{align}
Since a traceless tensor is orthogonal to all trace terms, it is sufficient to project either tensor in this contraction.

To express the Cartesian Hamiltonian in the spherical basis, we use the solid-harmonic tensors introduced in \cref{eq:cartYdef}.
These $2k+1$ tensors form a complete orthogonal basis of the symmetric traceless rank-$k$ tensors and satisfy
\begin{align}
    \label{eq:cartYprops}
    \big( \cartY{k}{q}\big)^* \, \cartY{k}{q'}
    =
    \frac{(2k+1)!!}{4\pi k!} \delta_{qq'}\,,
    \qquad
    \big( \cartY{k}{q}\big)^*
    =
    (-1)^q \, \cartY{k}{-q}\,.
\end{align}
Consequently the Cartesian contraction can be expressed as a sum over spherical components. 
Using \cref{eq:cartYprops,eq:QhatoM}, we obtain
\begin{align}
    \Hamilhat ^\krank
    &= \frac{4\pi}{(2k+1)!!}
    \sum_{q=-k}^k
    \left(
    \cartY{k}{q}\Vcarti{k}
    \right)^*
    \left(
    \cartY{k}{q}\Mcarti{k}
    \right)\nonumber\\
    &=
    \frac{4\pi}{(2k+1)!!}
    \sum_{q=-k}^k
    (-1)^q
    \left(
    \cartY{k}{q} \Vcarti{k}
    \right)
    \Qhat{k}{-q}\nonumber \\
    &= \Qsph{k}\sum_{q=-k}^k
    (-1)^q V_q^\krank \That{k}{-q}\,,
    \label{eq:Hsphapp}
\end{align}
where in the last step we have introduced the spherical EFG components
\begin{align}
    \label{eq:VsphVcart}
    V_q^\krank \equiv \frac{4\pi}{(2k+1)!!} \, \cartY{k}{q} \Vcarti{k}\,.
\end{align}
Using \cref{eq:sphYPk,eq:akdef}, and making use of the tracelessness of $\Vcarti{k}$, the axial ($q=0$) component can be expressed as
\begin{align}
    \label{eq:VsphVcartaxial}
    V_0^\krank = \frac{1}{k!} \sqrt{\frac{4\pi}{2k+1}} \Vcartz{k}\,.
\end{align}
For the ranks relevant here,
\begin{align}
    V_0^{(2)} = \sqrt{\frac{\pi}{5}}\Vcartzz\,, \qquad
    V_0^{(4)} = \frac{\sqrt{\pi}}{36}\Vcartzzzz\,.
\end{align}
%

\subsection{Cartesian Multipole Moments}
The Cartesian multipoles conventionally quoted in nuclear spectroscopy are rescaled versions of the irreducible Taylor moments $\Mcarti{k}$ introduced above (see \cref{eq:Mdef}).
A rank-dependent factor $b_k$ ensures that $b_kP_k(x)$ has the conventional integer coefficients (see \cref{eq:McartiPk})
\begin{align}
    \label{eq:QcartMcart}
    \Qcartz{k}
    &\equiv
    b_k \int \drth \rhohat(\rvec) r^k P_k\left(\frac{z}{r}\right)
    = a_k b_k \Mcartz{k}\,,\\
    \Qcarti{k} &\equiv a_k b_k \Mcarti{k}\,.
\end{align}
For the ranks relevant here,
\begin{align}
    P_2(x) 
    &= \frac{1}{2}\left( 3x^2 - 1\right)\,, & 
    a_2 &= \frac{3}{2}\,, & 
    b_2 &= 2\,, \\
    P_4(x) &= \frac{1}{8}\left(35x^4 - 30x^2 + 3\right) & 
    a_4 &= \frac{35}{8}\,, &
    b_4 &= 8\,, 
\end{align}
consequently,
\begin{align}
    \Qcartzz =& \int \drth \rhohat(\rvec) \left(3z^2 - r^2\right)= 3 \hat{M}_{zz} - \sum _a \hat{M}_{aa} \\
    \Qcartzzzz =& \int \drth \rhohat(\rvec) \left( 35 z^4 - 30 z^2 r^2 + 3 r^4\right) = 35 \hat{M}_{zzzz} - 30 \sum_a \hat{M}_{zzaa} + 3 \sum_{a,b} \hat{M}_{aabb}\,.
\end{align}
The relation between the Cartesian multipoles and their spherical counterparts follows immediately from \cref{eq:QhatOMzz,eq:QcartMcart},
\begin{align}
    \Qhat{k}{q=0} 
    \equiv \That{k}{0} \Qsph{k} 
    =a_k \sqrt{\frac{2k+1}{4\pi}}\Mcartz{k}
    =\frac{1}{b_k}\sqrt{\frac{2k+1}{4\pi}}\Qcartz{k}\,,
\end{align}
and the spectroscopic moments are related in the exact same way,
\begin{align}
    \label{eq:QcartsQsphs}
    \Qsphspec{k}{I} 
    = \frac{1}{b_k}\sqrt{\frac{2k+1}{4\pi}} \bra{I,I}\Qcart{z\ldots z} \ket{I,I}
    = \frac{1}{b_k}\sqrt{\frac{2k+1}{4\pi}}\Qcartspec{k}{I}\,,
\end{align}
where we introduced the Cartesian spectroscopic multipole moment $Q_{z\ldots z}^{(I)}\equiv \bra{I,I}\Qcart{z\ldots z} \ket{I,I}$.
For the ranks relevant here,
\begin{align}
    \Qsphspec{2}{I} = \sqrt{\frac{5}{16\pi}} \Qcartzzspec{I}\,, \qquad
    \Qsphspec{4}{I} = \frac{3}{16\sqrt{\pi}}\Qcartzzzzspec{I}\,.
\end{align}
When expressed in terms of the conventional Cartesian multipoles, the interaction in \cref{eq:Hcartapp} takes the form
(see also \cref{eq:Hcart})
\begin{align}
    \Hamilhat^\krank
    =\frac{1}{k!} \Vcarti{k} \Mcarti{k} = \frac{1}{k!a_k b_k} \Vcarti{k} \Qcarti{k} \,,
\end{align}
in particular,
\begin{align}
    \Hamilhat^{(2)} = \frac{1}{6} \Vcart{ij} \Qcart{ij}\,, \qquad 
    \Hamilhat^{(4)} = \frac{1}{840} \Vcart{ijkl} \Qcart{ijkl}\,.
\end{align}
By the Wigner--Eckart theorem (\cref{eq:WignerEckart}), only the $q=0$ spherical component has a nonzero stretched-state matrix element. 
Consequently, the contraction of the symmetric traceless EFG tensor with $\bra{I,I}\Qcarti{k}\ket{I,I}$ reduces to the product of their axial components, with the geometrical factor $a_k$ corresponding to the coefficient of the highest power in the Legendre polynomial $P_k$ (see \cref{eq:akdef}).
Using the Cartesian spectroscopic moment \smash{$Q_{z\ldots z}^{(I)}$} introduced in \cref{eq:QcartsQsphs}, we obtain
\begin{align}
    \label{eq:IIHIIcartQ}
    \bra{I,I}\Hamilhat^\krank \ket{I,I}
    =&\frac{1}{k! a_k b_k} \Vcarti{k} \bra{I,I} \Qcarti{k} \ket{I,I}\nonumber\\
    =& \frac{1}{k! b_k} \Vcartz{k} \Qcartspec{k}{I}\,.
\end{align}
For $k=2,4$,
\begin{align}
    \bra{I,I}\Hamilhat^{(2)}\ket{I,I} 
    =& V_0^{(2)} \Qsphspec{2}{I} 
    = \frac{1}{4} \Vcartzz \Qcartzzspec{I}\,, \\
    \bra{I,I}\Hamilhat^{(4)} \ket{I,I}
    =& V_0^{(4)} \Qsphspec{4}{I}
    = \frac{1}{192} \Vcartzzzz \Qcartzzzzspec{I}\,.
\end{align}

\subsection{Matrix Representations}
\label{sec:matrix-representations}

We define the matrix elements of the rank-$k$ Hamiltonian in the angular-momentum basis as
\begin{align}
    \mathcal{H}^\krank_{I,mm'} \equiv \bra{I,m} \Hamilhat^\krank \ket{I,m'}\,, 
\end{align}
where $I$ is either chosen to be the ground state angular momentum, $I=5/2$, or the excited state angular momentum $I=3/2$, and $-I \leq m,m'\leq I$.
The entries \smash{$\mathcal{H}^\krank _{I,mm'}$} are fixed by \cref{eq:IIHIIcartQ} and angular-momentum algebra, i.e. the Wigner--Eckart theorem (see also \cref{eq:WignerEckart}),
\begin{align}
    \langle I', m'|\That{k}{q}|I, m\rangle 
    = 
    (-1)^{I'-m'}
    \langle I' || \That{k}{}|| I\rangle 
    \begin{pmatrix}
        I' & k & I\\
        -m' & q & m
    \end{pmatrix}\,,
\end{align}
which relates the different elements. 
Combining \cref{eq:WignerEckart,eq:Qsphspecdef,eq:QcartsQsphs,eq:VsphVcartaxial} yields
\begin{align}
    \mathcal{H}^\krank_{I,mm'} 
    \equiv \bra{I,m} \Hamilhat^\krank \ket{I,m'}  
    &= \sum_{q=-k}^k (-1)^q \, V_q^\krank \bra{I,m} \That{k}{-q} \ket{I,m'} \Qsph{k} \nonumber
    \\
    &=  \frac{1}{k!\,b_k }
    \Vcartz{k} \Qcartspec{k}{I}
    \sum_{q=-k}^k
    (-1)^{I-m+q} 
    \xi^\krank_q
    \frac{
        \begin{pmatrix}
            I & k & I\\
            -m & -q & m'
        \end{pmatrix}
    }{
        \begin{pmatrix}
            I & k & I\\
            -I & 0 & I
        \end{pmatrix}
    }\,. 
    \label{eq:Hmmp}
\end{align}
where we introduced the dimensionless relations
\begin{align}
    \label{eq:xidef}
    \xi_q^\krank \equiv \frac{V_q^\krank}{V_0^\krank}\,,
    \qquad
    \xi_0^\krank = 1
\end{align}
Since the electrostatic potential is real and $\sphY{k}{-q}{\hat{\rvec}}=(-1)^q \left(\sphY{k}{q}{\hat{\rvec}}\right)^*$, its spherical components satisfy
\begin{align}
    \label{eq:Vreal}
    V_{-q}^\krank = (-1)^q \, \big( V_q^\krank \big)^*\,.
\end{align}
and hence
\begin{align}
    \label{eq:realxi}
    \xi_{-q}^\krank
    = 
    (-1)^q \, \big( \xi_q^\krank\big)^*\,.
\end{align}
While $\Vcart{z \ldots z} Q_{z\ldots z}^{(\Isp)}/(k! b_k)$ sets the overall interaction strength, the dimensionless quantities $\xi^\krank_q$ specify the shape of the EFG in a given crystal and a given coordinate frame.

\paragraph*{Rank Two ($k=2$)}

Combining the spherical-to-Cartesian transformation, \cref{eq:VsphVcart} with the definition of the dimensionless EFG ratios \smash{$\xi_q^\krank$} (\cref{eq:xidef}), we obtain
\begin{align}
    \label{eq:etacartgeneral}
    \xi_q^{(2)}=\begin{cases}
        1 & q=0 \,,\\
        \eta_q & q=1,2 \,,\\
        -\eta_{-q}^* & q=-1\,,\\
        \eta_{-q}^* & q=-2  \,,
    \end{cases}
\end{align}
with
\begin{align}
    \eta_1 = -\sqrt{\frac{2}{3}}\frac{\Vcart{xz}+i\Vcart{yz}}{\Vcartzz} \,, \qquad
    \eta_2 = \frac{\Vcart{xx}-\Vcart{yy}+2i\Vcart{xy}}{\sqrt{6}\Vcartzz}
\end{align}
Here and below, the Cartesian components refer to the symmetric traceless tensors defined in \cref{eq:Vcartidef}.
Since the rank-2 EFG tensor is real and symmetric, it can be diagonalised by a suitable choice of coordinate axes.
In this principle-axis frame, the off-diagonal elements of the EFG tensor are zero, $\Vcart{xy}=\Vcart{xz}=\Vcart{yz}=0$, so that 
\begin{align}
    \eta_1  = 0\,,
    \qquad
    \eta_2  = 
    \frac{\eta}{\sqrt{6}}\,,
\end{align}
where
\begin{align}
    \eta 
    \equiv 
    \frac{\Vcart{xx}-\Vcart{yy}}{\Vcart{zz}}\,.
    \label{eq:etadef_app}
\end{align}
The quadrupole EFG is therefore fully characterized by its overall scale $\Vcartzz$ and one single asymmetry parameter $\eta$.
In the basis $\ket{I,I}, \ket{I,I-1}, \ldots \ket{I,-I}$, the quadrupole Hamiltonian for $I=3/2$ is
\begin{align}
    \mathcal{H}^{(2)}_{I=3/2}
    = \frac{1}{4} \Vcartzz \Qcartzzspec{3/2}
        \begin{pmatrix}
            1 & 0 & \frac{\eta}{\sqrt{3}} & 0 \\
            0 & -1 & 0 & \frac{\eta}{\sqrt{3}} \\
            \frac{\eta}{\sqrt{3}} & 0 & -1 & 0 \\
            0 & \frac{\eta}{\sqrt{3}} & 0 & 1
        \end{pmatrix}\,,
\end{align}
while for $I=5/2$,
\begin{align}
    \mathcal{H}^{(2)}_{I=5/2}
    =
    \frac{1}{4}\Vcartzz\Qcartzzspec{5/2}
    \begin{pmatrix}
        1 & 0 & \dfrac{\eta}{\sqrt{10}} & 0 & 0 & 0 \\
        0 & -\dfrac15 & 0 & \dfrac{3\eta}{5\sqrt2} & 0 & 0 \\
        \dfrac{\eta}{\sqrt{10}} & 0 & -\dfrac{4}{5} & 0 & \dfrac{3\eta}{5\sqrt2} & 0 \\
        0 & \dfrac{3\eta}{5\sqrt2} & 0 & -\dfrac{4}{5} & 0 & \dfrac{\eta}{\sqrt{10}} \\
        0 & 0 & \dfrac{3\eta}{5\sqrt2} & 0 & -\dfrac{1}{5} & 0 \\
        0 & 0 & 0 & \dfrac{\eta}{\sqrt{10}} & 0 & 1
    \end{pmatrix}\,.
\end{align}
The vanishing matrix elements with odd $\Delta m$ reflect the absence of the $q=\pm 1$ EFG components in the principal-axis frame.
Consequently, the quadrupole Hamiltonian separates into the two independent subspaces 
\begin{align}
    \label{eq:subspaces}
    &I=\frac{3}{2}:  \quad m=\left\lbrace \frac{3}{2}, -\frac{1}{2}\right\rbrace\,,\; 
    \left\lbrace \frac{1}{2}, -\frac{3}{2}\right\rbrace\,,\nonumber\\
    &I=\frac{5}{2}:  \quad m=\left\lbrace \frac{5}{2}, \frac{1}{2}, -\frac{3}{2}\right\rbrace\,,\; 
    \left\lbrace \frac{3}{2}, -\frac{1}{2}, -\frac{5}{2}\right\rbrace\,,
\end{align}
where we have omitted the kets for brevity.
The reduced matrices used in \cref{sec:clockhamiltonians} are obtained by selecting either of these blocks. 
The second block is related to the first by time reversal and contains no additional spectral information.

\paragraph*{Rank Four ($k=4$)}

For the rank-four EFG we introduce four generally complex dimensionless parameters \smash{$\zeta_q\equiv \xi_q^{(4)}$}, $q=1,2,3,4$, whereas the negative-$q$ components are fixed by the reality condition \cref{eq:realxi}.
\begin{align}
    \xi_q^{(4)}
    \equiv
    \begin{cases}
        1 & q=0\,,\\
        \zeta_q  & q=1,2,3,4\,,\\
        -\zeta_{-q}^* & q=-1,-3\,, \\
        \zeta_{-q}^* & q=-2,-4\,.
    \end{cases}
\end{align}
They are related to the Cartesian components of the symmetric traceless EFG tensor by
\begin{align}
    \label{eq:zetacartgeneral}
    \begin{split}
    \zeta_1 &= -\frac{2}{\sqrt{5}} \frac{\Vcart{xzzz}+i\Vcart{yzzz}}{\Vcartzzzz}\,,\\
    \zeta_2 &= \sqrt{\frac{2}{5}}\frac{2\Vcart{xxzz}+2i \Vcart{xyzz} + \Vcartzzzz}{\Vcartzzzz}\\
    \zeta_3 &= -\frac{2}{\sqrt{35}}
    \frac{4\Vcart{xxxz}+4i\Vcart{xxyz}+3\Vcart{xzzz} + i\Vcart{yzzz}}{\Vcartzzzz}\,, \\
    \zeta_4 &= \frac{1}{\sqrt{70}} \frac{8\Vcart{xxxx}+8i\Vcart{xxxy}+8\Vcart{xxzz}+4i\Vcart{xyzz}+\Vcartzzzz}{\Vcartzzzz}
    \end{split}
\end{align}
The negative-$q$ components again follow from the reality condition \cref{eq:realxi}.
Equations~\eqref{eq:zetacartgeneral} provide the general connection between the spherical parameters entering the Hamiltonian and the Cartesian EFG components. 

\begin{align}
    \mathcal{H}^{(4)}_{I=5/2}
    = \frac{1}{192}\Vcartzzzz \Qcartzzzzspec{5/2}
    \begin{pmatrix}
        1
        & -2\zeta_1^*
        & 3\zeta_2^*
        & -\sqrt{14}\zeta_3^*
        & \sqrt{14}\zeta_4^*
        & 0
        \\
        -2\zeta_1
        & -3
        & \sqrt{10}\zeta_1^*
        & -\sqrt{5}\zeta_2^*
        & 0
        & \sqrt{14}\zeta_4^*
        \\
        3\zeta_2
        & \sqrt{10}\zeta_1
        & 2
        & 0
        & -\sqrt{5}\zeta_2^*
        & \sqrt{14}\zeta_3^*
        \\
        -\sqrt{14}\zeta_3
        & -\sqrt{5}\zeta_2
        & 0
        & 2
        & -\sqrt{10}\zeta_1^*
        & 3\zeta_2^*
        \\
        \sqrt{14}\zeta_4
        & 0
        & -\sqrt{5}\zeta_2
        & -\sqrt{10}\zeta_1
        & -3
        & 2\zeta_1^*
        \\
        0
        & \sqrt{14}\zeta_4
        & \sqrt{14}\zeta_3
        & 3\zeta_2
        & 2\zeta_1
        & 1
    \end{pmatrix}\,.
\end{align}

\subsection{Angular-Momentum Operator Representation}

The multipole Hamiltonians may equivalently be expressed in terms of angular-momentum operators.
For the $q=0$ component, the general expression \cref{eq:Hmmp} gives
\begin{align}
    \bra{I,m}\Hamilhat^{(2)}\ket{I,m}
    =& \frac{1}{4} \Vcartzz \Qcartzzspec{I}
    \frac{3m^2-I(I+1)}{I(2I-1)}\nonumber \\
    =& \frac{\Vcartzz \Qcartzzspec{I}}{4I(2I-1)}
    \bra{I,m} \left( 3 \hat{I}_z^2 - \hat{\mathbf{I}}^2 \right)\ket{I,m}
\,.
    \label{eq:Hquadmm}
\end{align}
The off-diagonal terms arise from $q=\pm 2$.
The corresponding matrix elements in \cref{eq:Hmmp} can be related via the conventional raising and lowering operators
\begin{align}
    \hat{I}_\pm \ket{I,m}
    = 
    \sqrt{I(I+1)-m(m\pm 1)}
    \ket{I,m\pm 1}\,.
\end{align}
For $m'=m\pm 2$, evaluating the $3j$ symbol in \cref{eq:Hmmp} and using $\xi_{\pm 2}^{(2)}=\eta/\sqrt{6}$ from \cref{eq:etadef_app} gives
\begin{align}
    \bra{I,m}\Hamilhat^{(2)} \ket{I,m \pm 2}
    =
    \frac{\Vcartzz\Qcartzzspec{I}}{4I(2I-1)}
    \frac{\eta}{2}
    \bra{I,m} \hat{I}_{\mp}^2 \ket{I,m \pm 2}\,.
    \label{eq:Hquadmmp}
\end{align}
Combining \cref{eq:Hquadmm} and \cref{eq:Hquadmmp} yields the Hamiltonian in \cref{eq:HQspec}
\begin{align}
    \Hamilhat^{(2)}_I = 
    \frac{\Vcartzz\Qcartzzspec{I}}{4I(2I-1)}
    \left[
        3\hat I_z^2-\hat{\mathbf I}^2
        + \frac{\eta}{2} \left(\hat I_+^2+\hat I_-^2 \right)
    \right]\,,
\end{align}
where $\hat I_\pm\equiv\hat I_x\pm i\hat I_y$.
Thus the conventional spin-operator form is obtained by evaluating the same $q=0,\pm2$ matrix elements that appear in the spherical-tensor representation and recognizing their action in the spin-$I$ basis.

\subsection{Intrinsic and Spectroscopic Moments}
\label{app:rotormultipoles}

The rotor model involves three distinct multipole objects: the multipole operator $\hat{Q}^\krank_q$, its \emph{intrinsic} moment \smash{$\Qsph{k}^\Kband$} in the body-fixed frame, and the \emph{spectroscopic} moment \smash{$\Qsphspec{k}{I,K}$} of a physical state of definite angular momentum $I$.

The multipole operator is the standard rank-$k$ electric multipole defined as in \cref{eq:Qhatdef}.
We use Greek indices for its components in the intrinsic frame and Latin indices for those in the laboratory frame.
For an axially symmetric intrinsic state $\ket{\chi_K}$, only the $\nu=0$ expectation value is nonzero, and we define
\begin{align}
    \label{eq:Qsphdef}
    \bra{\chi_K} \Qhat{k}{\nu} \ket{\chi_K}
    = \delta_{\nu 0} \Qsph{k}^\Kband\,,
\end{align}
in particular, $\Qsph{k}^\Kband = \bra{\chi_K} \Qhat{k}{0} \ket{\chi_K}$.\footnote{
This coincides with the convention of Bohr and Mottelson~\cite{BohrMottelsonVol2} [Eq.~(6-60)], defined through $\sphYdef{k}{\mu}$ for all ranks $k$. 
It differs from the intrinsic quadrupole moment $eQ_0=\langle K|\int \mathrm{d}\drth\rho_e(\rvec) r^2(3\cos^2\theta-1)|K\rangle$ commonly quoted in the experimental literature, and which Bohr and Mottelson give in [Eq.~(4-66)].
The two are related by \smash{$eQ_0=\sqrt{16\pi/5}\Qsph{2}^\Kband$}.
Since $Q_0$ is specific to the quadrupole, we define all multipole moments through the operator \cref{eq:Qhatdef}.
}
Although $\Qsph{k}^\Kband$ is defined in the intrinsic frame, it enters laboratory-frame observables through the rotation of the multipole operator,
\begin{align}
    \label{eq:QsphqQsphmu}
    \Qhat{k}{q} 
    = \sum_{\nu=-k}^k \mathcal{D}_{q\nu}^k(\omega) \Qhat{k}{\nu}\,.
\end{align}
where $\mathcal{D}_{q\nu}^k(\omega) $ is a Wigner rotation matrix and $\omega$ denotes the Euler angles relating the intrinsic and laboratory frames.

For a rigid rotor, the laboratory state of definite $I,M$ belonging to the intrinsic configuration $K$ is\footnote{Here we use the single-component rotor state.
For the static moments considered here this is exact, since the $\mathcal{R}$-conjugate ($K\to -K$) component contributes only when $2K\leq k$.}
\begin{align}
    \label{eq:rotorstate}
    |IMK\rangle 
    = \sqrt{\frac{2I+1}{8\pi^2}}
    \mathcal{D}_{MK}^{I}(\omega)
    |\chi_K\rangle\,,
\end{align}
where $M$ and $K$ are the projections of $\mathbf{I}$ on the laboratory and symmetry axes.

Inserting \cref{eq:Qsphdef,eq:QsphqQsphmu,eq:rotorstate} into the laboratory-frame matrix element gives an integral over three Wigner rotation matrices
\begin{align}
    \label{eq:DDDint}
    \bra{I'M'K'}\hat{Q}_q^\krank \ket{IMK}
    = \frac{\sqrt{(2I'+1)(2I+1)}}{8\pi^2} \Qsph{k}^\Kband
    \int \dd\omega ~
    \mathcal{D}_{M'K'}^{I'*}(\omega) \,
    \mathcal{D}_{q0}^{k}(\omega) \,
    \mathcal{D}_{MK}^{I}(\omega)\,.
\end{align}
Using $\mathcal{D}_{MK}^{I^*}=(-1)^{M-K}\mathcal{D}_{-M,-K}^I$ and the standard integral
\begin{align}
    \int \dd \omega \,
    \mathcal{D}_{m_1 n_1}^{j_1}(\omega) \,
    \mathcal{D}_{m_2 n_2}^{j_2}(\omega) \,
    \mathcal{D}_{m_3 n_3}^{j_3}(\omega)
    = 8\pi^2
    \begin{pmatrix}
        j_1 & j_2 & j_3\\
        m_1 & m_2 & m_3
    \end{pmatrix}
    \begin{pmatrix}
        j_1 & j_2 & j_3\\
        n_1 & n_2 & n_3
    \end{pmatrix}\,,
\end{align}
the matrix element factorizes into laboratory and intrinsic angular-momentum coefficients,
\begin{align}
    \label{eq:rotorQmat}
    \bra{I'M'K'} \Qhat{k}{q} \ket{IMK}
    = (-1)^{M'-K'}
    \sqrt{(2I'+1)(2I+1)}
    \begin{pmatrix}
        I' & k & I\\
        -M' & q & M
    \end{pmatrix}
    \begin{pmatrix}
        I' & k & I\\
        -K' & 0 & K
    \end{pmatrix}
    \Qsph{k}^\Kband\,.
\end{align}
The spectroscopic moment is the diagonal matrix element of the $q=0$ laboratory component in the stretched state $M=I$ (see \cref{eq:Qsphspecdef_app}). 
Within the rotor model, this matrix element is obtained by setting  $I'=I$, $M'=M=I$ and $q=0$ in \cref{eq:rotorQmat}, yielding
\begin{align}
    \label{eq:QspecQint}
    \Qsphspec{k}{I,K}
    \equiv
    \bra{I,I,K} \Qhat{k}{0} \ket{I,I,K}
    =
    (-1)^{I-K} (2I+1)
    \begin{pmatrix}
        I & k & I\\
        -K  & 0 & K
    \end{pmatrix}
    \begin{pmatrix}
        I & k & I\\
        -I & 0 & I
    \end{pmatrix} 
    \Qsph{k}^\Kband\,,
\end{align}
reproducing \cref{eq:QsphsQsph}. 
It is convenient to collect the purely geometrical factor into the intrinsic-to-spectroscopic projection factor
\begin{align}
    \mathcal{P}_k^{(I,K)} = (-1)^{I-K} (2I+1)
    \begin{pmatrix}
        I & k & I\\
        -K & 0 & K
    \end{pmatrix}
    \begin{pmatrix}
        I & k & I\\
        -I & 0 & I
    \end{pmatrix}\,.
\end{align}
The spherical spectroscopic moment is therefore simply
$\Qsphspec{k}{I,K} = \mathcal{P}_k^{(I,K)} \Qsph{k}^\Kband$.
Since the spherical-to-Cartesian conversion depends solely on the rank $k$, the same projection factor relates the corresponding Cartesian intrinsic and spectroscopic moments.
In summary,
\begin{align}
    \label{eq:QsphspecQsphint_app}
    \Qsphspec{k}{I,K} = \mathcal{P}_k^{(I,K)} \Qsph{k}^\Kband\,, \qquad
    \Qcartspec{k}{I,K} = \mathcal{P}_k^{(I,K)} \Qcartzval{k}^\Kband\,,
\end{align}
where $z$ denotes the body-fixed symmetry axis. 

Within a given rotational band, the intrinsic moment $\Qsph{k}^\Kband$ can therefore be factored out of the laboratory-frame multipole operator, \smash{$\Qhat{k}{q} = \Qsph{k}^\Kband\That{k}{q}$}, where \smash{$\That{k}{q}$} contains the rotational geometry. 
Comparing the dependence of the general matrix element in \cref{eq:rotorQmat} with the Wigner--Eckart theorem \cref{eq:WignerEckart} identifies the reduced matrix element. 
\begin{align}
    \label{eq:Tnorm}
    \langle I,K || \That{k}{} || I,K\rangle 
    = 
    (-1)^{I-K} (2I+1)
    \begin{pmatrix}
        I & k & I \\
        -K & 0 & K
    \end{pmatrix}\,.
\end{align}
This is the tensor entering the laboratory-frame Hamiltonian in \cref{eq:MultipolExpRotor}: $\Qsph{k}^\Kband$ contains the intrinsic nuclear-structure information common to the band, while \smash{$\That{k}{q}$} describes its rotational projection onto states of definite angular momentum.

The intrinsic state $\ket{\chi_K}$ is not a state of definite angular momentum, so its intrinsic moments are not restricted by the selection rule $k\leq 2I$.
This restriction applies to the spectroscopic moments of a spin-$I$ state and follows from the Wigner-$3j$ symbol in \cref{eq:WignerEckart}.

An overview of the derived model parameters and the multipole moments in the different normalization conventions is given in Table~\ref{tab:derivedinput}.

\begin{table}[t]
    \centering
    \begin{tabular}{c|c|c|c}
    Parameter & Value  & Equations & Dependence\\
    \hline
        $\Roo^\gs$ & $7.033(20)~\mathrm{fm}$ & \ref{eq:Roodef} & $\chrsq^\gs\, , \, a^\gs$
        \\
        $\Roo^\m$ & $7.034(20)~\mathrm{fm}$ & \ref{eq:Roodef} & $\chrsq^\gs\, , \delta \chrsq\, , \, a^\m$
        \\
        \hline
        $C_2$  & $11.143(59)~e\mathrm{b}$ & \ref{eq:Ckdef} & $\chrsq^\gs\, , a^\gs $
        \\
        $C_4$  & $6.342(64)~e\mathrm{b}^2$ & \ref{eq:Ckdef} & $\chrsq^\gs\, , \, a^\gs $
        \\
        \hline
        $\Qsph{2}^\Gband$ & $2.746(18)~e\mathrm{b}$ & \ref{eq:QsphsQsphNuclearClock}, \ref{eq:QsphspecQsphint_app} & $Q_{\rm g}$ 
        \\
        $\Qsph{2}^\Mband$ &   $2.796(18)~e\mathrm{b}$ & \ref{eq:QsphsQsphNuclearClock}, \ref{eq:QsphspecQsphint_app} &  $Q_{\rm g}\,,\, Q_{\rm is}/Q_{\rm g}$ 
        \\
        $\Qsph{2}^\Mband/\Qsph{2}^\Gband$ & $1.017911(18)$ & \ref{eq:QsphsQsphNuclearClock}, \ref{eq:QsphspecQsphint_app} & $Q_{\rm is}/Q_{\rm g}$ 
        \\
        $\Qsph{4}^\Gband$ & $1.56(30)~e\mathrm{b}^2$ & \ref{eq:QsphsQsphNuclearClock}, \ref{eq:QsphspecQsphint_app} & $Q_{40}$
        \\
        \hline
        $\beta_2^\gs$ & $0.2465(21)$ & \ref{eq:QkCk}, \ref{eq:Ckdef} & $Q_{\rm g}\,,\,\chrsq^\gs\, , \, a^\gs$
        \\
        $\beta_2^\m$ & $0.2509(21)$ & \ref{eq:QkCk}, \ref{eq:Ckdef} & $Q_{\rm g}\,,\, Q_{\rm is}/Q_{\rm g}\,,\,\chrsq^\gs\, , \, \delta\chrsq\, , \, a^\gs$ 
        \\
        $\beta_4^\gs$ & $0.246(48)$ & \ref{eq:QkCk}, \ref{eq:Ckdef} & $Q_{\rm g}\,,\, Q_{40}\,,\,\chrsq^\gs\, , \, a^\gs$ \\
         \hline
    \end{tabular}
    \caption{Radial parameters $\Roo$, coefficients $C_k$, intrinsic multipole moments $\Qsph{2}^\Kband$, and deformation parameters $\beta_k$ inferred from the data in Table~\ref{tab:expinput}.
    The quoted uncertainties reflect the uncertainties of the experimental inputs only and do not include model uncertainties. 
    In particular, the deformation parameters are obtained using the linearized relation \cref{eq:QkCk}, neglecting higher-order corrections in the deformations.
    }
    \label{tab:derivedinput}
\end{table}

\section{Crystal Symmetries}
\label{app:crystal_syms}

The EFG tensors introduced in \cref{app:multipoletranslations} are constrained by the local point-group symmetry of the crystal at the \Th site.
In this appendix we derive these constraints for the electric quadrupole ($k=2$) and hexadecapole ($k=4$) interactions considered in the main text, and give the corresponding symmetry-restricted Hamiltonians.

The spherical and Cartesian representations of the EFG are useful for complementary purposes.
The spherical components $V_q^\krank$ transform according to the rank-$k$ representation of spatial rotations.
The irreducible Cartesian tensor \smash{$\Vcarti{k}$}, on the other hand, provides a direct geometrical description relative to the crystal axes.
We therefore first establish the transformation properties of the EFG components and then impose the crystal symmetries.

As established in \cref{eq:Vreal}, the reality of the electrostatic potential $V(\rvec) = V^*(\rvec)$ implies
\begin{align}
    V_{-q}^\krank
    = (-1)^q \big(V_q^\krank\big)^*\,.
\end{align}
This follows from the complex-conjugation properties of the spherical harmonics and holds independently of the crystal point-group symmetry.
In the following we will refer to the EFGs satisfying only this relation as belonging to the generic ($\zerosym$) symmetry class.

\subsection{Spherical and Cartesian EFG Components}
\label{app:spherical-Cartesian_crystals}

The spherical EFG components transform under crystal point-group operations in the same way as the spherical harmonics $\sphY{k}{q}{\hat{\rvec}}$.
Invariance under an operation $g$ of the local point group $G$ requires
\begin{align}
    \label{eq:V_under_point-group}
    V_q^\krank
    = 
    \sum_{q'=-k}^k D^\krank_{qq'}(g) V_{q'}^\krank\,,
\end{align}
where $D^\krank(g)$ denotes the rank-$k$ representation of $g$.

For rotations around the $z$ axis this relation is diagonal.
In particular,
\begin{align}
    \label{eq:CnVtransf}
    C_n(z): \qquad
    V_q^\krank
    \to
    e^{-2\pi i q/n } V_q^\krank\,.
\end{align}
For continuous axial symmetry about the $z$ axis, invariance under the rotations in \cref{eq:CnVtransf} must hold for arbitrary angles $\phi$, rather than only the discrete angles $2\pi/n$.
This requires $V_q^\krank=0$ for all $q\neq 0$, consequently, only the $q=0$ component survives, corresponding to $\eta=0$ (see \cref{eq:etadef_app}) at rank two and $\zeta_1=\zeta_2=\zeta_3=\zeta_4=0$ (see \cref{eq:zetacartgeneral}) at rank four.

Reflections can instead relate components with opposite $q$.
For the coordinate conventions used here,
\begin{align}
    \sigma_x: \qquad &
    (x, y, z) \to (-x, y, z)\,,
    & 
    V_q^\krank & \to (-1)^q V_{-q}^\krank 
    \label{eq:sigmaxtransf}
    \\
    \sigma_y: \qquad &
    (x, y, z) \to (x, -y, z)\,,
    &
    V_q^\krank & \to V_{-q}^\krank
    \label{eq:sigmaytransf}
    \\
    \sigma_z: \qquad &
    (x, y, z) \to (x, y, -z)\,,
    &
    V_q^\krank & \to (-1)^{k+q} V_q^\krank\,.
    \label{eq:sigmaztransf}
\end{align}
These relations follow directly from the corresponding transformations of the spherical harmonics $\sphYdef{k}{q}$.
The parity transformation corresponds to spatial inversion, equivalent to the successive application of the three coordinate reflections $\sigma_x$, $\sigma_y$ and $\sigma_z$.
\begin{align}
    P: \qquad &
    (x, y, z) \to (-x, -y, -z)\,,
    &
    V_q^\krank & \to (-1)^k V_q^\krank
\end{align}
For the even-rank tensors considered here, $P$ symmetry does not impose additional constraints.

\subsection{Symmetry Classes}
\label{app:crystalsymmetries}

For the even-rank EFG tensors considered here, several crystallographically distinct point groups impose identical constraints on the $V_q^\krank$ components.
For the purposes of the multipole Hamiltonian they can therefore be grouped into three effective symmetry classes, which we refer to as twofold ($\twosym$), tetragonal ($\tetrasym$), and cubic ($\cubsym$).
These classes describe the constraints on the EFG tensors and do not imply an equivalence of the corresponding crystallographic point groups.

The general multipole Hamiltonians and their matrix representations were established in \cref{app:multipoletranslations}.
Here we apply the symmetry constraints to the ground ($I=5/2$) and isomeric ($I=3/2$) states of \Th.
For the three symmetry classes considered below, only even-$q$ components survive such that the Hamiltonians separate into the two time-reversal-related subspaces of \cref{eq:subspaces}.
It is therefore sufficient to give one block for each nuclear state.

\subsection*{Twofold Symmetry ($\twosym$)}

We define the effective twofold symmetry class by invariance under a $\pi$ rotation about the local $z$ axis and a mirror plane $\sigma_x$ containing this axis.
The rotation and mirror plane impose, respectively,
\begin{equation}
    \label{eq:twosymconds}
    \begin{split}
    C_2(z):\qquad &
    V_q^\krank = e^{-iq \pi}V_q^\krank \,,\\
    \sigma_x : \qquad &
    V_q^\krank = (-1)^q V_{-q}^\krank\,.
    \end{split}
\end{equation}
Invariance under the rotation implies that only even values of $q$ can be nonzero, while the mirror plane reduces these to \smash{$V_q^\krank= V_{-q}^\krank$}.
Together with the reality condition \cref{eq:Vreal}, this implies that the surviving components are real.

The combination of $C_2(z)$ and $\sigma_x$ also generates the second vertical mirror plane $\sigma_y$ (\cref{eq:sigmaytransf}), which therefore imposes no additional constraints. 

At rank two, the allowed components are therefore \smash{$V_0^{(2)}$} and \smash{$V_{\pm 2}^{(2)}$}, with \smash{$V_2^{(2)}=V_{-2}^{(2)}\in \mathbb{R}$}.
The EFG is characterized by the overall scale $\Vcartzz$ and the asymmetry parameter $\eta$ (see \cref{eq:etadef_app}).

In Cartesian language, the defining operations act as
\begin{align}
    C_2(z): \quad & (x,y,z) \to (-x,-y,z)\,,\\
    \sigma_x: \quad & (x,y,z) \to (-x,y,z)\,.
\end{align}
Their combination requires any nonzero Cartesian EFG component to contain an even number of each Cartesian index.
At rank two this gives $\Vcart{xy} = \Vcart{xz} = \Vcart{yz} = 0 $, so that the crystal axes form a principal-axis frame of the quadrupole EFG.

At rank four, $q=0, \pm 2, \pm 4$ survive, and the independent parameters may be chosen as $\Vcartzzzz$, $\zeta_2$ and $\zeta_4$.
The general Cartesian relations of \cref{eq:zetacartgeneral} reduce to 
\begin{equation}
    \label{eq:zetatwosym}
    \begin{split}
    \zeta_1 =& \zeta_3 = 0\,,\\
    \zeta_2 =& \sqrt{\frac{2}{5}}\frac{2 \Vcart{xxzz} + \Vcartzzzz}{\Vcartzzzz}\,,
    \\
    \zeta_4 =& \frac{8\Vcart{xxxx}+8\Vcart{xxzz}+\Vcartzzzz}{\sqrt{70}\Vcartzzzz}\,,
    \end{split}
\end{equation}
assuming $\Vcartzzzz\neq 0$.
Thus $\eta$, $\zeta_2$ and $\zeta_4$ are real in the symmetry-adapted coordinate frame.

The corresponding Hamiltonians follow from the general matrix representation \cref{eq:Hmmp}.
In the ground-state subspace $m=\lbrace 5/2, 1/2, -3/2\rbrace$, they are
\begin{align}
    \label{eq:Hquadgstwosym}
    \mathcal{H}^{(2)}_{I=5/2,\twosym} 
    &= 
    \frac{1}{4} \Qcartzzspec{\gs} \Vcartzz
    \begin{pmatrix}
         1 & \frac{\eta }{\sqrt{10}} & 0 \\
         \frac{\eta }{\sqrt{10}} & -\frac{4}{5} & \frac{3\eta}{5 \sqrt{2}} \\
         0 & \frac{3 \eta }{5\sqrt{2}} & -\frac{1}{5}
    \end{pmatrix}\,,
    \\ 
    \label{eq:Hexagstwo}
    \mathcal{H}^{(4)}_{I=5/2,\twosym}  
    &= 
    \frac{1}{192} \Qcartzzzzspec{\gs} \Vcartzzzz 
    \begin{pmatrix}
        1 & 3\zeta_2 & \sqrt{14} \zeta_4 \\
        3\zeta_2 & 2 & -\sqrt{5}\zeta_2 \\
        \sqrt{14}\zeta_4 & -\sqrt{5} \zeta_2 & -3
    \end{pmatrix}
    \,,
\end{align}
and in the $m =\{ \frac{3}{2}, -\frac{1}{2}\}$ subspace for the isomer
\begin{align}
    \label{eq:Hquadmtwo}
    \mathcal{H}^{(2)}_{I=3/2,\twosym} 
    = 
    \frac{1}{4} \Qcartzzspec{\m} \Vcartzz
    \begin{pmatrix}
        1 & \frac{\eta }{\sqrt{3}} \\
        \frac{\eta }{\sqrt{3}} & -1 \\
    \end{pmatrix}\,.
\end{align}
The hexadecapole interaction vanishes identically for the isomer because $k=4>2I=3$.

As derived in \cref{eq:IIHIIcartQ}, the Hamiltonians are normalized to their stretched-state matrix elements, which, in the chosen bases, correspond to the upper-left entry of each matrix.
The quadrupole Hamiltonian is identical to the general case, \cref{eq:Hquadmgeneral}. 
This reflects the fact that any symmetric traceless rank-2 tensor may be diagonalized by a suitable choice of principal-axis frame, eliminating the $q=\pm1$ components. 
Consequently, twofold symmetry imposes no additional constraints on the quadrupole interaction beyond those already implied by this choice of coordinates.

The isomeric hyperfine splitting follows directly from \cref{eq:Hquadmtwo},
\begin{align}
    \label{eq:nu_m_twosym}
    [\nu_{\m}]_{\twosym} &= \frac{1}{2} \Qcartzzspec{\m} \Vcartzz \sqrt{1+\frac{\eta^2}{3}}\,.
\end{align}
For the ground state, we treat the hexadecapole interaction as a perturbation of the quadrupole Hamiltonian.
To characterize their relative strength, we introduce the dimensionless parameter $h$ (see also \cref{eq:hdef})
\begin{align}
    h\equiv \frac{\Qcartzzzzspec{\gs}\,\Vcartzzzz}{\Qcartzzspec{\gs}\,\Vcartzz}
\end{align} 
and assume $|h|\ll 1$.
Denoting the eigenstates and eigenvalues of the quadrupole Hamiltonian \cref{eq:Hquadgstwosym} by $\ket{I=\tfrac{5}{2},n}$ and \smash{$E_{I=5/2,n}^{(2)}$}, respectively, where $n=1,2,3$ labels the three Kramers-degenerate levels, the ground-state energies to first order in the hexadecapole interaction are
\begin{align}
    E_{5/2,n} = E_{5/2,n}^{(2)} + \delta E_{5/2,n}^{(4)} + \cO(|\Qcartzzspec{\gs} \Vcartzz |h^2)\,,\qquad
    \delta E_{5/2,n}^{(4)} = \mel{\tfrac{5}{2},n}{\Hamilhat^{(4)}}{\tfrac{5}{2},n}\,.
\end{align}
The hyperfine splittings are obtained from differences between these energies.
If the hierarchy $h\ll1$ breaks down, the perturbative treatment is no longer justified, and the full Hamiltonian $\Hamilhat^{(2)}+\Hamilhat^{(4)}$ must be diagonalized.

\subsection*{Tetragonal Symmetry ($\tetrasym$)}

The tetragonal symmetry class additionally imposes invariance under a fourfold rotation around the $z$ axis,
\begin{align}
    \label{eq:tetrasymcond}
    C_4(z): \qquad V_q^\krank = e^{-iq \pi/2} V_q^\krank\,.
\end{align}
Invariance therefore requires $q=0\pmod 4$.
At rank two, only $q=0$ survives. 
Consequently,
\begin{align}
    \label{eq:etateta}
    \eta = 0\,,
\end{align}
implying that the quadrupole EFG is axially symmetric. 
In Cartesian form, $\Vcart{xx} = \Vcart{yy} $, while tracelessness requires $\Vcart{xx} + \Vcart{yy} + \Vcartzz=0$.

At rank four, $q=0,\pm 4$ survive, and the independent parameters are conveniently chosen as $\Vcartzzzz$ and $\zeta_4$.
The fourfold rotation requires $    \Vcart{xxxx} = \Vcart{yyyy}$ and $\Vcart{xxzz} = \Vcart{yyzz}$.
Together with tracelessness, $\Vcart{xxzz} + \Vcart{yyzz} + \Vcartzzzz=0$, this gives
\begin{align}
    \label{eq:zetatetra}
    \zeta_1 = \zeta_2 = \zeta_3 = 0\,, \qquad 
    \zeta_4 = \frac{8\Vcart{xxxx}-3\Vcartzzzz}{\sqrt{70}\Vcartzzzz}\,,
\end{align}
assuming $\Vcartzzzz\neq 0$.
The tetragonal Hamiltonians follow directly from \cref{eq:Hquadgstwosym,eq:Hexagstwo,eq:Hquadmtwo}.
The quadrupole Hamiltonians become diagonal in the $|I,m\rangle$ basis, implying that the quadrupole eigenstates have definite $m$.
The hexadecapole Hamiltonian contains both diagonal ($q=0$) elements and off-diagonal elements, proportional to $\zeta_4$, that couple states with $\Delta m=\pm 4$.
\begin{equation}
    \label{eq:Htetra}
    \begin{split}
    \mathcal{H}^{(2)}_{I=5/2,\tetrasym} 
    = 
    \frac{1}{4} \Qcartzzspec{\gs} \Vcartzz
    \begin{pmatrix}
         1 & 0 & 0 \\
         0 & -\frac{4}{5} &0 \\
         0 & 0 & -\frac{1}{5}
    \end{pmatrix}\,,
    \qquad 
    & \mathcal{H}^{(4)}_{I=5/2,\tetrasym} 
    = 
    \frac{1}{192} \Qcartzzzzspec{\gs} \Vcartzzzz 
    \begin{pmatrix}
        1 & 0 & \sqrt{14} \zeta_4 \\
        0 & 2 & 0 \\
        \sqrt{14} \zeta_4 & 0 & -3
    \end{pmatrix}\,,
    \\
    \mathcal{H}^{(2)}_{I=3/2,\tetrasym} 
    = &
    \frac{1}{4}\Qcartzzspec{\m} \Vcartzz
    \begin{pmatrix}
        1 & 0 \\
        0 & -1 \\
    \end{pmatrix}\,.
    \end{split}
\end{equation}
To first order in $h$, the hyperfine splittings are 
\begin{align}
    \label{eq:necD4h}
    [\nu_{\m}]_{\tetrasym} 
    &= 
    \frac{1}{2}  \Qcartzzspec{\m} \Vcartzz \, , \\
    \label{eq:ndbD4h}
    [\nu_+]_{\tetrasym} 
    &= 
    \frac{1}{2}  \Qcartzzspec{\gs} \Vcartzz \left(\frac{3}{5}  + \frac{1}{24} h\right) + \cO(h^2) \, , \\
    \label{eq:ncaD4h}
    [\nu_-]_{\tetrasym} 
    &= 
    \frac{1}{2} \Qcartzzspec{\gs} \Vcartzz \left(\frac{3}{10}  - \frac{5}{96} h\right) + \cO(h^2) \,.
\end{align}
The linear in $h$ hexadecapole energy corrections are given by the diagonal matrix elements $\mel{I,m}{\Hamilhat^{(4)}}{I,m}$.
For a nonzero quadrupole interaction, the terms proportional to $\zeta_4$ couple nondegenerate states and therefore do not contribute to the linear in $h$ hexadecapole energy corrections, but enter only at second order.
The above relations can be inverted to obtain, \cref{eq:hfstetra}.

\subsection*{Cubic Symmetry ($\cubsym$)}

Finally, the cubic symmetry class imposes additional rotations that interchange the three Cartesian axes.
In addition to the tensor constraints above, it is sufficient to impose a threefold rotation about a body diagonal, represented by
\begin{align}
    \label{eq:cubsymcond}
    x \to y \to z \to x\,.
\end{align}

At rank two this requires $\Vcart{xx}=\Vcart{yy}=\Vcartzz$.
Together with tracelessness, i.e. $\Vcart{xx}+\Vcart{yy}+\Vcartzz=0$ this implies
$\Vcart{xx}=\Vcart{yy}=\Vcartzz=0$.
Thus no rank-2 EFG is allowed by cubic symmetry, and
\begin{align}
    \mathcal{H}^{(2)}_{\cubsym} = 0\,.
\end{align}
The rank-4 tensor, however, contains a nonvanishing cubic invariant.
Cubic symmetry requires
\begin{align}
    \Vcart{xxxx}=\Vcart{yyyy} = \Vcart{zzzz}\,, \qquad
    \Vcart{xxyy} = \Vcart{xxzz} = \Vcart{yyzz}\,.
\end{align}
Tracelessness relates these components through
\begin{align}
    \Vcartzzzz + 2\Vcart{xxzz} = 0\,.
\end{align}
Substitution into \cref{eq:zetatetra} fixes the relative weight of the surviving $q=0$ and $q=\pm 4$ components,
\begin{align}
    \label{eq:zetacub}
    \zeta_4 = \frac{5}{\sqrt{70}}\,.
\end{align}
The corresponding cubic invariant has angular dependence proportional to $\sphYdef{4}{0} + 5/\sqrt{70} (\sphYdef{4}{4}+\sphYdef{4}{-4})$.
Thus the cubic rank-4 EFG contains a single independent parameter, which we choose to be its overall scale $\Vcartzzzz$.
\medskip

The cubic hexadecapole Hamiltonian follows from the tetragonal hexadecapole Hamiltonian in \cref{eq:Htetra} by setting $\zeta_4=5/\sqrt{70}$.

\begin{equation}
    \label{eq:Hexacub}
    \mathcal{H}^{(4)}_{I=5/2,\cubsym} 
    = 
    \frac{1}{192} \Qcartzzzzspec{\gs} \Vcartzzzz 
    \begin{pmatrix}
        1 & 0 & \sqrt{5} \\
        0 & 2 & 0 \\
        \sqrt{5} & 0 & -3
    \end{pmatrix}\,.
\end{equation}
Diagonalizing this reduced Hamiltonian yields two distinct energy levels,
\begin{align}
    E_1  = -\frac{1}{48} \Qcartzzzzspec{\gs}\Vcartzzzz\,,\qquad
    E_2 = E_3 = \frac{1}{96} \Qcartzzzzspec{\gs}  \Vcartzzzz\,.
\end{align}
The sign of $\Qcartzzzzspec{\gs}\Vcartzzzz$ determines the ordering of these levels: for a positive product, the fourfold-degenerate level lies above the twofold-degenerate level, and below if the product is negative.
In this limit, the two ground-state splittings $\nu_+$ and $\nu_-$ coincide, yielding
\begin{align}
    [\nu_\pm]_{\cubsym} = \frac{1}{32}
    \left| \Qcartzzzzspec{\gs} \Vcartzzzz \right|\,.
\end{align}
Since both the quadrupole and hexadecapole interactions vanish for the $I=3/2$ isomer, the isomeric manifold remains unsplit in cubic symmetry, i.e. $[\nu_{\m}]_{\cubsym}=0$.

\section{Coulomb Energy of a Deformed Fermi Distribution}
\label{app:WS}

In the following, we derive the nuclear Coulomb energy using a deformed Fermi model, which parametrizes the nuclear shape through its radial size, surface diffuseness and multipole deformations.
\begin{align}
    \rho(r,\Omega) = \frac{\rhoo}{1+e^{\frac{r-R(\Omega)}{a}}}\,,
    \qquad 
    R(\Omega) = \Roo \bigg (1+\sum_{k=1}^\infty \beta_k \sphY{k}{0}{\Omega} \bigg)
    \label{eq:rho_WS_app}
\end{align}
where $R_0$ is the radius parameter, $a$ the surface diffuseness, and $\beta_k$ are the deformation parameters. 
We restrict the spherical-harmonic expansion to $m=0$, corresponding to axially symmetric intrinsic charge distributions. 
This approximation is motivated by the description of the low-lying spectrum of \Th in terms of rotational bands with approximately axially symmetric intrinsic configurations (see \cref{sec:particle-rotor}).
For the spherical harmonics, we use the normalization $\int \mathrm{d}\Omega\, Y_{l m}^*(\Omega) \, Y_{l'm'}(\Omega) = \delta_{l l'}\delta_{m m'}$.

The normalization factor $\rho_0$ is fixed by the nuclear charge via the relation
\begin{align}
    \int \! \drth \, \rho(\rvec) =  Ze
    \label{eq:charge_cons}
\end{align}
with $Z=90$ for thorium. 
Moreover, since the dipole deformation parameter $\beta_1$ corresponds to a translation of the charge distribution relative to the origin, we can set $\beta_1 = 0$ by choosing the origin at the center of charge. 
In other words, we choose our coordinate system such that
\begin{align}
    \int \! \drth \, \rvec \rho(\rvec) = 0 \,.
\end{align}

\subsection{Expansion in Deformation Parameters}
\label{app:rho_exp}

Assuming the deviations from spherical symmetry are small, $\beta_k\ll 1$ for all $k$,  the charge density may be expanded in $\beta_k$.
At zeroth order, the Fermi profile has the parameters, $\rhoo$, $\Roo$ and $a$.
The first two parameters can be related to the total charge (see \cref{eq:charge_cons}), the mean-square radius, while for the surface diffuseness we adopt $a^\gs=a^\m=0.5$~fm, a representative value for heavy nuclei based on electron-scattering experiments~\cite{DeVriesDeVries}.  
See Table~\ref{tab:expinput} for the current best constraints. 
Once deformations are introduced, however, the predictions for these observables acquire $\mathcal{O}(\beta^2)$ corrections.

We expand the charge density and the Fermi parameters in powers of the deformation parameters:
\begin{align}
    \label{eq:rho_exp}
    \rho(r,\Omega) &= \rho^{(0)}(r) + \rho^{(1)}(r,\Omega)
                    + \rho^{(2)}(r,\Omega) + \mathcal{O}(\beta^3) \,, \\
    \label{eq:rhoo_exp}
    \rhoo &= \rhoo^{(0)} + \rhoii + \mathcal{O}(\beta^3)\, , \\
    \label{eq:Roo_exp}
    \Roo &= \Roo^{(0)} + \Roii + \mathcal{O}(\beta^3) \, .
\end{align}
The surface diffuseness $a$ is treated as an intrinsic property of the nuclear surface and as an independent input parameter. It is therefore not expanded. 
For readability, we write \smash{$\rhoo^{(0)} \to \rhoo$} and \smash{$\Roo^{(0)}\to \Roo$} from here on, retaining the superscript only for $\mathcal{O}(\beta^2)$ quantities.
The second-order monopole corrections \smash{$\rhoii$} and \smash{$\Roii$} ensure that the total nuclear charge and mean-square charge radius $\chrsq$ remain fixed as the deformation parameters are varied. 
Their explicit forms are derived in \cref{app:rhoRo}.
The choice to keep these corrections explicit, rather than absorbing them into the model parameters is a matter of bookkeeping rather than accuracy.

Plugging \cref{eq:rhoo_exp,eq:Roo_exp} into \cref{eq:rho_WS_app} and expanding in $\beta_k$, we obtain 
\begin{equation}
    \label{eq:rho_exp_explicit}
    \begin{aligned}
        \rho^{(0)}(r) &= \rhoo f(r) \,, \\[4mm]
        \rho^{(1)}(r,\Omega) &= - \rhoo \Roo f'(r) \sum_{\ell} \beta_{\ell} \sphY{\ell}{0}{\Omega} \,, \\
        \rho^{(2)}(r,\Omega) &=  \frac{1}{2} \rhoo \Roo^2 f''(r)  \sum_{\ell, l} \beta_{\ell} \beta_{l} \sphY{\ell}{0}{\Omega} \sphY{l}{0}{\Omega} + \rhoii f(r) - \rhoo \Roii f'(r) \,.
    \end{aligned}
\end{equation}
Here, we have used as an abbreviation the Fermi distribution function
\begin{align}     
    \label{eq:fdef}
    f(r) \equiv& \frac{1}{1 + e^{\frac{r - \Roo}{a}}} \,.
\end{align}
Note that $f(r)$ satisfies the relations
\begin{align}
    \label{eq:flims}
    \lim_{r\to 0}r^n f(r) = 0\,,\qquad 
    \lim_{r\to \infty}r^n f(r) = 0\,. \qquad
    n > 0 \,.
\end{align}
In working with the expansion of the charge density in the deformation parameters, we often encounter the moments of $f(r)$ and its derivatives. 
The following integrals will therefore be useful:
\begin{align}
    \label{eq:Ilk}
    \begin{split}   
        I_n^{(0)} &\equiv \int_0^\infty \dr r^n f(r)
                  = -a^{n+1} \Gamma(n+1) \, \Li{n+1}{-b}\\
        I_n^{(1)} &= \int_0^\infty \dr r^n f'(r) 
            = \big[r^n f(r) \big]_{r=0}^\infty - n\int_0^\infty\!\dr r^{n-1}f(r)
            = -n I_{n-1}^{(0)}\\
        \vdots \quad & \qquad \qquad\vdots\\
        I_{n}^{(k)} &= \int_0^\infty \dr r^n f^{(k)}(r)
            = (-1)^{k}\frac{n!}{(n-k)!} I_{n-k}^{(0)}
            = (-1)^{k+1} n! \, a^{n-k+1}\Li{n-k+1}{-b}\,,
    \end{split}
\end{align}
with the notation
\begin{align}
    \label{eq:bdef}
    b \equiv e^{\Roo/a}\,.
\end{align}
In \cref{eq:Ilk} we have used integration by parts; the boundary terms vanish thanks to the asymptotic behavior from \cref{eq:flims}. 
The leading dependence on the surface diffuseness $a$ is obtained by applying the Sommerfeld expansion~\cite{Sommerfeld:1928xx, AshcroftMermin}, an asymptotic expansion in $a/\Roo \ll 1$ about the hard-shell limit ($a\to 0$) of the Fermi distribution $f(r)$. 
More concretely, we write the Fermi function as $f(r) = \Theta(\Roo-r) + \big[f(r)-\Theta(\Roo-r)\big]$ where $\Theta$ is the Heaviside function and the difference $g(r)=f(r)-\Theta(\Roo-r)$ is sharply localized near the surface. 
It is an odd function of $(r-\Roo)/a$ up to exponentially small corrections, and it decays on the scale $a$.
The step-function integrates trivially, $\int_0^\infty \dr r^n \Theta(\Roo - r) = \Roo^{n+1}/(n+1)$ and gives a hard-shell ($a=0$) result. The correction is $\int_0^\infty \dr r^n g(r)$. 
Because $g(r)$ is concentrated within a few $a$ of $r=\Roo$, the smooth test function $r^n$ can be expanded in a Taylor series about $r=\Roo$ and integrated term by term against $g(r)$.
Because $g(r)$ is odd in $(r-\Roo)/a$, only odd powers of $(r-\Roo)$ contribute when the lower integration limit is extended to $-\infty$, up to exponentially suppressed corrections.
This yields an expansion in even powers of $a/\Roo$, which terminates for non-negative integer $n$ because $r^n$ is a polynomial.
Explicitly,
\begin{align}
    \label{eq:SommerfeldExp}
    I_{n}^{(0)} = \int_0^\infty \dr r^n f(r)
        \approx \frac{\Roo^{n+1}}{n+1}
              + \frac{(\pi a)^2}{6} n \Roo^{n-1}
              + \frac{7(\pi a)^4}{360}n(n-1)(n-2) \Roo^{n-3}
              + \ldots
              + \mathcal{O}\big(e^{-\Roo/a}\big)\,.
\end{align} 
We will also need the cumulative radial moments, which are defined as
\begin{align}
    \label{eq:Fnkr}
    F_n^{(k)}(r) &= \int_0^r \mathrm{d}s\, s^n f^{(k)}(s) \,.
\end{align}
They satisfy $F_n^{(k)}(\infty) = I_n^{(k)}$ and $\frac{\mathrm{d}}{\mathrm{d}r}F_n^{(k)}(r) = r^n f^{(k)}(r)$.

In the numerical analysis we use the Sommerfeld expansion once, to relate $\Roo$ and $a$ to the measured mean-square charge radius $\langle r^2\rangle$ (see Appendix~\ref{app:rhoRo}). 
Elsewhere the expansion is used only for analytic cross-checks and to make the dependence on the surface diffuseness $a$ transparent. 
The leading term of the Sommerfeld expansion corresponds to the hard-shell result ($a=0$), where the charge density takes the form
\begin{align}
    \label{eq:rho_hard-shell}
    \rho(r,\Omega) =  \rhoo \, \Theta\big(R(\Omega)-r\big)\,.
\end{align}
Note that the relations derived here differ from the volume-conserving treatment of the liquid-drop model, which holds $\tilde{V}$ rather than $\chrsq$ fixed. 
The two are related by a change of constraint (see Appendix~\ref{app:scheme-dependence}).
The explicit relations between \smash{$\rhoo$, $\rhoii$, $\Roo$, $\Roii$}, $a$ and $Ze$, as well as $\chrsq$ are derived in the following section.

\subsection{Relating $\rho_0$, $\Roo$ and $a$ to Experiment}
\label{app:rhoRo}

We now aim to express as many parameters as possible of the deformed Fermi model in terms of experimental observables, using the expansion in terms of deformation parameters $\beta_k$ introduced in \cref{eq:rho_exp}. 
We have already argued in \cref{eq:charge_cons} that $\rho_0$ is fixed by charge conservation. 
Plugging \cref{eq:rho_exp} into \cref{eq:charge_cons} and using the integrals from \cref{eq:Ilk}) leads to
\begin{align}
    \label{eq:Zexp}
    Ze &= \int_0^\infty \dr r^2 \int \dOmega \Big[
              \rho^{(0)}(r) + \rho^{(1)}(r,\Omega) + \rho^{(2)}(r, \Omega)
          \Big]
        + \mathcal{O}(\beta^3)
    \notag \\
    &= 4\pi \rhoo I_2^{(0)} 
     + \frac{1}{2} \rhoo \Roo^2 I_2^{(2)}\sum_k \beta_k^2
     + 4\pi \rhoii I_2^{(0)} 
     - 4\pi \rhoo \Roii I_2^{(1)}
     + \mathcal{O}(\beta^3) \, .
\end{align}
As explained in \cref{app:rho_exp}, we are free to trade $\mathcal{O}(\beta_k^2)$ contributions between the zeroth-order terms in $\rho_0$, $R_0$, and the second-order monopole corrections \smash{$\Roii$} and \smash{$\rhoii$}. 
We choose to fix the zeroth-order value of $\rho_0$ such that 
\begin{align}
    \label{eq:rhoodef}
    \rhoo &= \frac{Ze}{4\pi I_2^{(0)}}
           = -\frac{Ze}{8\pi a^3\Li{3}{-b}} \, .
\end{align}
The $\mathcal{O}(\beta^2)$ contributions in \cref{eq:Zexp} then has to vanish:
\begin{align}
    \label{eq:b2_charge_cons}
    \frac{1}{2} \rhoo \Roo^2 I_2^{(2)}\sum_{\ell} \beta_\ell^2
    + 4\pi \rhoii I_2^{(0)} 
    - 4\pi \rhoo \Roii I_2^{(1)} 
    = 0 \,.
\end{align}
We will use this condition to determine $\rho_0^{(2)}$ and $R_0^{(2)}$ below.  

The nuclear radius parameter $\Roo$ can be related to the measured \smash{$\ev{r^2}$} by
\begin{align}
    \ev{r^2} &= \frac{\int \drth \, r^2 \rho(\rvec)}{\int\drth\,\rho(\rvec)} 
        =  \frac{1}{Ze} \bigg( 
               4\pi \rhoo I_4^{(0)} 
             + \frac{1}{2}\rhoo \Roo^2 I_4^{(2)}\sum_k \beta_k^2
             + 4 \pi \rhoii I_4^{(0)}
             - 4\pi \rhoo \Roii I_4^{(1)}
             + \mathcal{O}(\beta^3)
           \bigg) \, .
    \label{eq:avg-rsq-expansion}
\end{align}
We again make the choice that already the zeroth-order terms on the right-hand side of \cref{eq:avg-rsq-expansion} should match the measured \smash{$\ev{r^2}$} on the left-hand side, which implies choosing \smash{$R_0^{(0)}$} such that
\begin{align}
    \label{eq:b0_rsq}
    \chrsq &= \frac{I_4^{(0)}}{I_2^{(0)}}
            = 12 a^2 \frac{\Li{5}{-b}}{\Li{3}{-b}} \,.
\end{align}
The deformation-induced corrections to $\ev{r^2}$ then need to cancel order-by-order in $\beta_k$. 
At $\mathcal{O}(\beta^2)$ we have
\begin{align}
    \label{eq:b2_R0_cons}
    \frac{1}{2} \rhoo \Roo^2 I_4^{(2)} \sum_k \beta_k^2
    + 4 \pi \rhoii I_4^{(0)}
    - 4 \pi \rhoo \Roii I_4^{(1)}
    = 0 \,.
\end{align}
Applying the Sommerfeld approximation, \cref{eq:SommerfeldExp}, to \cref{eq:b0_rsq} yields 
\begin{align}    
    \label{eq:rsqlo}
    \chrsq = \frac{3}{5}\Roo^2 + \frac{7}{5} (\pi a)^2 
    + \mathcal{O}\big(e^{-\Roo/a}\big)\,,
\end{align}
or, equivalently,
\begin{align}
    \label{eq:Roodef}
    \Roo = \sqrt{\frac{5}{3} \chrsq - \frac{7}{3} \left(\pi a\right)^2}
         + \mathcal{O}\big(e^{-\Roo/a}\big)\,.
\end{align}
As desired, this expression for $\Roo$ is free of deformation-induced corrections, directly relating $\Roo$ to the experimentally accessible mean-square charge radius and surface diffuseness.

The second-order monopole corrections $\Roii$ and $\rhoii$ can be determined by solving \cref{eq:b2_charge_cons,eq:b2_R0_cons}, and applying \cref{eq:Ilk} to convert all integrals \smash{$I_n^{(k)}$} with $k>0$ to integrals of the form \smash{$I_{m}^{(0)}$}. 
We find
\begin{align}
    \label{eq:rhoiidrhoo}
    \frac{\rhoii}{\rhoo} &= \frac{\Roo^2}{2\pi}
        \frac{3 I_1^{(0)} I_2^{(0)} - I_0^{(0)} I_3^{(0)}}
             {2 I_2^{(0)} I_3^{(0)} - I_1^{(0)} I_4^{(0)}} \sum_k \beta_k^2
    = \frac{(\log b)^2}{8\pi\mathcal{W}(b)} \Big(
          \Li{2}{-b} \Li{3}{-b} + \Li{4}{-b} \log(b+1)
      \Big) \sum_k \beta_k^2 \,, \\
    \label{eq:RoiidRoo}
    \frac{\Roii}{\Roo} &= \frac{\Roo}{8\pi}
        \frac{I_0^{(0)}I_4^{(0)} - 6\big(I_2^{(0)}\big)^2}
             {2I_2^{(0)}I_3^{(0)} - I_1^{(0)} I_4^{(0)}} \sum_k \beta_k^2
    = -\frac{\log b}{8\pi\mathcal{W}(b)} \Big(
           \Li{3}{-b}^2 + \Li{5}{-b} \log(b+1)
       \Big) \sum_k \beta_k^2 \, ,
\end{align}
with the abbreviation
\begin{align}
    \label{eq:Wdef}
    \mathcal{W}(x) 
    = 
    \Li{3}{-x} \Li{4}{-x} - \Li{2}{-x} \Li{5}{-x} \, .
\end{align}
%

\subsection{Coulomb Energy Expansion}
\label{app:EC-Fermi}

The electrostatic (Coulomb) energy of a charge distribution $\rho(\rvec)$ can be computed using
\begin{align}
    \label{eq:CoulombE}
    E_{\rm C} = \frac{1}{2} \int\!\drth \int\!\dsth
        \frac{\rho(\rvec) \rho(\svec)}{4\pi |\rvec-\svec|} \,.
\end{align}
To compute the Coulomb energy of an atomic nucleus in the deformed Fermi mode, we apply the expansion of the charge density in \cref{eq:rho_exp} to both $\rho(\rvec)$ and $\rho(\svec)$:
\begin{align}
    \label{eq:CoulombE-expanded}
    E_{\rm C} = \sum_{a+b \leq 2} E^{(a,b)} + \mathcal{O}(\beta^3)\,,
    \qquad
    E^{(a,b)} \equiv 
    \frac{1}{2} \int\!\drth \int\!\dsth
    \frac{\rho^{(a)}(\rvec) \rho^{(b)}(\svec)}{4\pi |\rvec-\svec|} \,.
\end{align}
The second equality defines $E^{(a,b)}$ as the contribution of $\rho^{(a)}(r,\Omega)$ and $\rho^{(b)}(r,\Omega)$ to the electrostatic energy. 
In evaluating \cref{eq:CoulombE-expanded} further, it is moreover convenient to use the Laplace expansion
\begin{align}
    \label{eq:LaplacExp}
    \frac{1}{|\rvec-\svec|} &= \sum_{\ell=0}^\infty \frac{4\pi}{2\ell+1}\frac{s^{\ell}}{r^{\ell +1}} \sum_{m=-\ell}^\ell \left(-1\right)^m \sphY{\ell}{-m}{\Omegar}\sphY{\ell}{m}{\Omegas}
    \qquad\qquad\text{($s \leq r$)}\,.
\end{align}
Splitting the radial integration domain into $s \leq r$ and $r\leq s$, and exchanging the integral variables in the second region, we obtain
\begin{multline}
    \label{eq:Eab}
    E^{(a,b)} = \frac{1}{2} \sum_{\ell=0}^\infty \frac{1}{2\ell+1}
        \int_0^\infty\!\mathrm{d}r \, r^{1-\ell}
        \int_0^r \! \mathrm{d}s \, s^{2 + \ell}
        \times \sum_{m=-\ell}^\ell \left(-1\right)^m\int \mathrm{d}\Omega_r \, \int \mathrm{d}\Omega_s \\
    \times \Big[
        \rho^{(a)}(r,\Omega_r) \, \sphY{\ell}{-m}{\Omega_r} \, \rho^{(b)}(s,\Omega_s) \, \sphY{\ell}{ m}{\Omega_s}
      + \rho^{(b)}(r,\Omega_r) \, \sphY{\ell}{ m}{\Omega_r} \, \rho^{(a)}(s,\Omega_s) \, \sphY{\ell}{-m}{\Omega_s} 
    \Big]\,,
\end{multline}
with $a,b \in \lbrace 0,1,2\rbrace$ and $\rho^{(a)}$ and $\rho^{(b)}$ as defined in \cref{eq:rho_exp}.

Integrating over all directions, the $\mathcal{O}(\beta)$ contributions ($(a,b) = (1,0)$ or $(0,1)$) cancel out,
\begin{align}
    \label{eq:E01}
    E^{(0,1)}
    \propto & 
    \sum_{\ell, m} \sum_{k>0}
    \beta_k
    \left(
    \int \dOmegar \sphYdef{\ell}{-m}
    \right)
    \left(
    \int \dOmegas \sphYdef{k}{0}\sphYdef{\ell}{m}
    \right) + \left( r \leftrightarrow s\right)\notag \\
    \propto & \sum_{\ell, m} \delta_{m0}\delta_{\ell 0}\sum_{k>0}\beta_k\delta_{k\ell }
    = 0
\end{align}
and similarly for $E^{(1,0)}$.
The first angular integral enforces $\ell=0$, while the second one requires $k=\ell$ due to the orthogonality of the spherical harmonics. This implies that all terms encoding deformations of the charge distribution ($k>0$) vanish.
Consequently, to leading order in the deformation parameters, the Coulomb energy takes the form
\begin{align}
    E_{\rm C} = E^{(0,0)} + E^{(0,2)} + E^{(1,1)} + E^{(2,0)} + \mathcal{O}(\beta^3)
\end{align}

\subsection{Spherical Contribution to the Coulomb Energy}
\label{app:EC-spherical}

The spherical contribution to the Coulomb energy, $E^{(0,0)}$, is obtained from \cref{eq:Eab} with $a,b=0$. 
Only the monopole term ($\ell=0$) contributes, yielding
\begin{align}
    E^{(0,0)} &= 4\pi \rhoo^2
    \int_0^\infty \dr \, r f(r) \, F_2^{(0)}(r)
    \,,
\end{align}
with $F_2^{(0)}(r)$ as defined in \cref{eq:Fnkr}. 
Substituting \cref{eq:rhoodef} for $\rho_0$, evaluating the inner radial integral, introducing the variable
 $u=e^{-(r-\Roo)/a}$, and integrating by parts yields
\begin{align}
    \label{eq:E00}
    E^{(0,0)} &= 4 \pi \rhoo^2 a^5\mathcal{A}(b)
               = \frac{3}{5}\frac{(Ze)^2}{4\pi \Roo} \mathcal{C}(b)\,,  
\end{align}
with $b$ as defined in \cref{eq:bdef} and
\begin{align}
   \mathcal{A}(x) &= 2 S_{3,2}(-x)+Z(x)\,, \qquad 
   Z(\xi) = \int_0^\xi \frac{\Li{2}{-u}^2}{u}\mathrm{d}u 
   \,, \label{eq:Acaldef}
  \\
  \mathcal{C}(x) =& \frac{5\log x \, \mathcal{A}(x)}{12\Li{3}{-x}^2}\,, \qquad \lim_{x\to \infty}\mathcal{C}(x)= 1 \,, \qquad \lim_{x\to \infty}\mathcal{C}'(x)= 0 \,.\label{eq:Ccaldef}
\end{align}
In the second expression for $E^{(0,0)}$, we have used \cref{eq:rhoodef,eq:bdef} to arrive at a form that makes it evident how the standard expression for the Coulomb energy of a hard sphere, $\frac35 (Ze)^2 / (4\pi R_0)$, is obtained in the $a \to 0$ limit.
In the definition of $\mathcal{A}(x)$, $S_{3,2}$ is a Nielsen generalized polylogarithm, defined through
\begin{align}
    \SLi{n,p}{z} = \frac{(-1)^{n+p-1}}{(n-1)!p!}\int_0^1 \frac{\mathrm{d}t}{t}(\log t)^{n-1} (\log(1-zt))^p\,.
    \label{eq:NielsenGeneralizedPolylog}
\end{align}
In the hard-shell limit $a\to 0$ (\cref{eq:rho_hard-shell}), $E^{(0,0)}$ reduces to the familiar Coulomb energy of a uniformly charged sphere. 
In this limit the surface diffuseness parameter is no longer present.

\subsubsection*{Derivatives}

The derivatives of $E^{(0,0)}$ with respect to the Fermi parameters are required in the calculation of the enhancement factor in \cref{sec:K}. 
Differentiating \cref{eq:E00} gives
\begin{align}
    \frac{\partial E^{(0,0)}}{\partial \Roo}
    =& \frac{(Ze)^2}{16\pi a^2 \Li{3}{-b}^2}\left(
    \mathcal{A}_1(b) - 2\mathcal{A}(b) \frac{\Li{2}{-b}}{\Li{3}{-b}}    \right)\label{eq:dE0dR0}
    \\
    \frac{\partial E^{(0,0)}}{\partial a}
    =& -\frac{(Ze)^2}{16\pi a^2}\left(\frac{\mathcal{A}(b)+\log(b) \mathcal{A}_1(b)}{\Li{3}{-b}^2}- \frac{2\log(b)\mathcal{A}(b)\Li{2}{-b}}{\Li{3}{-b}^3} \right)
    \label{eq:dE0da}
\end{align}
with $\mathcal{A}(x)$ as given in \cref{eq:Acaldef} and
\begin{align}
     \mathcal{A}_1(x) &= 2 S_{2,2}(-x)+\Li{2}{-x}^2\,,
   \label{eq:Acal1def}
\end{align}
These results follow from the identities
\begin{align}
    \frac{\mathrm{d}}{\mathrm{d}x}\Li{n}{-x} = \frac{\Li{n-1}{-x}}{x}\,,
    \qquad
    \frac{\mathrm{d}}{\mathrm{d}x}\SLi{n,p}{-x} = \frac{\SLi{n-1,p}{-x}}{x}\,,
    \qquad
    Z'(x) = \frac{\Li{2}{-x}^2}{x}\,   .
\end{align}

\subsection{Deformation Correction to the Coulomb Energy}
\label{app:EC-deformation}

Since the linear-order deformation contributions $E^{(0,1)}$, $E^{(1,0)}$ vanish identically after angular integration (see discussion around \cref{eq:E01}), the leading correction to the Coulomb energy in \cref{eq:E00} appears at $\mathcal{O}(\beta_k^2)$.
As already discussed in \cref{app:rho_exp,app:rhoRo}, the $\mathcal{O}(\beta^2)$ correction to the electrostatic energy receives contributions from two types of sources
\begin{enumerate}[label=(\Alph*)]
    \item \label{it:b2Y} a direct contribution arising from the deformation of the charge distribution (see \eqref{eq:rho_exp}),
    \item \label{it:b2CTs} two monopole corrections $\Roii$ and $\rhoii$, which enforce that the total nuclear charge and the charge radius $\ev{r^2}$ depend only on leading-order quantities (see \cref{eq:charge_cons,eq:b0_rsq}).
\end{enumerate}
Using the decomposition introduced in \cref{eq:Eab}, $E^{(0,2)}$ and $E^{(2,0)}$ contain contributions of both types, whereas $E^{(1,1)}$ arises solely from the direct contribution (\ref{it:b2Y}).

We begin by evaluating $E^{(0,2)}+E^{(2,0)}$.

\subsubsection{Evaluation of $E^{(0,2)}+E^{(2,0)}$}

The term $E^{(0,2)}+E^{(2,0)}$ receives its $\mathcal{O}(\beta^2)$-contributions from $\rho^{(2)}$ which according to \cref{eq:rho_exp_explicit} decomposes as
\begin{align}
    \rho^{(2)}(r,\Omega) &= \underbrace{\frac{1}{2} \rhoo \Roo^2 f''(r)
        \sum_{\ell,l} \beta_{\ell} \beta_{l} \, \sphY{\ell}{0}{\Omega} \, \sphY{l}{0}{\Omega}}_{\ref{it:b2Y}} 
    + \underbrace{\vphantom{\sum_{\ell,l}}
        f(r) \rhoii- \rhoo  f'(r)\Roii}_{\ref{it:b2CTs}}\,.
    \label{eq:rho-2}
\end{align}

\paragraph*{Direct Contributions (Type~\ref{it:b2Y} Terms)} 

Plugging \cref{eq:rho-2} into \cref{eq:Eab} and noting that $\rho^{(0)}$ is spherically symmetric so that only the monopole $\sphY{0}{0}{\Omega_r}$ survives the integration over $\Omega_r$, we see that the angular integration reduces to the Gaunt coefficient
\begin{align}
    \mathcal{G}_{ll'0} = \int \dOmega \sphY{l}{0}{\Omega} \,
        \sphY{l'}{0}{\Omega} \, \sphY{0}{0}{\Omega}
    = \frac{\delta_{ll'}}{\sqrt{4\pi}}\,.
\end{align}
In particular, the angular integration eliminates mixed deformation terms ($\ell \neq l$), yielding the following contribution from the type~\ref{it:b2Y} terms:
\begin{align}
    E^{(0,2)} + E^{(2,0)} \Big\vert_{\ref{it:b2Y}} &=
        \frac{1}{2}\Roo^2 \rhoo^2
        \sum_{\ell>0} \beta_\ell^2 
            \left[ 
                \int_0^\infty \dr r f(r) F_2^{(2)}(r)
                + \int_0^\infty \dr r f''(r) F_2^{(0)}(r)
            \right]
    \label{eq:E0220Ydef}
\end{align}
where $F_{2}^{(2)}(r)$ and $F_{2}^{(0)}(r)$ are defined in \cref{eq:Fnkr}.

Evaluating $F_{2}^{(2)}(r)$ and changing variables to $t=e^{-(r-\Roo)/a}$, we obtain
\begin{align}
    \int_0^\infty \dr r f(r) F_2^{(2)}(r) 
    &=\int_0^1 \mathrm{d}t
    \frac{a^3 b \log t }{(1 + b t)^3}\left(2 (1 + b t) \log t + bt ( \log t)^2 - 
   2(1 + bt)^2 \log\left(\frac{(1 + b)t}{1 + bt}\right)\right)\notag\\
    &= a^3 \Big\lbrace 
     7 \Li{3}{-b} 
    - (7 + 2 \log(1 + b)) \Li{2}{-b}
    - 2 \SLi{1,2}{-b}
    \Big\rbrace \,,
    \label{eq:I02i}
\end{align}
where $S_{1,2}$ is again a Nielsen generalized polylogarithm, see \cref{eq:NielsenGeneralizedPolylog}.

The second term in \cref{eq:E0220Ydef} can be evaluated by integrating by parts twice (with the boundary terms vanishing), using \smash{$\mathrm{d}_r F_2^{(0)}(r) = r^2 f(r)$}, and combining the derivative terms through the relation $f^2 = f + a f'$. Then, using the integrals from \cref{eq:Ilk} yields
\begin{align}
    \int_0^\infty \dr r f''(r) F_2^{(0)}(r)
    &=
    \frac{5}{2} \int_0^\infty \dr r^2 f(r)^2 
    =
    5a^3 \left[ \Li{2}{-b} - \Li{3}{-b}
    \right]\,.
    \label{eq:I02ii}
\end{align}
We obtain
\begin{align}
    E^{(0,2)} + E^{(2,0)} \Big\vert_{\ref{it:b2Y}}
    &=  \frac{(Ze)^2}{4\pi \Roo}\frac{(\log b)^3}{16\pi \Li{3}{-b}^2} \sum_{\ell>0}\beta_\ell^2 \Big[
    \Li{3}{-b} 
    - (1 + \log(1 + b)) \Li{2}{-b}
    - \SLi{1,2}{-b}
    \Big]
\end{align}

\paragraph*{Second-Order Monopole Corrections (Type~\ref{it:b2CTs} Terms)} 

The corrections proportional to $\Roii$ and $\rhoii$ (type~\ref{it:b2CTs} terms) can be written as
\begin{align}
    f(r) = \left.\frac{\partial \rho^{(0)}}{\partial \rhoo}\right\vert_{\Roo}\,, \qquad
    -\rhoo f'(r) = \left.\frac{\partial \rho^{(0)}}{\partial \Roo}\right\vert_{\rhoo}\,.
\end{align}
This can be seen directly from the definition of $\rho(r,\Omega)$ in \cref{eq:rho_WS_app}. Therefore,
\begin{align}
    \left.E^{(0,2)}+E^{(2,0)}\right\vert_{\ref{it:b2CTs}} =& \left.\frac{\partial E^{(0,0)}}{\partial \rhoo}\right\vert_{\Roo}\rhoii 
    + \left.\frac{\partial E^{(0,0)}}{\partial \Roo}\right\vert_{\rhoo} \Roii \,,
    \label{eq:E0220CTdef}
\end{align}
with $E^{(0,0)}$ from \cref{eq:E00}, $\rhoii/\rhoo$ from \cref{eq:rhoiidrhoo}, $\Roii/\Roo$ from \cref{eq:RoiidRoo}, and
\begin{align}
    \left.\frac{\partial E^{(0,0)}}{\partial \rhoo} \right\vert_{\Roo} &= 2 \frac{E^{(0,0)}}{\rhoo}
    = - \frac{(Ze) a^2 \mathcal{A}(b)}{\Li{3}{-b}} \,,
    \label{eq:dE00drhoR0}
    \\
    \left.\frac{\partial E^{(0,0)}}{\partial \Roo}\right\vert_{\rhoo} &= 4\pi \rhoo^2 a^4 \mathcal{A}_1(b)
    \label{eq:dE00dR0rho}
\end{align}
where $\mathcal{A}$ and $\mathcal{A}_1$ have been defined in \cref{eq:Acaldef,eq:Acal1def}, respectively.

Combining all contributions to $E^{(0,2)}+E^{(2,0)}$ and using \cref{eq:rhoodef,eq:Wdef,eq:Acaldef,eq:Acal1def} yields
\begin{align}
    E^{(0,2)}+E^{(2,0)} = \frac{(Ze)^2}{4\pi\Roo} \frac{(\log b)^3}{16\pi \Li{3}{-b}^2}\mathcal{D}(b)\sum_{\ell>0}\beta_\ell^2
    \label{eq:E0220}
\end{align}
with
\begin{align}
    \mathcal{D}(x) = & \Li{3}{-x}- \Big(1+\log(1+x)\Big)\Li{2}{-x}-S_{1,2}(-x) \notag\\
    &+ \frac{\mathcal{A}(x)}{\mathcal{W}(x)}\Big(\Li{2}{-x}\Li{3}{-x}+\log(1+x)\Li{4}{-x}\Big)\notag\\
    &-\frac{\mathcal{A}_1(x)}{2\mathcal{W}(x)}\Big(\Li{3}{-x}^2+\log(1+x)\Li{5}{-x}\Big)\,.
    \label{eq:Dcaldef}    
\end{align}
In the hard-shell limit $a \to 0$, the leading order terms cancel between the geometric contribution and the monopole corrections, leaving $\mathcal{D}(b)= (\log b)^2/2$ rather than $\mathcal{O}(\log^3 b)$. 
The asymptotic behavior of $E^{(0,2)}+E^{(2,0)}$ is therefore $(E^{(0,2)}+E^{(2,0)})/( \sum\beta_\ell^2 E^{(0,0)})\sim a/\Roo$, implying that at $a= 0$, the entire deformation energy is carried by $E^{(1,1)}$.

\subsubsection{Evaluation of $E^{(1,1)}$}

For $E^{(1,1)}$, both charge density factors in \cref{eq:Eab} take the form
\begin{align}
    \rho^{(1)}(r,\Omega) &= - \rhoo \Roo f'(r) \sum_{l} \beta_{l} \sphY{l}{0}{\Omega}\,.
\end{align}
(For the second $\rho^{(1)}$ factor, we call the summation variable $l'$.) 
Orthogonality of the spherical harmonics then enforces $l=l'=\ell$ and $m=0$, so that all mixed products $\beta_l\beta_{l'}$, $l\neq l'$ vanish.
The remaining radial integrals can now be reduced to
\begin{align}
    E^{(1,1)}
    =& \Roo^2\rhoo^2 \sum_{\ell > 0} \frac{ \beta_\ell ^2}{2\ell+1}\int_0^\infty \dr \, r^{1-\ell} f'(r) F_{2+\ell}^{(1)}(r)
\end{align}
with $F_{2+\ell}^{(1)}(r)$ from \cref{eq:Fnkr}.
$F_{2+\ell}^{(1)}(r)$ can be reduced by integration by parts,
\begin{align}
    F_{2+\ell}^{(1)}(r) 
    = r^{2+\ell} f(r) - (2+ \ell) F_{1+\ell}^{(0)}(r)\,.
\end{align}
After the substitution $t(s)=e^{-\frac{s-\Roo}{a}}$,
\begin{align}
    F_{1+\ell}^{(0)}(r)
    =\int_0^r \ds \frac{s^{1+\ell}}{1+e^{\frac{s-\Roo}{a}}}
    =
    -a \int_{t(0)}^{t(r)} \mathrm{d}t \frac{(\Roo-a\log t)^{1+\ell}}{1+t}\,.
\end{align}
Using the binomial theorem to expand the polynomial, the integral can be evaluated and yields
\begin{align}
    F_{1+\ell}^{(0)}(r)
    &=
    a \sum_{n=0}^{\ell+1}
    \begin{pmatrix}
        \ell+1\\
        n
    \end{pmatrix}
    \Roo^{\ell+1-n}(-a)^n
    \Bigl[
        J_n(t_0)-J_n(t_r)
    \Bigr],
\end{align}
where we have defined
\begin{align}
    J_n(t)
    \equiv&
    \int^t \ds\frac{\left(\log s\right)^n}{1+s}
    =
    (\log t)^n\log(1+t)
    +
    \sum_{j=1}^{n}
    (-1)^{j+1}
    \frac{n!}{(n-j)!}
    (\log t)^{n-j}
    \Li{j+1}{-t}.
    \label{eq:Jk}
\end{align}
Since $n$ runs only up to $\ell+1$, the highest-order polylogarithm appearing is $\Li{\ell+2}{-t}$.
Substituting \cref{eq:Jk} into the expression for \smash{\(F_{1+\ell}^{(0)}(r)\)} and subsequently into the one for \smash{\(E^{(1,1)}\)}, and using the explicit expression for $\rhoo$ from \cref{eq:rhoodef} yields
\begin{align}
    E^{(1,1)}
    &= \frac{(Ze)^2}{64\pi^2\Roo}
    \frac{(\log b)^3 }{\Li{3}{-b}^2}
    \sum_{\ell>0}
    \frac{\beta_\ell^2}{2\ell+1} \mathcal{B}_\ell (b) \,,
    \label{eq:E11}
\end{align}
where
\begin{align}
    \mathcal{B}_\ell (b) = b\int_0^1 \mathrm{d}t\,
    \left(
        \frac{\log t}{1+bt}
    \right)^2
    \Bigg[
    \frac{bt\log t}{1+bt}
    +
    (2+\ell)
    \left(
        \frac{\log b}{\log t}
    \right)^{\ell+1}
    \sum_{n=0}^{\ell+1}
    (-1)^{\ell+1+n}
    \begin{pmatrix}
        \ell+1\\
        n
    \end{pmatrix}
    \frac{
        J_n(b)-J_n(bt)
    }{
        (\log b)^n
    }
    \Bigg].
    \label{eq:Blcaldef}
\end{align}
For the physically relevant regime $b = e^{\Roo/a}\gg 1$, the alternating terms in the inner sum of \cref{eq:Blcaldef} partially cancel. 
It is therefore beneficial to evaluate the polylogarithms at extended working precision to preserve numerical accuracy.

In the hard-shell limit $a\to 0$, $\mathcal{B}_\ell(b)\to (\log b)^3/2$ and the multipolarity dependence of $E^{(1,1)}$ resides entirely in the explicit $1/(2\ell+1)$ factor in \cref{eq:E11}. 
The contributions $E^{(0,2)}+E^{(2,0)}$ and $E^{(1,1)}$ therefore behave differently to linear order in $a$:
\begin{align}
    \frac{E^{(1,1)}}{E^{(0,0)}}
    &= \frac{15}{8\pi}\sum_{\ell>0}\beta_\ell^2\left(\frac{1}{2\ell+1}-\frac{a}{\Roo}\right)
      + \mathcal{O}\Big(\beta^2 \frac{a^2}{\Roo^2}\Big) \,,
    \qquad
    \frac{E^{(0,2)}+E^{(2,0)}}{E^{(0,0)}}
    = \frac{15}{8\pi}\frac{a}{\Roo}\sum_{\ell>0}\beta_\ell^2
      + \mathcal{O}\Big(\beta^2 \frac{a^2}{\Roo^2}\Big) \,.
\end{align}
Thus $E^{(1,1)}$ dominates as $a\to0$, while $E^{(0,2)}+E^{(2,0)}$ cancels the
$\mathcal{O}(a/\Roo)$ correction to $E^{(1,1)}$, so that
\begin{align}
    \frac{E^{(0,2)}+E^{(1,1)}+E^{(2,0)}}{E^{(0,0)}}
    = \frac{15}{8\pi}\sum_{\ell>0}\frac{\beta_\ell^2}{2\ell+1}
      + \mathcal{O}\Big(\beta^2 \frac{a^2}{\Roo^2}\Big) \,.
    \label{eq:E2dE0}
\end{align}
\vspace{1ex}\noindent
In summary,
\begin{align}
    E_{\rm C} &= E^{(0,0)} + E^{(0,2)} + E^{(1,1)} + E^{(2,0)} + \mathcal{O}(\beta^3)\notag \\
    &= E^{(0,0)} + \sum_{k>0} \beta_k^2 \mathcal{E}_k^{(2)}+ \mathcal{O}(\beta^3)
    \,,
    \label{eq:EC_summary}
\end{align}
where
\begin{align}
    \label{eq:Ecaldef}
    E^{(0,0)}=\frac{(Ze)^2}{4\pi \Roo}\frac{3}{5}\mathcal{C}(b)\,,
    \qquad
    \mathcal{E}_k^{(2)}=
    \frac{(Ze)^2}{4\pi \Roo} \frac{(\log b)^3}{16\pi \Li{3}{-b}^2} \left[\mathcal{D}(b)
    + 
    \frac{\mathcal{B}_k (b)}{2k+1}\right]
\end{align}
with the auxiliary functions defined in \cref{eq:Ccaldef,eq:Dcaldef,eq:Blcaldef}.

\subsubsection{Relation to the Liquid-Drop Model}
\label{app:scheme-dependence}

We note that $\mathcal{E}^{(2)}_k$ in \cref{eq:Ecaldef} is positive, in contrast to the prediction of the volume-conserving liquid-drop model, which is $-\beta_2^2/(4\pi)$ for a quadrupole deformation~\cite{BohrMottelsonVol2,HasseMyers}. 
We now discuss the origin of this qualitative discrepancy.

We adopt, in this subsection only, the conventional parametrization in which the radius parameter itself carries the deformation dependence $R=R(\beta)$, rather than keeping the second-order corrections to $\Roo$ and $\rhoo$ explicit.
This is equivalent to our parametrization through $\mathcal{O}(\beta^2)$, as follows from \cref{eq:rho_exp,eq:Roo_exp,eq:rho_exp_explicit}.
Also, in this subsection only, we set $a= 0$. 

At $a=0$ the nuclear volume and mean-square charge radius are
\begin{align}
    \label{eq:tilVtilchrsqdef}
    \tilde{V} &= \frac{4\pi}{3} R^3 \Big(1 + \frac{3}{4\pi}\sum_\ell \beta_\ell^2 \Big) \,, \qquad
    \tilchrsq = \frac{3}{5} R^2 \Big(1 + \frac{7}{4\pi}\sum_\ell \beta_\ell^2 \Big)\,,
\end{align}
These are the uncorrected geometric relations. 
The $\mathcal{O}(\beta^2)$ term in $\tilchrsq$ is precisely what the monopole corrections of Appendix~\ref{app:rhoRo} cancel in our scheme. 
In the liquid-drop model the volume $\tilde{V}$ is held fixed (by adjusting $R$) as the deformation is switched on, whereas we hold the total mean-square charge radius $\chrsq$ fixed. 
The reason for our choice is that $\chrsq$ is measured for both clock states, whereas the nuclear volume is not, see \cref{tab:schemes}.
The total charge $Ze$ always has to be fixed. 

Regarding $E_{\rm C}$ as a function of $R$ and $\beta_\ell^2$ at fixed $Ze$, the derivative along a curve of constant $X = \langle r^2 \rangle$ or constant $X = \tilde{V}$ is
\begin{align}
    \label{eq:constraintchange}
    \left.\frac{\mathrm{d} E_{\rm C}}{\mathrm{d}\beta_\ell^2} \right\vert_X
    = \left.\frac{\partial E_{\rm C}}{\partial \beta_\ell^2}\right\vert_R 
    + \left.\frac{\partial E_{\rm C}}{\partial R}\right\vert_{\beta_\ell^2} 
    \left.\frac{\partial R}{\partial \beta_\ell^2}\right\vert_X\,.
\end{align}
Taking the difference between the expressions for $X=\langle r^2\rangle$ and $X=\tilde{V}$, the $\partial E_{\rm C}/\partial \beta_\ell^2\vert_{\Roo}$ terms cancel, leaving
\begin{align}
    \left.\frac{\mathrm{d} E_{\rm C}}{\mathrm{d} \beta_\ell^2}\right\vert_{Ze,\chrsq} 
    - \left.\frac{\mathrm{d} E_{\rm C}}{\mathrm{d} \beta_\ell^2}\right\vert_{Ze,V} 
    &=
    \left.\frac{\partial E_{\rm C}}{\partial R}\right\vert_{\beta_\ell^2} \bigg(\!\left. \frac{\partial R}{\partial \beta_\ell^2}\right\vert_{\chrsq}
    - \left.\frac{\partial R}{\partial \beta_\ell^2}\right\vert_V \bigg)
\end{align}
The two terms in parentheses follow from \cref{eq:tilVtilchrsqdef} by implicit differentiation, giving at leading order in $\beta_l^2$,
\begin{align}
    \left. \frac{\partial R}{\partial \beta_\ell^2}\right\vert_{\tilchrsq} 
    = -\left.\frac{\partial \tilchrsq}{\partial \beta_\ell^2}\right\vert_{\Roo}
      \bigg(\!\left.\frac{\partial \tilchrsq}{\partial \Roo}\right\vert_{\beta_\ell^2}\bigg)^{-1}
    = -\frac{7R}{8\pi}\,,
    \label{eq:dR-dbeta-r2const}
    \\
    \left. \frac{\partial R}{\partial  \beta_\ell^2}\right\vert_{\tilde{V}} 
    = -\left.\frac{\partial \tilde{V}}{\partial \beta_\ell^2}\right\vert_{\Roo}
      \bigg(\!\left.\frac{\partial \tilde{V}}{\partial \Roo}\right\vert_{\beta_\ell^2}\bigg)^{-1}
    = -\frac{R}{4\pi}\,.
    \label{eq:dR-dbeta-Vconst}
\end{align}
Both expressions are independent of $\ell$, since $\tilde{V}$ and $\tilchrsq$ depend on the deformation only through $\sum_\ell \beta_\ell^2$.
\Cref{eq:dR-dbeta-r2const} coincides with the second-order correction \smash{$\Roii$} of \cref{eq:RoiidRoo}.
Together with \smash{$\partial E_{\rm C}/\partial R\vert_{Ze, \beta_{\ell}^2}=-E^{(0,0)}/R$} (see \cref{eq:Ecaldef}), we obtain
\begin{align}
    \left.\frac{\mathrm{d} E_{\rm C}}{\mathrm{d} \beta_\ell^2}\right\vert_{Ze,\chrsq} 
    - \left.\frac{\mathrm{d} E_{\rm C}}{\mathrm{d} \beta_\ell^2}\right\vert_{Ze,V} 
    &= \frac{5}{8\pi}E^{(0,0)}
    \,.
\end{align}
This shows that a distribution deformed at fixed mean-square radius is more compressed than one deformed at fixed volume. 
The reason is that $\chrsq$ responds more strongly to deformation than $\tilde{V}$ does ($7/4\pi$ against $3/4\pi$, see \cref{eq:tilVtilchrsqdef}).
The associated increase in Coulomb energy outweighs the reduction produced by the deformation itself, leading to the positive sign in \cref{eq:E2dE0}.

Adding this shift to the volume-conserving result from Refs.~\cite{BohrMottelsonVol2,HasseMyers},
\begin{align}
    \label{eq:E2dE0Vcons}
    \left. \frac{\mathcal{E}^{(2)}}{E^{(0,0)}}\right\vert_{Ze, \tilde{V}} = -\frac{5}{4\pi} \frac{\ell-1}{2\ell +1}\,,
\end{align}
reproduces the limit $a\to 0$ of \cref{eq:E2dE0}:
\begin{align}
\label{eq:E2dE0translation}
    \left. \frac{\mathcal{E}^{(2)}}{E^{(0,0)}}\right\vert_{Ze, \chrsq} 
    = \frac{5}{8\pi} \left(1-\frac{2(\ell-1)}{2\ell+1}\right)
    = \frac{15}{8\pi}\frac{1}{2\ell +1}\,.
\end{align}
The two schemes are compared in Table~\ref{tab:schemes}.

\begin{table}[t]
    \centering
    \setlength{\tabcolsep}{12pt}
    \begin{tabular}{l l l}
      \toprule
                                          & Liquid-Drop          & This Work          \\
      \midrule
        Held fixed                        & $Ze$, $\tilde{V}$    & $Ze$, $\chrsq$     \\
        $\delta \chrsq$                   & predicted            & input from experiment \\
        $\delta \tilde{V}$                & zero by construction & predicted           \\
        $\mathcal{E}^{(2)}_\ell/E^{(0,0)}$  & $-\frac{5}{4\pi} \frac{\ell - 1}{2\ell +1}$ (\cref{eq:E2dE0Vcons}) & $\frac{15}{8\pi} \frac{1}{2\ell+1}$ (\cref{eq:E2dE0,eq:E2dE0translation})\\
        \bottomrule
    \end{tabular}
    \caption{Comparison of the constant volume liquid-drop treatment with the scheme described in this Appendix.}
    \label{tab:schemes}
\end{table}

\subsection{Relation to Observables}
\label{app:relation_to_observables}

\Cref{eq:EC_summary} expresses the Coulomb energy through the parameters of the Fermi model, $\rho_0$, $\Roo$, $a$ and $\beta_k$. 
We now relate these to measurable quantities. 
In the following, we denote the difference between a property $X$ of the isomer and the corresponding property of the ground state as:
\begin{align}
    \label{eq:dX}
    \delta X \equiv X^\m - X^\gs
\end{align}

The available observables are the nuclear charge $Ze$, the mean-square charge radius $\chrsq$, and the rank-$k$ spectroscopic multipole moments \smash{$\Qsphspec{k}{\Isp}$} of a given nuclear state with spin $I$.
The first two fix the monopole sector: for fixed surface diffuseness $a$, \cref{eq:rhoodef,eq:Roodef} can be solved for $\rhoo$ and $\Roo$.
The diffuseness is poorly constrained by the available data. 
We fix it to the value
\begin{align}
    a^\gs = a^\m = \SI{0.5}{fm}
\end{align}
typical of heavy nuclei~\cite{DeVriesDeVries}. 
We discuss the effect of the surface diffuseness in more detail in \cref{app:K}.
Since the two clock states are band heads of different rotational bands, we allow them independent radial parameters $\Roo$ and deformation parameters $\beta_k$.

Absolute charge radii are obtained from electron scattering and muonic-atom data~\cite{Angeli:2013epw}, whereas the isomer shift $\delta \chrsq$ is measured by laser spectroscopy~\cite{Thielking:2017qet,Safronova:2018fue} and is correspondingly known far more precisely (see \cref{tab:expinput}).
Thanks to the way we have constructed the corrections $\Roii$ and $\rhoii$ in \cref{app:rhoRo}, the isomer shift $\delta \langle r^2\rangle$ is independent of the deformation parameters $\beta_k$, and
\begin{align}
    \delta \chrsq = \frac{\partial \chrsq}{\partial \Roo} \delta \Roo\,.
\end{align}
The deformation parameters are determined by the intrinsic multipole moments $\Qsph{k}$, which are related to the measured spectroscopic moments through the axial rotor model (see \cref{sec:rotorspec,sec:particle-rotor,app:rotormultipoles}).
Since the deformation enters $E_{\rm C}$ quadratically, only the leading-order relation between $\beta_k$ and $\Qsph{k}$ is needed. $\mathcal{O}(\beta^2)$ corrections to $\Qsph{k}$ would affect $E_{\rm C}$ only at $\mathcal{O}(\beta^3)$, which we neglect.
We define the intrinsic multipole moments in the same spherical-harmonic basis as the Fermi density in \cref{eq:rho_WS_app}, the Laplace expansion in \cref{eq:LaplacExp}, and the multipole Hamiltonian of \cref{eq:MultipolExp}, so that no rank-dependent normalization factors appear. 
We thus write
\begin{align}
    \label{eq:QkCk}
    \Qsph{k>0} = \int \drth \rho(\rvec) r^k \sphY{k}{0}{\Omega}
    = \beta_k C_k + \mathcal O(\beta^2)\,,
\end{align}
with
\begin{align}
    \label{eq:Ckdef}
    C_k = -(k+2)! \rho_0 a^{k+3} \Li{k+2}{-b} \log (b) 
    = (k+2)!\frac{ Ze}{8\pi }  a^{k}\frac{\Li{k+2}{-b}\log (b)}{\Li{3}{-b}} \,,
\end{align}
where we have used \cref{eq:Ilk}. 
At $a= 0$, $C_k\to 3/(4\pi) Ze \Roo^k$.
\cref{eq:QkCk} can then be combined with \cref{eq:QspecQint,eq:QcartsQsphs} to obtain the relation between the deformation parameters $\beta_k$ and the Cartesian spectroscopic moments commonly quoted in the experimental literature.

\subsection{Expressing the Sensitivity Factor in Terms of Observables}
\label{app:K}

Having related the model parameters to observables, we now express the new physics sensitivity factor $\KNP$ (\cref{eq:KNPsectorsum}) in terms of observables as well, and we quantify the modeling approximations and experimental uncertainties that enter it.

Linearizing the Coulomb energy in the small differences between the two clock states gives
\begin{align}
    \label{eq:Kbdexp}
    \KNP \approx \KNP_{\chrsq} + \KNP_a + \sum_k \KNP_{\Qsph{k}} + \mathcal{O}(\delta^2, \beta^3)\,,
\end{align}
where
\begin{align}
    K_{X} = \frac{1}{\Delta E} \frac{\partial E_{\rm C}}{\partial X}\delta X\,, \qquad X \in \lbrace \chrsq, a, \lbrace \Qsph{k}\rbrace_k\rbrace\,.
\end{align}
The derivatives are taken at fixed $Ze$, the observables $\chrsq$, $\Qsph{k}$, $a$ being the independent variables. 
The model parameters $\rhoo$, $\Roo$, $\beta_k$ follow from \cref{eq:rhoodef,eq:Roodef,eq:QkCk} (see also Table~\ref{tab:derivedinput} for an overview of the parameter interdependence).
Using $\partial \Roo/\partial \chrsq \vert_a = 5/(6\Roo)$ and $\partial \Roo/\partial a \vert_{\chrsq} = -7\pi^2 a/(3\Roo)$ from \cref{eq:rsqlo}, together with \cref{eq:dE0dR0,eq:dE0da,eq:Ckdef},
\begin{align}
    \frac{\partial E_{\rm C}}{\partial \langle r^2\rangle} 
    &= \;
    \frac{5}{6\Roo}\frac{\partial E^{(0,0)}}{\partial \Roo} + \mathcal{O}\left(\beta^2\right)
    = \frac{5}{6\Roo}\frac{(Ze)^2}{16\pi a^2 \Li{3}{-b}^2}\bigg(
    \mathcal{A}_1(b) - 2\mathcal{A}(b) \frac{\Li{2}{-b}}{\Li{3}{-b}} \bigg)
    \,,
    \label{eq:dEC-dr2}\\
    \frac{\partial E_{\rm C}}{\partial \Qsph{k}}
    &=  \;
    2 \frac{\Qsph{k}}{C_k^2}\mathcal{E}_k^{(2)} + \mathcal{O}(\beta^2) 
    = \frac{2 \Qsph{k}}{(k+2)!^2 a^{2k+1} \Li{2+k}{-b}^2}\bigg(\mathcal{D}(b)+\frac{\mathcal{B}_k(b)}{2k+1}\bigg) + \mathcal{O}(\beta^2)
    \label{eq:dEC-dQk} \\
    \frac{\partial E_{\rm C}}{\partial a}
    &=  \;
    \frac{\partial E^{(0,0)}}{\partial a } -\frac{7\pi^2 a}{3\Roo}\frac{\partial E^{(0,0)}}{\partial \Roo} + \mathcal{O}\left(\beta^2\right)  \\
    &=  -\frac{(Ze)^2}{16\pi a^2}\bigg(\frac{\mathcal{A}(b)+\log(b) \mathcal{A}_1(b)}{\Li{3}{-b}^2}- \frac{2\log(b)\mathcal{A}(b)\Li{2}{-b}}{\Li{3}{-b}^3} \bigg)\notag \\
    &\quad -\frac{7\pi^2 a}{3\Roo} \frac{(Ze)^2}{16\pi a^2 \Li{3}{-b}^2}\bigg(
    \mathcal{A}_1(b) - 2\mathcal{A}(b) \frac{\Li{2}{-b}}{\Li{3}{-b}}    \bigg)  \,,
    \label{eq:dEC-da}
\end{align}
with $\mathcal{A}$ and $\mathcal{A}_1$ from \cref{eq:Acaldef,eq:Acal1def}. 
All explicit $\Roo$-dependence cancels in $\partial E_{\rm C}/\partial \Qsph{k}$, which depends on $\Roo$ only through $b=e^{\Roo/a}$. 
The multipole moments enter $\KNP$ through the deformation parameters $\beta_k$, which are related to $\Qsph{k}$ by \cref{eq:QkCk}.
In this linearized form of $\KNP$, the $\beta_k$ of both states are defined with respect to a common $C_k$, evaluated at the ground state radius, so that $\delta(\beta_k^2)$ isolates the change in deformation. 
The radius shift is carried separately by $\delta \chrsq$.

The intrinsic multipole moments $\Qsph{k}$ appearing in \cref{eq:dEC-dr2,eq:dEC-dQk,eq:dEC-da} can be related to the measured Cartesian ground state quadrupole moment $\Qcartzzspec{\gs}$ and quadrupole ratio $\RQ \equiv \Qcartzzspec{\m}/\Qcartzzspec{\gs}$ (see \cref{tab:expinput}) using the rotor-model to project the intrinsic multipole moments $\Qsph{k}$ onto the measured spectroscopic moments (see \cref{eq:QsphsQsphNuclearClock}) and relating these to the Cartesian multipole moments (see \cref{eq:QcartsQsphs,eq:Qcarts24def}).

Writing each input as a fractional deviation from its value in \cref{tab:expinput}, and setting $\delta \Qsph{4}, \delta a=0$ at the reference point, the linearized form of $K$ in \cref{eq:Kbdexp} gives
\begin{align}
    \label{eq:Kscaling}
    \KNP \simeq  \; 7.9 \times 10^3 
    & - 3.8 \times 10^4 \left(\frac{\chrsq}{(5.7557~\mathrm{fm})^2}-1 \right)
     - 1.8 \times 10^4 \left(\frac{\delta \chrsq}{0.0105~\mathrm{fm}^2}-1\right)
    \notag \\
    & + 5.2 \times 10^4 \left(\frac{\Qcartzzspec{\gs}}{3.11~e\mathrm{b}}-1 \right)
     + 1.5 \times 10^6 \left(\frac{\RQ}{0.57003}-1 \right)
    \notag \\
    &+\ldots
\end{align}
The ellipsis denotes higher-order contributions.
The numerical coefficients multiplying the fractional deviations $X/\bar{X}-1$ are the logarithmic sensitivities $\partial \KNP/\partial \log X=\bar{X} \partial \KNP/\partial X$, listed in \cref{tab:K_sigprop}.
The fact that they exceed $\KNP$ itself reflects the monopole-quadrupole cancellation discussed in the main text. 

\paragraph*{Hexadecapole Moment}
No analogue of the quadrupole ratio $\mathcal{R}_Q$ exists at rank four since the $I=3/2$ isomer cannot have a spectroscopic hexadecapole moment $\Qcartzzzzspec{\m}$, leaving its intrinsic moment $\Qsph{4}^\Mband$ unconstrained.
We therefore parametrize its contribution to the sensitivity through $\delta \Qsph{4} \equiv (\Qsph{4}^\Mband - \Qsph{4}^\Gband)$, as in \cref{eq:K_result}.
Relating the multipole moments to the deformation parameters introduces an additional complication: the leading-order relation $\Qsph{4}=\beta_4 C_4$ in \cref{eq:QkCk} can receive significant higher-order corrections.
For the benchmark intrinsic hexadecapole moment
$\Qsph{4}^\Gband=1.56~e\mathrm{b}^2$ inferred from Ref.~\cite{CEBemis_1988}, the leading-order relation gives $\beta_4^\gs\simeq 0.246$, compared with $\beta_4^\gs \simeq 0.171$ obtained by solving for the deformation parameters using the full Fermi charge distribution. 
Thus, assuming equal intrinsic hexadecapole moments for the ground state and isomer, such that $\delta \Qsph{4}=0$, does not generally imply equal deformation parameters, $\delta \beta_4=0$.
Constraining $\beta_4^\m$ would require an observable other than $\Qcartzzzzspec{\m}$ (which vanishes), such as a rank-four spectroscopic moment of a higher-spin member of the isomer's rotational band, or a Coulomb-excitation measurement such as the one performed by Ref.~\cite{CEBemis_1988} for the ground-state band.

\subsection{Evaluation Procedures and Uncertainties}

Equation~\eqref{eq:KNPsectorsum} may be evaluated either through the linearized form of \cref{eq:Kbdexp}, or from the difference of the two states' Coulomb energies without linearizing in the parameter differences $\delta X$ (see \cref{eq:dX}),
\begin{align}
    \label{eq:Kexpbdiff}
    K = \frac{E_{\rm C}^\m - E_{\rm C}^\gs}{\Delta E}\,, \qquad
    E_{\rm C}^{\gs, \m} = E^{(0,0)}(\Roo^{\gs, \m}, a^{\gs, \m}) + \sum_k \left(\beta_k^{\gs, \m}\right)^2 \mathcal{E}^{(2)}_k (\Roo^{\gs, \m}, a^{\gs, \m})\,,
\end{align}
with $E^{(0,0)}$ and $\mathcal{E}^{(2)}$ from \cref{eq:EC_summary}. 
Both forms use the same $\mathcal{O}(\beta^2)$ Coulomb energy. 
\cref{eq:Kexpbdiff} additionally retains the $\mathcal{O}(\delta ^2)$ terms, namely $(\delta\beta_k)^2$ and the variation of $\mathcal{E}_k^{(2)}$ between the two states. 
To quantify the approximations, we therefore evaluate $K$ along six routes, obtained by combining
\begin{itemize}
    \item the linearised form  (\cref{eq:Kbdexp}), or the exact difference (\cref{eq:Kexpbdiff})
    \item  the evaluation of $E_{\rm C}$  through the semi-analytic forms  (\cref{eq:EC_summary}), by direct numerical integration of the radial integrals (\cref{eq:Eab}) or without any expansion in $\beta$, using the full Fermi charge density (\cref{eq:rho_WS_app}).
\end{itemize}
In the last case the model parameters are obtained by solving $\{\chrsq, \Qsph{2}, \Qsph{4}\}$ for $\{\Roo, \beta_2, \beta_4\}$ nonperturbatively, rather than through \cref{eq:Roodef,eq:QkCk}.
The results are collected in Table~\ref{tab:Kroutes} for three values of the intrinsic ground state hexadecapole moment, assuming $\delta \Qsph{4}=0$ throughout.

At vanishing deformation, all six routes agree to $2\times 10^{-4}$, validating the monopole sector. 
The semi-analytic and numerically integrated evaluations of the finite Coulomb energy difference agree to better than $10^{-7}$ in relative terms, confirming \cref{eq:E00,eq:E0220,eq:E11}. 
The corresponding linearized evaluations agree at the level of $2\times 10^{-4}$.

The residual differences quantify the two theory approximations. 
For vanishing intrinsic hexadecapole moment, linearizing in the state differences $\delta X$ shifts $\KNP$ by $\approx 2\%$ within the quadratic approximation.
Dispensing with the $\beta$-expansion altogether changes $\KNP$ by approximately $-8\%$ relative to the quadratic result, providing an estimate of the $\mathcal{O}(\beta^2)$ truncation uncertainty in the quoted quadrupole-sector sensitivity $\KNP_{\leq 2} \sim 8\times 10^3$.

The dependence on the intrinsic hexadecapole moment is considerably stronger in the full-density calculation. 
While the quadratic approximation changes only weakly with $\Qsph{4}^\Gband$, the full-density result decreases from $\KNP=7.4\times 10^3$ at $\Qsph{4}^\Gband=0$ to $1.4\times 10^3$ at the benchmark motivated by Ref.~\cite{CEBemis_1988}, even with $\delta \Qsph{4}=0$.
This reflects higher-order deformation effects both in the Coulomb energy and in the relation between intrinsic moments and deformation parameters.

\begin{table}[t]
    \centering
    \setlength{\tabcolsep}{6pt}
    \begin{tabular}{l | cc | cc | cc}
    \hline
    & \multicolumn{6}{c}{$\KNP/10^3$, assuming $\Qsph{4}^\Gband=\Qsph{4}^\Mband\equiv \Qsph{4}$} \\
    \cline{2-7}
      & \multicolumn{2}{c|}{$\Qsph{4}=0~e\mathrm{b}^2$}
      & \multicolumn{2}{c|}{$\Qsph{4}=0.39~e\mathrm{b}^2$}
      & \multicolumn{2}{c}{$\Qsph{4}=1.56~e\mathrm{b}^2$}\\
      & Lin. in $\delta$ & Exact in $\delta$ & Lin. in $\delta$ & Exact in $\delta$ & Lin. in $\delta$ & Exact in $\delta$\\
    \hline
    Semi-analytic
      & 7.9 & 8.1 & 7.9 & 8.1 & 7.9 & 8.0 \\
    Num. radial integration
      & 7.9 & 8.1 & 7.9 & 8.1 & 7.9 & 8.0 \\
    Full Fermi density
      & 7.7 & 7.4 & 6.3 & 6.1 & 1.7 & 1.4 \\
    \hline
    \end{tabular}
    \caption{The sensitivity factor $\KNP$, in units of $10^3$, for three values of the intrinsic spherical hexadecapole moment $\Qsph{4}^\Gband=\Qsph{4}^\Mband\equiv\Qsph{4}$, in units of $e\mathrm{b}^2$.
    The final benchmark corresponds to the ground-band moment inferred by Ref.~\cite{CEBemis_1988}.
    ``Lin. in $\delta$'' denotes linearization in the state differences $\delta X$, while ``Exact in $\delta$'' retains their nonlinear effects.
    The first two rows compare the quadratic Coulomb-energy expansion evaluated semi-analytically and by numerical radial integration. 
    The final row uses the full deformed Fermi charge distribution with nonperturbatively determined parameters.}
    \label{tab:Kroutes}
\end{table}

\section{Halo--Model Calculation}
\label{sec:halo-calc}

In this appendix, we derive the spin--orbit contributions to the mean-square charge radius and to the spectroscopic quadrupole and hexadecapole moments quoted in \cref{subsec:halo-model}.
We use the one-neutron wave function $\Phi_{Ilm,\alpha}=f(r)\,\Omega_{Ilm,\alpha}(\theta,\phi)$ of \cref{eq:halo-wavefunction} and the spin--orbit charge density of \cref{eq:halo-charge-density}~\cite{Ong:2010gf}.
Repeated spinor indices are summed, and we suppress the labels $Ilm$ where no confusion can arise.
With Condon--Shortley conventions, the angular spinors of the stretched states, $m=I$, are
\begin{align}
    \Omega_{\frac52,2,\frac52}
    &=\begin{pmatrix}Y_{22}\\0\end{pmatrix},
    &
    \Omega_{\frac32,2,\frac32}
    &=\frac{1}{\sqrt{5}}\begin{pmatrix}-Y_{21}\\2\,Y_{22}\end{pmatrix}.
    \label{eq:halo-app-stretched-spinors}
\end{align}
Since the \iso{Th}{228} core is spinless, the nuclear spin is $\vec{I}=\vec{L}+\vec{S}$ with $\vec{S}=\vec{\sigma}/2$, so that $\vec{\sigma}\cdot\vec{L}=\vec{I}^2-\vec{L}^2-\vec{S}^2$.
The angular spinors are therefore eigenfunctions of $\vec{\sigma}\cdot\vec{L}$, with an eigenvalue that depends on $I$ and $l$ but not on $m$,
\begin{align}
    \vec{\sigma}\cdot\vec{L}\,\Omega_{Ilm}
    =\langle\vec{\sigma}\cdot\vec{L}\rangle\,\Omega_{Ilm},
    \qquad
    \langle\vec{\sigma}\cdot\vec{L}\rangle
    =I(I+1)-l(l+1)-\frac{3}{4}
    =\begin{cases}
        \phantom{-}2, & I=5/2,\\
        -3, & I=3/2.
    \end{cases}
    \label{eq:halo-app-sigma-l-eigenvalues}
\end{align}
The quantities in \cref{eq:halo-charge-radius-def,eq:halo-quadrupole-def,eq:halo-hexadecapole-def} are moments of $\rho_{\mathrm{SO}}$ with weight $r^n\,W(x)$, $x=\cos\theta$: the radius has $n=2$ and $W=1$, the quadrupole moment $n=2$ and $W(x)=3x^2-1$, and the hexadecapole moment $n=4$ and $W(x)=35x^4-30x^2+3$, the latter two evaluated at $m=I$.
Inserting $\Phi_{Ilm}=f\,\Omega_{Ilm}$ into \cref{eq:halo-charge-density}, each such moment factorizes into the radial and angular integrals
\begin{equation}
\begin{aligned}
    \mathcal R_n
    &\equiv\int_0^\infty \mathrm{d}r\,r^nf(r)^2,
    \qquad
    A[W]\equiv\int \mathrm{d}\Omega\,W(x)\,\Omega^\dagger\Omega,
    \\
    C[W]
    &\equiv\int_0^\pi \mathrm{d}\theta\int_0^{2\pi}\mathrm{d}\phi\,W(x)\,
      i\vec{\sigma}_{\alpha\beta}\cdot\hat{\rvec}
      \left(
          \partial_\theta\Omega_\alpha^*\,\partial_\phi\Omega_\beta
          -\partial_\phi\Omega_\alpha^*\,\partial_\theta\Omega_\beta
      \right).
\end{aligned}
\label{eq:halo-app-master-integrals}
\end{equation}
Here $A[W]$ collects the terms with a radial derivative $f'$, whose spinor structure reduces to $\langle\vec{\sigma}\cdot\vec{L}\rangle\,\Omega^\dagger\Omega$ by $\rvec\times\vec{\nabla}=i\vec{L}$ and \cref{eq:halo-app-sigma-l-eigenvalues}, while $C[W]$ collects the angular derivatives of $\Omega$.
Integrating the former by parts in $r$, $\int_0^\infty\mathrm{d}r\,r^{n+1}ff'=-\frac{n+1}{2}\mathcal R_n$, gives
\begin{align}
    \frac{1}{Z}\int\mathrm{d}^3\rvec\,r^n\,W(\cos\theta)\,\rho_{\mathrm{SO}}[\Phi_{Ilm}](\rvec)
    =\frac{\mu_n}{Zm_n^2}\,\mathcal R_n
    \left[\frac{n+1}{2}\,\langle\vec{\sigma}\cdot\vec{L}\rangle\,A[W]+\frac{1}{2}\,C[W]\right].
    \label{eq:halo-app-master}
\end{align}
The normalization in \cref{eq:halo-wavefunction} fixes $\mathcal R_2=1$, so that the radius and the quadrupole moment do not depend on $f(r)$, while $\mathcal R_4=\tilde p\,\mathrm{fm}^2$ is the parameter of \cref{eq:halo-radial-moment}.
For the stretched states of \cref{eq:halo-app-stretched-spinors}, the $\phi$ integration is trivial, and the angular integrals reduce to
\begin{equation}
\begin{aligned}
    A^{(5/2)}[W]&=\frac{15}{16}\int_{-1}^{1}\mathrm{d}x\,W(x)\,(1-x^2)^2,
    &\quad
    C^{(5/2)}[W]&=-\frac{15}{2}\int_{-1}^{1}\mathrm{d}x\,W(x)\,x^2(1-x^2),
    \\
    A^{(3/2)}[W]&=\frac{3}{4}\int_{-1}^{1}\mathrm{d}x\,W(x)\,(1-x^2),
    &\quad
    C^{(3/2)}[W]&=\frac{3}{2}\int_{-1}^{1}\mathrm{d}x\,W(x)\,(2-3x^2),
\end{aligned}
\label{eq:halo-app-stretched-integrals}
\end{equation}
where the superscript labels the state.

\subsection{Mean-Square Charge Radius}

Inserting \cref{eq:halo-charge-density} into \cref{eq:halo-charge-radius-def} and integrating by parts, with $\vec{\nabla}r^2=2\rvec$ and $\rvec\times\vec{\nabla}=i\vec{L}$, yields the diagonal matrix element
\begin{align}
    \delta\langle r^2\rangle_{\mathrm{SO}}^{(I)}
    =\frac{\mu_n}{Zm_n^2}\,\langle\Phi_{Ilm}|\,\vec{\sigma}\cdot\vec{L}\,|\Phi_{Ilm}\rangle
    =\langle\vec{\sigma}\cdot\vec{L}\rangle\,\frac{\mu_n}{Zm_n^2}.
    \label{eq:halo-app-radius-sigma-l}
\end{align}
By \cref{eq:halo-app-sigma-l-eigenvalues}, this is independent of $m$, and with $\langle\vec{\sigma}\cdot\vec{L}\rangle=2$ and $-3$ it reproduces \cref{eq:delta-r2-halo-52-32,eq:halo-charge-radius-results}.

\subsection{Spectroscopic Quadrupole Moment}

For the quadrupole moment, \cref{eq:halo-quadrupole-def}, $n=2$ and $W(x)=3x^2-1$
With \cref{eq:halo-app-stretched-integrals} and $\mathcal R_2=1$, \cref{eq:halo-app-master} gives $q_{zz}^{(5/2)}=-2\,\mu_n/(Zm_n^2)$ and $q_{zz}^{(3/2)}=\frac{3}{5}\,\mu_n/(Zm_n^2)$, as quoted in \cref{subsec:halo-model}; their difference is \cref{eq:halo-quadrupole-results}.

These are matrix elements in the stretched states.
By the Wigner--Eckart theorem, \cref{eq:WignerEckart}, the diagonal matrix element of a rank-two operator in the substate $m$ is the stretched one multiplied by $[3m^2-I(I+1)]/[I(2I-1)]$, cf.\ \cref{eq:Hquadmm}
At $m=1/2$, this factor is $-4/5$ for $I=5/2$ and $-1$ for $I=3/2$, so the $m=1/2$ matrix elements differ by $-\frac{11}{5}\,\mu_n/(Zm_n^2)=+2.1\times10^{-3}\,\mathrm{fm}^2$.
Multiplied by $Z$, this gives $0.185\,\mathrm{fm}^2$, the halo--model value of Ref.~\cite{Caputo:2024doz}, see below \cref{eq:halo-quadrupole-ratio}.

\subsection{Spectroscopic Hexadecapole Moment}

For the hexadecapole moment, \cref{eq:halo-hexadecapole-def}, $n=4$ and $W(x)=35x^4-30x^2+3=8P_4(x)$, with the Legendre polynomial $P_4$,
With \cref{eq:halo-app-stretched-integrals}, \cref{eq:halo-app-master} gives $q_{zzzz}^{(5/2)}=\frac{24}{7}\,\mathcal R_4\,\mu_n/(Zm_n^2)$ and $q_{zzzz}^{(3/2)}=0$, as quoted in \cref{subsec:halo-model}.
The isomer moment vanishes for any $f(r)$, since $P_4$ is orthogonal to its angular integrands in \cref{eq:halo-app-stretched-integrals}, which are polynomials of degree two.
The difference, $\Delta q_{zzzz}=-\frac{24}{7}\,\mathcal R_4\,\mu_n/(Zm_n^2)$, gives \cref{eq:halo-hexadecapole-results} for $\mathcal R_4=\tilde p\,\mathrm{fm}^2$.

\bibliography{bib}
\end{document}